\documentclass[12pt]{iopart}

\usepackage{amsopn}
\usepackage{iopams}
\usepackage{setstack}
\usepackage{graphicx}
\usepackage[T1]{fontenc}
\usepackage[english]{babel}
\usepackage[pdfpagelabels,plainpages=false]{hyperref}   
\usepackage{amssymb}
\usepackage{braket}
\usepackage{bbm}
\usepackage{fancyhdr}
\usepackage{units}
\usepackage{dsfont}
\DeclareMathOperator{\argg}{arg}
\newcommand{\unity}{\ensuremath{\mathds{1}}}

\newcommand{\mpstm}[2]{\ensuremath{\mathcal{T}_{#1}^{#2}}}
\newcommand{\otm}[1]{\ensuremath{\mathcal{J}_{#1}}}
\newcommand{\cmpstm}[2]{\ensuremath{\mathcal{P}_{#1}^{#2}}}
\newcommand{\Fig}[1]{figure~\ref{#1}}
\newcommand{\Sec}[1]{section~\ref{#1}}
\newcommand{\rbra}[1]{\ensuremath{\left(#1\right|}}
\newcommand{\rket}[1]{\ensuremath{\left|#1\right)}}
\newcommand{\nrm}[1]{\ensuremath{\left\Vert #1 \right\Vert}}
\newcommand{\ic}{\ensuremath{{\rm i}}}
\newcommand{\ec}{\ensuremath{{\rm e}}}
\newcommand{\hilbert}{\ensuremath{\mathcal{H}}}
\newcommand{\eq}[1]{\eqref{#1}}
\newcommand{\skippart}[1]{}
\newcommand{\eqref}[1]{(\ref{#1})}

\def\DXXZ{100} 
\def\figpath{.} 
\def\one{\ensuremath{\hbox{$\mathrm I$\kern-.6em$\mathrm 1$}}}
\def\TMscale{1}

\DeclareMathOperator{\dist}{dist}
\DeclareMathOperator{\diam}{diam}

\begin{document}

\title{Transfer Matrices and Excitations with Matrix Product States}

 \author{V~Zauner$^{1}$, D~Draxler$^{1}$, L~Vanderstraeten$^{2}$, M~Degroote$^{2}$, J~Haegeman$^{2}$, M~M~Rams$^{1,3}$, V~Stojevic$^{4}$, N~Schuch$^{5}$ and F~Verstraete$^{1,2}$}
  \address{$^{1}$Vienna Center for Quantum Technology, University of Vienna, Boltzmanngasse 5, 1090 Wien, Austria}
 \address{$^{2}$Ghent University, Krijgslaan 281, 9000 Gent, Belgium}
 \address{$^{3}$Institute of Physics, Krak\'ow University of Technology, Podchorazych 2, 30-084 Krak\'ow, Poland}
 \address{$^{4}$London Centre for Nanotechnology, University College London, Gordon St., London, WC1H 0AH, United Kingdom}
 \address{$^{5}$Institut f\"ur Quanteninformation, RWTH Aachen University, D-52056 Aachen, Germany}
\ead{valentin.zauner@univie.ac.at}

\begin{abstract}
We investigate the relation between static correlation functions in the ground state of local quantum many-body Hamiltonians and the dispersion relations of the corresponding low energy excitations using the formalism of tensor network states. In particular, we show that the Matrix Product State Transfer Matrix (MPS-TM) -- a central object in the computation of static correlation functions -- provides important information about the location and magnitude of the minima of the low energy dispersion relation(s) and present supporting numerical data for one-dimensional lattice and continuum models as well as two-dimensional lattice models on a cylinder. We elaborate on the peculiar structure of the MPS-TM's eigenspectrum and give several arguments for the close relation between the structure of the low energy spectrum of the system and the form of static correlation functions. Finally, we discuss how the MPS-TM connects to the exact Quantum Transfer Matrix (QTM) of the model at zero temperature. We present a renormalization group argument for obtaining finite bond dimension approximations of MPS, which allows to reinterpret variational MPS techniques (such as the Density Matrix Renormalization Group) as an application of Wilson's Numerical Renormalization Group along the virtual (imaginary time) dimension of the system.
\end{abstract}

\noindent{Keywords: Strongly Correlated Systems, Static Correlations, Dispersion Relations, Transfer Matrices, Tensor Network States, Renormalization Group}

\pacs{03.65.-w,05.30.-d,05.10.Cc}
\submitto{\NJP}

\maketitle

\section{Introduction}
\label{s:intro}
Determining the vacuum of an interacting field theory or the ground state of a strongly interacting quantum system described by a local translational invariant Hamiltonian is one of the most fundamental and challenging tasks in quantum many-body physics. Once obtained -- possibly in some variational way -- how much information about the Hamiltonian is then encoded within the ground state? We will demonstrate that it is possible to extract many low-energy features of the Hamiltonian by just having access to the ground state. This is possible due to the Hamiltonian being a sum of (quasi-) local terms; this locality is the key to uncovering the mysteries of quantum many-body systems, such as the presence of a finite group velocity in quantum lattice systems, known as the Lieb-Robinson bound \cite{LR,Nachtergaele}, and the relation between the spectral gap and correlation length \cite{HastingsLR}. The latter result connects a single characteristic of the static correlation functions of the ground state to one 
particular excitation energy.


This work continues along this line by investigating to what extent information about the \emph{full} dispersion relations of the different elementary excitations of the model is encoded within the ground state and its correlations. Throughout the paper we assume translation-invariant Hamiltonians, such that excited states can always be characterized by momentum. Any statement regarding the spectrum of a Hamiltonian is to be interpreted up to an overall energy scaling and a constant energy shift. The shift is typically chosen such that the ground state energy $E_0=0$. The overall energy scale is represented by a characteristic velocity (e.g. the Lieb-Robinson velocity related to the norm of the Hamiltonian terms, or some spin-wave velocity) in the system.

In theory, the full dispersion relation can be reproduced from the ground state if the map between a local Hamiltonian and its corresponding ground state is bijective. For \textit{strictly} $n$-local Hamiltonians, i.e. Hamiltonians for which every term acts only on a finite number $n$ of neighboring sites, such a bijective relation is generically obtained. There the $n$-site reduced density matrices (RDMs) of ground states represent extreme points in the convex set of all possible $n$-site RDMs. The Hamiltonian can then be represented as a hyperplane in the space of such RDMs, and the energy will necessarily be minimized for an extreme point in this set. Each of these points uniquely determines an $n$-local parent Hamiltonian via the tangent space to the boundary at this point, if the boundary is smooth there \cite{faithful}. This argument is however of very limited practical use as it is computationally virtually infeasible to characterize this convex set \cite{Nrepre}. Also, the uniqueness is only obtained by restricting to a class of $n$-local Hamiltonians and there might exist other $(n+k)$-local (with $k\geq 1$) or quasi-local Hamiltonians for which this is the exact ground state. One of the main goals of this paper is thus to identify which features of all those Hamiltonians can be captured in the ground state and its correlations.

We follow a more practical approach based on local information contained within the ground state, which is naturally accessible through a tensor network representation of the same.
A central local object arising in tensor network simulations is the tensor network transfer matrix (defined in \Sec{s:TM}). Indeed, the main motivation for this work originates from numerical results obtained from tensor network simulations of ground states of strictly local translation invariant Hamiltonians in the thermodynamic limit. There it is observed that the spectrum of the transfer matrix exhibits a very peculiar structure, from which certain information about the low energy excitation spectrum of the underlying Hamiltonian can be extracted. These results are presented and discussed in \Sec{s:numerics} for a set of prototypical quantum models on lattices in one and two dimensions, as well as (1+1)-dimensional field theories.

We provide several arguments for explaining these observations in \Sec{s:statcorr}. There we argue how the structure of the eigenvalue spectrum of the transfer matrix allows to reproduce the expected form of correlation functions in gapped quantum states and use the single mode approximation to relate these eigenvalues to excited states of the Hamiltonian. We also show the converse, i.e. that the excited states of the Hamiltonian affect the static correlations functions, either by employing arguments from relativistic theories or by using momentum filtering to refine the celebrated proof of Hastings in \cite{HastingsLR} for the relation between gap and correlation length.

In \Sec{s:qtm} we follow an alternative approach 
by directly connecting the transfer matrix in the context of tensor network states to the exact quantum transfer matrix (QTM) \cite{Suzuki,Betsuyaku} at zero temperature, which appears in path-integral formulations of partition functions or ground states of quantum systems. Tensor network methods for studying such transfer matrices have been successful since the invention of the Transfer Matrix Renormalization Group (TMRG) method to simulate classical models in two \cite{Nishino_2D1,Nishino_2D2} and higher dimensions \cite{Nishino_3D}. Invoking a quantum-to-classical mapping, this method has later been generalized and used to simulate one-dimensional quantum models at finite temperature \cite{Bursill,WangXiang,Shibata} and recently to also include real time evolution \cite{Sirker,Huang}. In these methods, the object which is approximated by a tensor network is the (quantum) transfer matrix itself. In this work however, we investigate 
the transfer matrix at zero temperature generated by a tensor network approximation of the \textit{ground state}. We also explain how the renormalization group (RG) allows to interpret the MPS-TM as a compressed version of the QTM. More specifically, in \Sec{ss:RG} we demonstrate how Wilson's Numerical Renormalization Group (NRG) for impurity systems -- or its recent reformulation using the Multi-scale Entanglement Renormalization Ansatz (MERA) \cite{MERA,MERA_rev,MERA_criticality,MERA_impurity1,MERA_impurity2} -- allows to build an MPS approximation of the ground state with finite bond dimension $D$ from the QTM. This construction yields a novel connection between tensor network states and RG methods.


\section{Tensor Network Transfer Matrices}
\label{s:TM}
In this section we define the regular and mixed Transfer Matrix (TM) for Matrix Product States (MPS) \cite{FNW,MPS-P,MPS-V,MPS-Scholl2} on one dimensional lattice systems and continuous Matrix Product States (cMPS) \cite{cMPS-F,cMPS-O,cMPS-J} on (1+1)-dimensional field theories respectively. In the context of higher-dimensional lattice systems described by Projected Entangled Pair States (PEPS) \cite{PEPS,MPS-V} we consider two-dimensional lattice systems on cylinders. There we obtain an effective one-dimensional 
lattice system by blocking sites on rings around the cylinder as described in \Sec{ss:tm}.

As we are interested in bulk properties of quantum systems, we will generally work in the thermodynamic limit, where for gapped one-dimensional quantum lattice systems, a good approximation of the ground state can be obtained by using a uniform MPS ansatz with finite bond dimension $D$
\begin{equation}
 \ket{\psi[A]}=\sum_{\{\mathbf{s}\}}\bi{v}_{L}^{\dagger}\Big(\prod_{j\in\mathbb{Z}}A^{s_{j}}\Big)\bi{v}_{R}\ket{\{\mathbf{s}\}},
 \label{eq:umps}
\end{equation} 
where $A^{s_{j}}$ is a set of $d$ matrices $\in\mathbb{C}^{D\times D}$ containing all variational parameters defining the state, $s_{j}$ labels states within the $d$-dimensional local Hilbert space on each site and $j$ labels sites on the lattice. $\bi{v}_{L}$ and $\bi{v}_{R}$ are boundary vectors which have no effect on bulk properties. An optimal MPS representation of the ground state can readily be calculated using variational uniform MPS techniques \cite{iTEBD,TDVP}. For ground states of higher-dimensional lattice systems similar techniques can be used for uniform PEPS \cite{iPEPS}. 

Equivalently, ground states of (1+1)-dimensional field theories in the thermodynamic limit can be well approximated by uniform cMPS, where, e.g., a one-flavor bosonic cMPS of finite bond dimension $D$ is given by
\begin{equation}
 \ket{\psi[Q,R]}=\bi{v}^{\dagger}_{L}\mathbb{P}\, \exp\Big(\int_{-\infty}^{\infty}\rmd x [Q(x)\otimes\unity + R(x)\otimes\hat{\psi}^{\dagger}(x)]\Big)\bi{v}_{R}\ket{\Omega},
\end{equation} 
where again matrices $Q(x), R(x)\in \mathbb{C}^{D\times D}$ contain all variational parameters defining the state. Here $\hat{\psi}^{\dagger}(x)$ are bosonic creation operators, $\mathbb{P}$ is the path ordering operator, $\ket{\Omega}$ is the vacuum of the field theory and $\bi{v}_{L}$ and $\bi{v}_{R}$ are again boundary vectors having no effect on bulk properties. To obtain cMPS ground state approximations, the algorithm of \cite{TDVP} can be adapted accordingly.

\subsection{Regular Transfer Matrix}
\label{ss:tm}
It is well known (and reiterated in \Sec{ss:oz}) that static correlation functions with respect to a uniform MPS ground state are obtained using the regular \textit{MPS transfer matrix} (MPS-TM) \cite{MPS-V}, which is given by 
\begin{equation}
\mpstm{A}{}=\sum_s \bar{A}^{s}\otimes A^{s},
\label{eq:mpstm}
\end{equation}
with $\bar{A}^{s}$ the complex conjugate of $A^{s}$. To simplify notation we will generally omit the subscript $A$ denoting the MPS matrix if it is not necessary to avoid confusion.

For continuum results we define the (generator of a) uniform cMPS transfer matrix
\begin{equation}
\cmpstm{Q,R}{}=\bar{Q}\otimes \unity + \unity \otimes Q+\bar{R}\otimes R,
\label{eq:cmpstm}
\end{equation}
where again $\bar{Q}$ and $\bar{R}$ denote the complex conjugates of $Q$ and $R$ respectively. We will again omit subscripts whenever they are not necessary. The relation to the lattice transfer matrix is given by
\begin{equation}
\cmpstm{}{}=\lim_{\epsilon\to 0} \frac{1}{\epsilon} \log \mathcal{T}\label{eq:relationmpscmpstm}
\end{equation}
with $\epsilon$ the lattice spacing of an underlying lattice discretization and $\mathcal{T}$ the transfer matrix of the corresponding MPS defined on the discretized lattice.

\begin{figure}
\centering
\includegraphics[width=0.8\columnwidth]{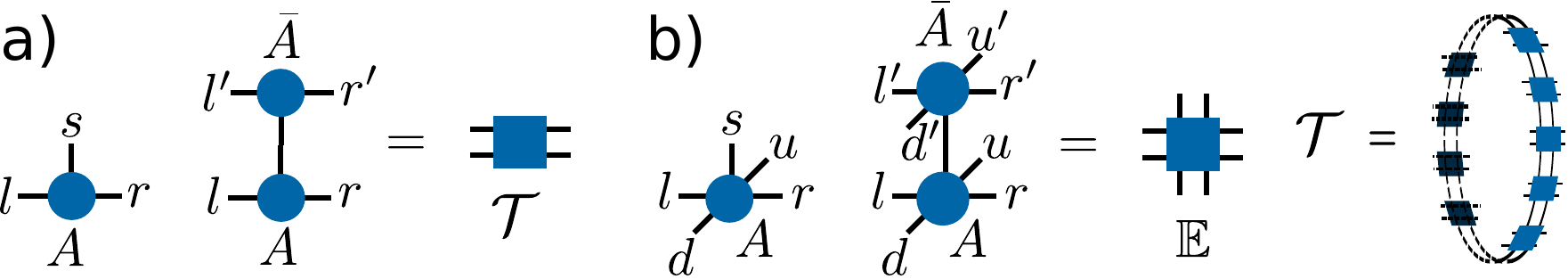}
\caption{
\textbf{a)} Graphical representation of the the MPS matrix $A^{s}_{lr}$ and the regular MPS-TM $\mpstm{A}{}$ constructed from it.
\textbf{b)} Graphical representation of the PEPS tensors $A^{s}_{udlr}$, the elementary PEPS-TM $\mathbb{E}$ and the quasi one-dimensional transfer matrix $\mpstm{}{}$ constructed by blocking $\mathbb{E}$ on a ring around the cylinder for the case of a square lattice.
}
\label{fig:peps_tm}
\end{figure}

Finally, for two-dimensional systems studied using PEPS, we work in the setting of infinitely long cylinders. By blocking the PEPS tensors $A^{s}_{udlr}$ on a ring along the (finite) transversal $y$-direction of the cylinder we can then interpret this contracted object as a uniform MPS along the (infinite) longitudinal $x$-direction of the cylinder and we define the longitudinal transfer matrix as in \eq{eq:mpstm}. For a square lattice geometry this MPS has bond dimension $D^{N_{y}}$ and physical dimension $d^{N_{y}}$ where $N_{y}$ is the number of sites along the circumference of the cylinder. Equivalently, an elementary TM $\mathbb{E}$ can be constructed from the individual PEPS tensors $A^{s}_{udlr}$ and the TM along a ring is obtained by contracting these elementary TMs along a ring. A graphical representation of the obtained TM is given in \Fig{fig:peps_tm}.

\subsection{Symmetries and the Mixed Transfer Matrix}
\label{ss:syms_mixedtm}
If a uniform MPS defined by a set of matrices $A^{s}_{1}$ is invariant under a local unitary symmetry operation $u$, one can show \cite{syms} that
\begin{equation}
 A_{2}^{s}=\sum_{k}\braket{s|u|k}A_{1}^{k}=\rme^{\rmi\theta}V^{\dagger}A_{1}^{s}V,
 \label{eq:mpssym}
\end{equation} 
where $A_{2}^{s}$ defines the transformed state. Here $V$ is a unitary gauge transformation on the auxiliary space and $\rme^{\rmi\theta}$ is the dominant eigenvalue with magnitude one of the \textit{mixed
transfer matrix}
\begin{equation}
 \mathcal{T}^{A_{2}}_{A_{1}}=\sum_{s} \bar{A}_{2}^{s} \otimes A_{1}^{s}.
 \label{eq:mixedmpstm}
\end{equation}
In fact, the MPS is invariant under the local symmetry $u$ if and only if the spectral radius of the mixed transfer matrix $\rho(\mathcal{T}^{A_{2}}_{A_{1}})$ is one, i.e. the fidelity per lattice site is one. In the case of higher dimensional lattice systems, a relation similar to \eq{eq:mpssym} holds for PEPS \cite{syms}.

For phases with a spontaneously broken symmetry, the ground state is degenerate and the variationally best ground state approximations within the manifold of MPS of bond dimension $D$ are minimally entangled states which exhibit maximal symmetry breaking. Such states can be transformed into each other by applying symmetry operations of the broken symmetry.

In the ground state of a one-dimensional quantum system, continuous symmetries for which the order parameter does not commute with the Hamiltonian cannot be spontaneously broken \cite{MerminWagner,Coleman}.
Nevertheless, close to or within a gapless phase with a continuous symmetry, it is sometimes energetically favorable for a variational (c)MPS approximation of the ground state to break this symmetry and to approximate an excited state with very small excitation energy and much smaller entanglement instead. This pseudo symmetry breaking is purely an effect of finite bond dimension and also gives rise to a pseudo order parameter \cite{WangBreak,DamianExcitations}. The symmetry is restored in the limit $D\to\infty$.

In a phase with broken symmetry on a lattice, let $A^{s}_{1}$ and $A^{s}_{2}$ be MPS approximations of two ground states with the same variational energy but different order parameters and maximally broken symmetry. The orthogonality of these states requires that the fidelity per lattice site must be strictly smaller than one, i.e. the spectral radius of the mixed transfer matrix $\rho(\mathcal{T}_{A_{1}}^{A_{2}})<1$. 

Equivalently, for continuum systems we define the (generator of the) mixed cMPS transfer matrix as
\begin{equation}
\cmpstm{Q_{1},R_{1}}{Q_{2},R_{2}}=\bar{Q}_{2}\otimes \unity + \unity \otimes Q_{1}+\bar{R}_{2}\otimes R_{1}
\label{eq:mixedcmpstm}
\end{equation} 
where $Q_{1},R_{1}$ and $Q_{2},R_{2}$ are two different cMPS representations. Similar to the lattice case, if $Q_{1},R_{1}$ and $Q_{2},R_{2}$ describe two equally good ground state approximations with the same variational energy, but with different order parameters and maximum symmetry breaking, the spectrum $\cmpstm{Q_{1},R_{1}}{Q_{2},R_{2}}$ has strictly negative real parts.

The degeneracy of the ground state in phases with broken symmetries gives rise to topologically nontrivial excitations (kinks or domain walls), which typically correspond to the elementary excitations of the model. The mixed (c)MPS-TM of type \eq{eq:mixedmpstm} or \eq{eq:mixedcmpstm} of these symmetry broken ground states plays a crucial role in obtaining a variational approximation for such excitations, whereas the regular (c)MPS-TM of type \eq{eq:mpstm} or \eq{eq:cmpstm} is the central object for topologically trivial excitations \cite{DamianExcitations,JuthoExcitations}.

\section{Numerical Results}
\label{s:numerics}
This section illustrates and discusses typical spectra of the regular and mixed (c)MPS-TM of obtained (c)MPS ground state approximations and compares it to low energy excitations for several 
quantum models of interest. For the eigenvalues of the (c)MPS-TM we write 
\begin{equation}
\lambda_{j}=\rme^{-\varepsilon_{j}+\rmi\phi_{j}}
\label{eq:EV_form}
\end{equation} 
where $\varepsilon_{j}=-\log|\lambda_{j}|$ and $\phi_{j}=\argg{\lambda_j}$ is the complex argument. This form already suggests that the $\varepsilon_{j}$ will be related to some characteristic energies of the model, as motivated throughout the remainder of this paper.

Low lying variational excitation energies for one-dimensional models are obtained by means of both topologically trivial and nontrivial uniform (c)MPS ansatzes \cite{JuthoExcitations,DamianExcitations} and -- if applicable -- are shown together with exact solutions. 

For two-dimensional models we exploit the observed relation between eigenvalues of the transfer matrix and location and magnitude of energy dispersion relations to give a first estimate of the dispersion of elementary excitations.

\subsection{One-dimensional Lattice Models}
\label{ss:1dlattice}


We will first focus on three prototypical one-dimensional lattice models. We start with the spin-1/2 XY model in an external magnetic field
\begin{equation}
 H_{\rm XY}=-\sum_{j}(1+\gamma)S_{j}^{x}S_{j+1}^{x} + (1-\gamma)S_{j}^{y}S_{j+1}^{y}+gS^{z}_{j},
 \label{eq:XY}
\end{equation} 
which can be solved exactly 
\cite{XY_Katsura, XY_Barouch, XY_BunderMcKenzie}. Here $S^{\alpha}_{j}$ denote spin-1/2 operators defined on site $j$. We consider the gapped ferromagnetic regime $0<\gamma<1$ and $0<g<1$, where the system is in a symmetry broken phase and the ground state is twofold degenerate with local order parameter $m_{x}=\braket{S^{x}_{j}}$. Here the elementary excitations are domainwall-like and therefore well approximated by a topologically nontrivial MPS ansatz. Specifically, we consider the incommensurate phase $\gamma^{2} + g^{2}<1$, where correlations oscillate with arbitrary wave vectors.

As a second example we consider the spin-1/2 XXZ model in an external magnetic field
\begin{equation}
H_{\rm XXZ}=-\sum_{j}S_{j}^{x}S_{j+1}^{x} + S_{j}^{y}S_{j+1}^{y} + \Delta S_{j}^{z}S_{j+1}^{z}+hS^{z}_{j}.
\label{eq:XXZ}
\end{equation}
This model is solvable as well and the ground state and elementary excitations in the thermodynamic limit can be obtained via Bethe ansatz \cite{Bethe, XXZBethe1, Takahashi}. Here we consider the antiferromagnetic gapless incommensurate phase specified by $-1<\Delta<0$ and $0<|h|<1-\Delta$, where there are gapless excitations at multiples of the Fermi-momentum $k_{\rm F}=(\frac{1}{2} - m_{z})\pi$ with $m_{z}=\braket{S^{z}_{j}}$ the ground state magnetization. In this phase there is no spontaneous symmetry breaking, however due to criticality the finite $D$ MPS ground state approximation breaks the continuous rotational symmetry in the XY plane (c.f. \Sec{ss:syms_mixedtm}). This makes it possible to use a topologically non-trivial variational MPS ansatz for excitations.

As a third example we study the (extended) Fermi-Hubbard model
\begin{eqnarray}
 H_{\rm HUB}=&-\sum_{\sigma j}c^{\dagger}_{\sigma j}c_{\sigma j+1} - c_{\sigma j}c^{\dagger}_{\sigma j+1} + V n_{j}n_{j+1}\nonumber\\
&+ \sum_{j}U\Big(n_{\uparrow j}-\frac{1}{2}\Big)\Big(n_{\downarrow j}-\frac{1}{2}\Big) - \mu n_{j},
\label{eq:HUB}
\end{eqnarray}
where $c^{\dagger}_{\sigma},c_{\sigma}$ denote creation and annihilation operators of fermions with spin $\sigma$, $n_{\sigma}=c^{\dagger}_{\sigma}c_{\sigma}$ and $n=n_{\uparrow} + n_{\downarrow}$. For $V\neq0$ this model is non-integrable. We consider the repulsive regime, where $U,V>0$, away from half filling ($\mu\neq0$), which again corresponds to a gapless incommensurate phase. There is no spontaneous symmetry breaking in this phase and we consider topologically trivial excitations only.

The last example being studied is the Kondo Lattice model (KLM) \cite{KLM}
\begin{equation}
 H_{\rm KLM} = -\sum_{\sigma j}c^{\dagger}_{\sigma j}c_{\sigma j+1} - c_{\sigma j}c^{\dagger}_{\sigma j+1} - \mu n_{j} + J \bi{S}^{c}_{j} \cdot\bi{S}^{d}_{j}
 \label{eq:KLM}
\end{equation} 
where $c^{\dagger}_{\sigma},c_{\sigma}$ denote creation and annihilation operators of conduction electrons (c) with spin $\sigma$, $n=c^{\dagger}_{\uparrow}c_{\uparrow} + c^{\dagger}_{\downarrow}c_{\downarrow}$, and $\bi{S}^{c}_{j}$ and $\bi{S}^{d}_{j}$ are the spin operators for conduction electrons (c) and localized electrons (d) respectively. We consider the paramagnetic metallic phase away from half filling ($\mu\neq0$). This model in this phase has also been studied at finite temperature with TMRG techniques in \cite{KLM_kfermi}.

\begin{figure}
 \centering
 \includegraphics[width=\TMscale\linewidth,keepaspectratio=true]{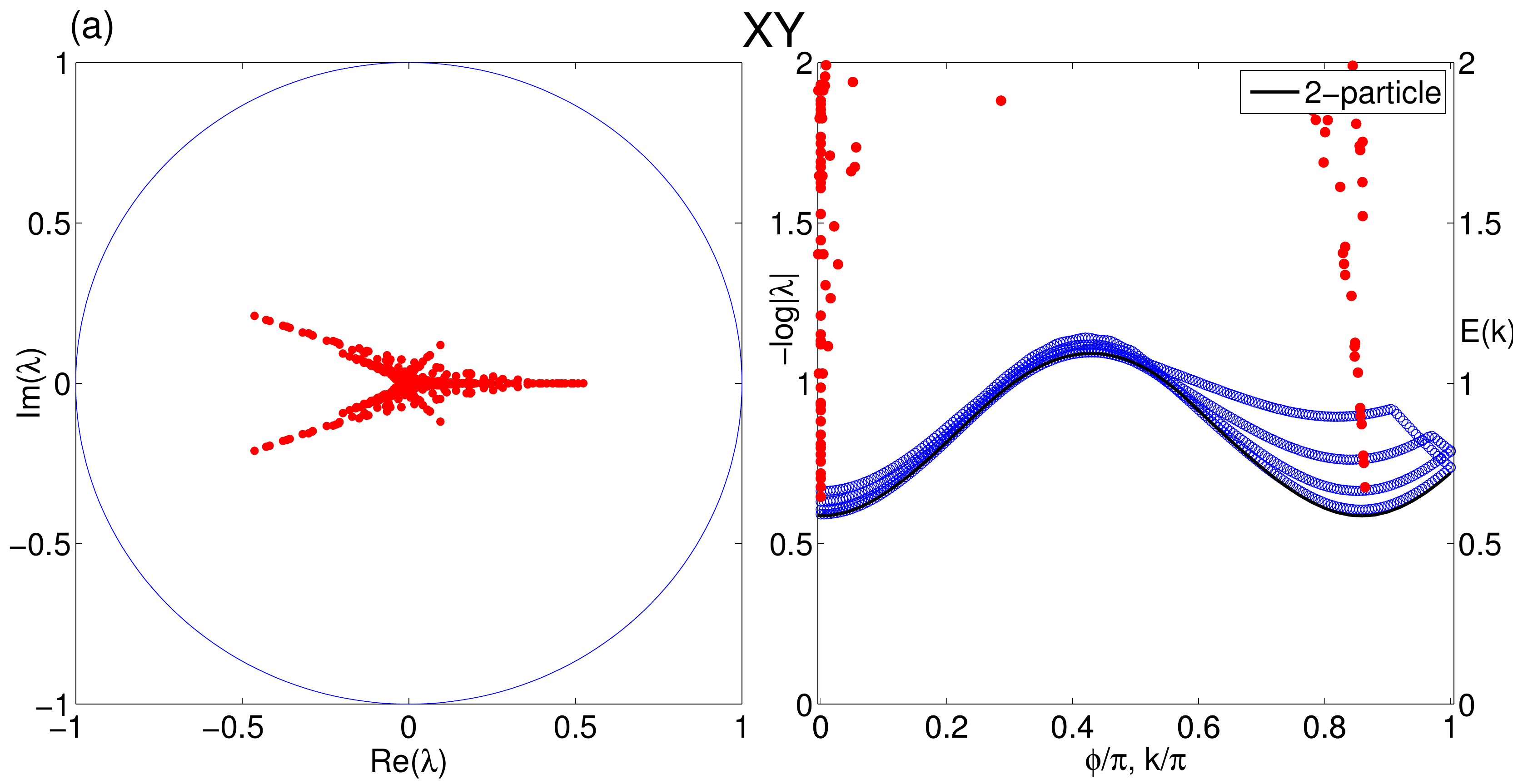}
 \includegraphics[width=\TMscale\linewidth,keepaspectratio=true]{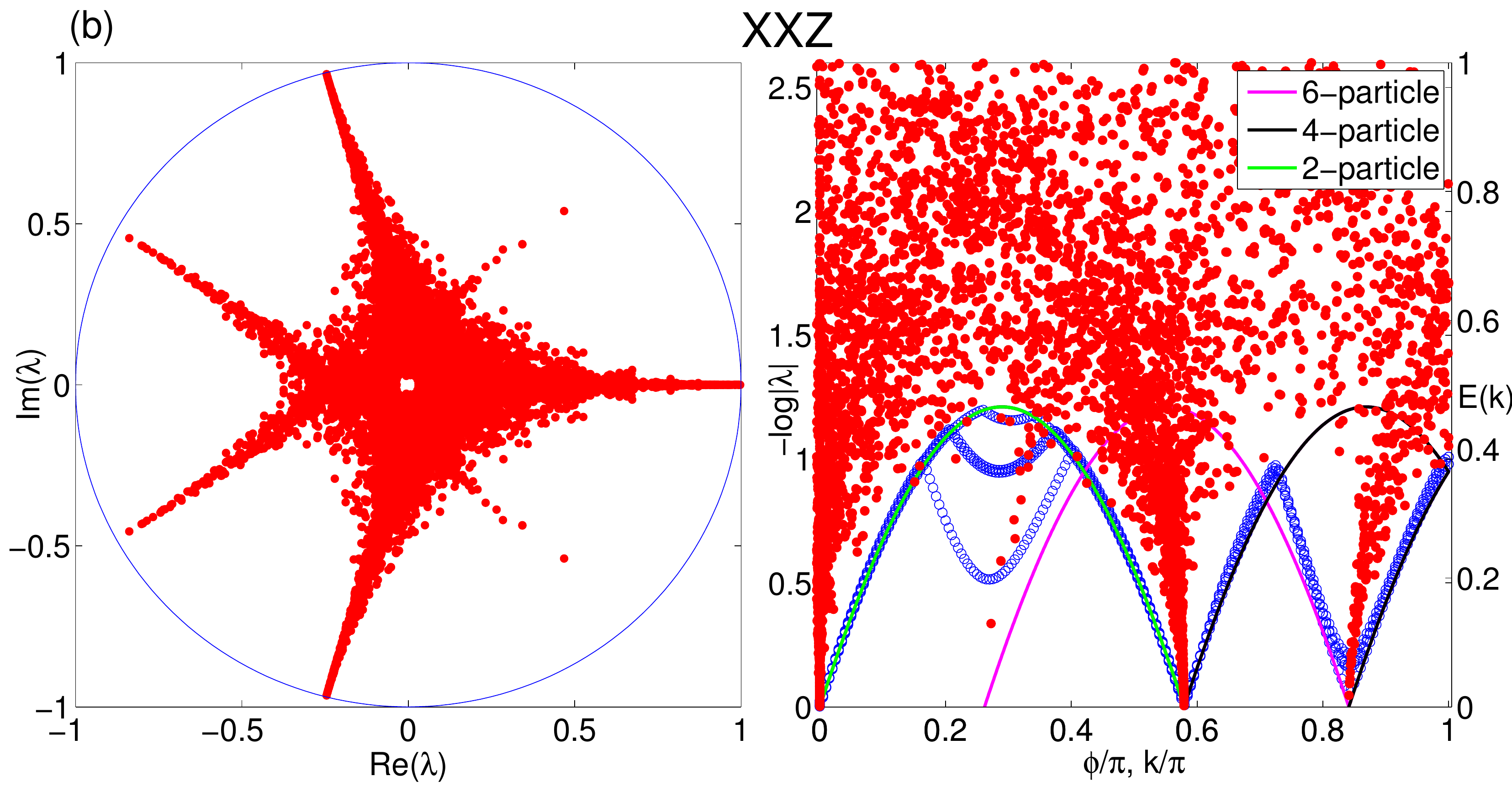}

 \caption{Eigenvalues of the \textit{regular} MPS-TM and topologically trivial excitations for (a) the XY model at $\gamma=0.3$, $g=0.2$ and $D=40$ and (b) the XXZ model at $\Delta=-0.5$, $h=1$ and $D=\DXXZ$.
 \textit{Left column}: Eigenvalues $\lambda_{j}=\rme^{-\varepsilon_{j}+\rmi\phi_{j}}$ of the MPS-TM on the complex plane within the unit circle. \textit{Right column}: $\varepsilon_{j}=-\log|\lambda_{j}|$ vs. $\phi_{j}$ (red symbols, left vertical axis) along with the lowest excitation energies obtained from a topologically trivial variational MPS ansatz (blue symbols, right vertical axis). We also show the exact lower boundaries of multi-particle continua (solid lines, right vertical axis) for reference.}
\label{fig:TMvsE1}
\end{figure}

\begin{figure}
 \centering
 \includegraphics[width=\TMscale\linewidth,keepaspectratio=true]{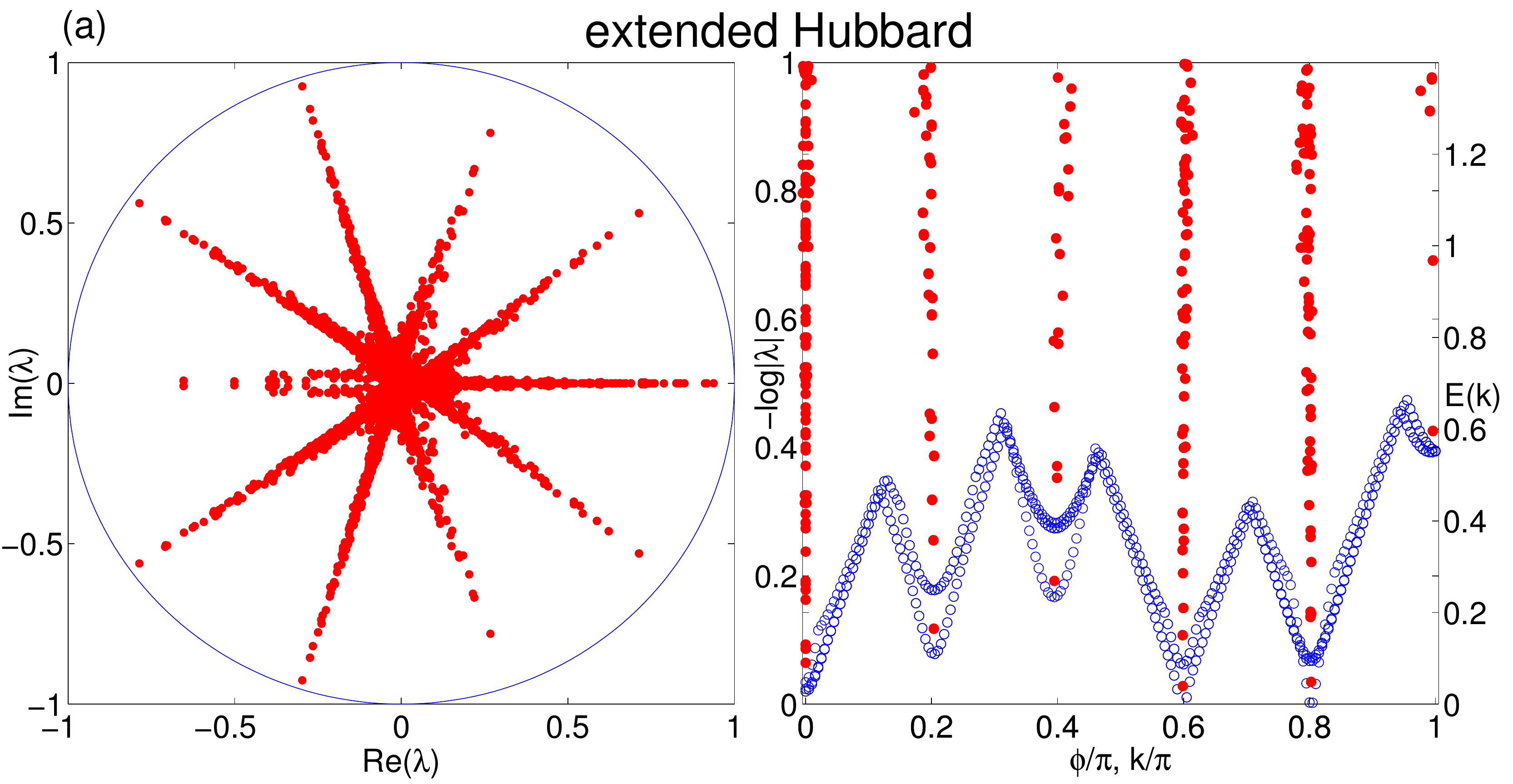}
 \includegraphics[width=\TMscale\linewidth,keepaspectratio=true]{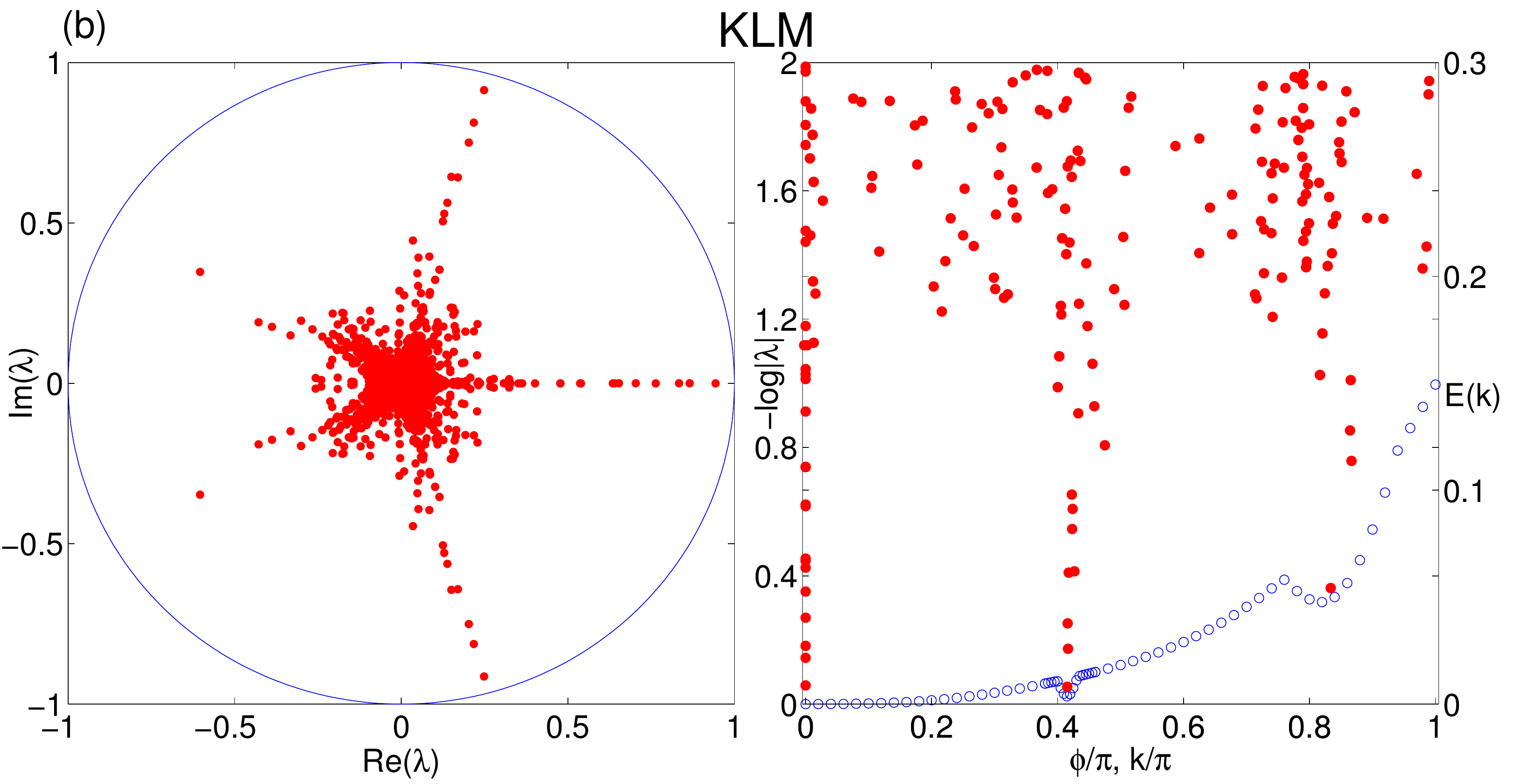}

 \caption{The same quantities as in \Fig{fig:TMvsE1} for (a) the extended Hubbard model at $U=5$, $V=1$, $\mu=2$ and $D=100$ and (b) the Kondo Lattice model at $J=2$, $\mu=-1$ and $D=120$. As there are no exact solutions for these models in these parameter regimes we only show variational excitation energies.}
 
\label{fig:TMvsE2}
\end{figure}

\begin{figure}
\centering
\includegraphics[width=\TMscale\linewidth,keepaspectratio=true]{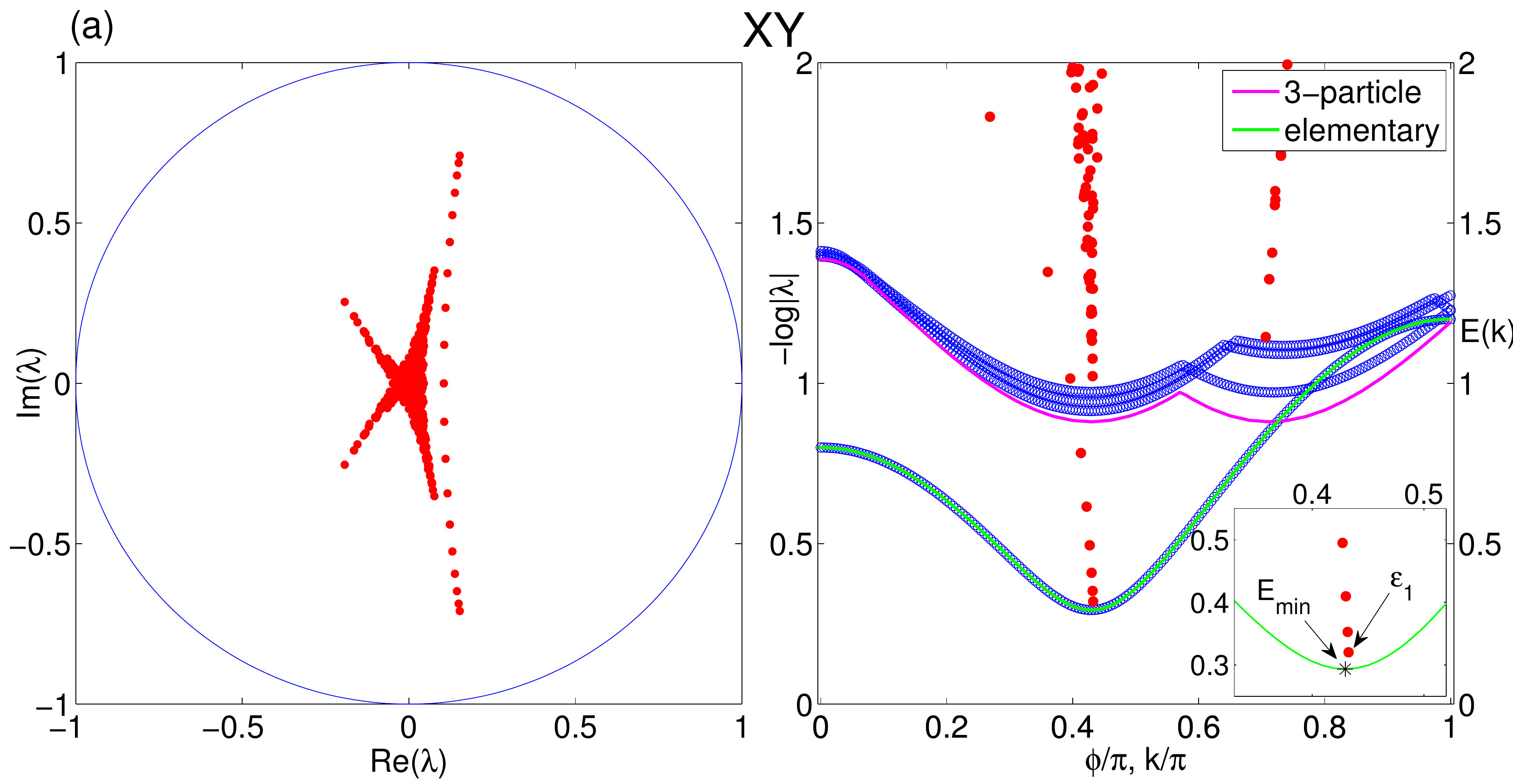}
\includegraphics[width=\TMscale\linewidth,keepaspectratio=true]{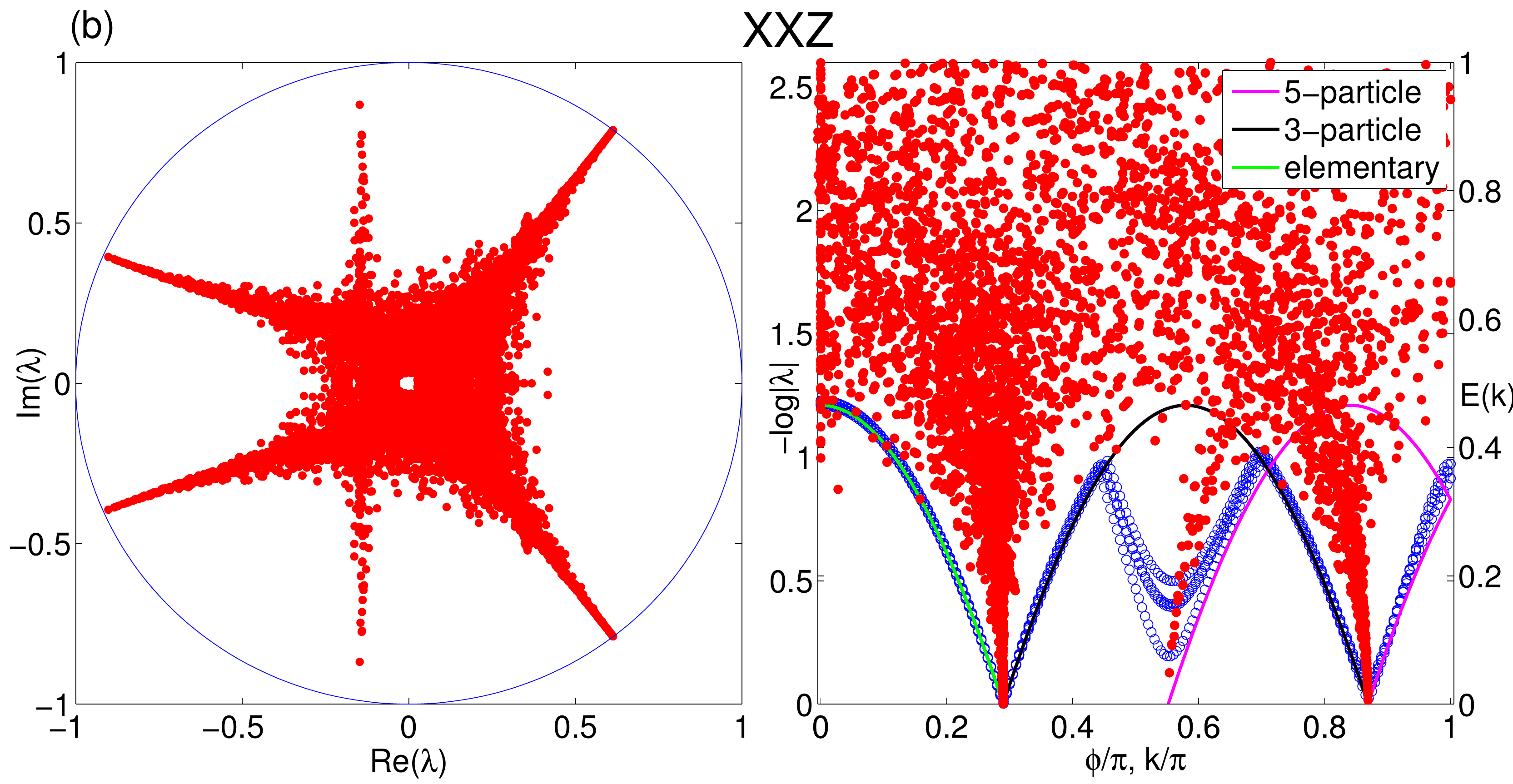}

\caption{ 
The same quantities for the same models and parameters as in \Fig{fig:TMvsE1} for the \textit{mixed} MPS-TM and topologically non-trivial excitations.
For reference we also show the exact dispersion of elementary excitations and the lower boundaries of multi-particle continua (solid lines, right vertical axis). \textit{(a) inset:} Magnification of the plot around the minimum $E_{\rm min}$ of the exact elementary excitation energies, marking the smallest non-zero $\varepsilon_{1}$ used to estimate the characteristic velocity $v_{1}=E_{\rm min}/\varepsilon_{1}$. 
Notice that in (a) we have chosen the same scale for $\varepsilon_{j}=-\log|\lambda_{j}|$ and $E(k)$ to emphasize the energy scale dictated by the characteristic velocity $v_{1}$.}
\label{fig:TMmixedvsE}
\end{figure}

In \Fig{fig:TMvsE1} and \Fig{fig:TMvsE2} we show results for (1) the XY model at $\gamma=0.3$, $g=0.2$ and $D=40$, (2) the XXZ model at $\Delta=-0.5$, $h=1$ and $D=\DXXZ$, (3) the extended Hubbard model at $U=5$, $V=1$, $\mu=2$ and $D=100$, and (4) the Kondo Lattice model at $J=2$, $\mu=-1$ and $D=120$. 
On the left we plot the eigenvalues $\lambda_{j}=\rme^{-\varepsilon_{j}+\rmi\phi_{j}}$ of the regular MPS-TM on the complex plane within the unit circle, whereas on the right we plot $\varepsilon_{j}=-\log|\lambda_{j}|$ vs. complex argument $\phi_{j}$, along with the  lowest variational excitation energies obtained from a topologically trivial variational MPS ansatz \cite{JuthoExcitations}. We do not plot the dominant eigenvalue $\lambda_{0}=1$. In \Fig{fig:TMmixedvsE} we show results for the mixed MPS-TM and topologically non-trivial excitations for the XY model and XXZ model only, with the same parameters as above. If available we also plot the exact dispersion of the elementary excitations as well as the lower boundaries of multi-particle continua for reference.

From \Fig{fig:TMmixedvsE} it is apparent that the topologically non-trivial variational ansatz captures elementary excitations with high accuracy. The two particle continuum consists of combinations of two elementary excitations and is thus partially captured by a topologically trivial ansatz (c.f. \Fig{fig:TMvsE1} and \Fig{fig:TMvsE2}), which is consistent with results in \cite{DamianExcitations}. The full continuum can be recovered by using a variational MPS ansatz including scattering states of elementary excitations \cite{LaurensScattering}. Consequently, low lying states with higher odd particle number are partially captured by a topologically non-trivial ansatz, whereas low lying states with even particle number are partially captured by a topologically trivial ansatz. In the case where there are no topologically non-trivial excitations, there is no such distinction.

Concerning the eigenvalues of the MPS-TM we can now make the following remarkable observations. In the plots on the left of \Fig{fig:TMvsE1}, \Fig{fig:TMvsE2} and \Fig{fig:TMmixedvsE} we see that most of the eigenvalues arrange themselves along several lines with approximately constant complex argument $\phi_{j}=\phi_{\alpha}$, where $\alpha$ labels distinct lines. This fact is reflected in the arrangement of the $\varepsilon_{j}$ in columns in the plots on the right, where low lying $\varepsilon_{j}$ correspond to eigenvalues $\lambda_{j}$ close to the unit circle. Denote the \textit{lowest lying} $\varepsilon_{j}$ for each line $\alpha$ as $\varepsilon_{\alpha}$.

We can also observe that the complex arguments $\phi_{\alpha}$ of these  $\varepsilon_{\alpha}$ coincide very precisely with the momenta $k_{\rm min}$ of the minima in the dispersion of excitation energies. This fact has been exploited in \cite{KLM_kfermi} to locate the Fermi-momentum $k_{\rm F}$ in the KLM. While there was no justification given there as to how the phase $\phi$ of the second largest eigenvalue of the TM is related to low energy excitations, we contribute several arguments for this fact in section \ref{s:statcorr}. Connections between peaks in the static structure factor and the logarithm of the absolute value and the complex argument of TM eigenvalues have also been exploited in \cite{Sirker,Kemper} to study the temperature-dependence of static correlation functions.

The values of the lowest lying $\varepsilon_{\alpha}$ are related to minima $E_{\rm min}^{\alpha}$ in the dispersion of elementary excitation energies via some characteristic velocities $v_{\alpha}$ \footnote{As $\mpstm{}{}$ essentially ``evolves'' the system in real space (as opposed to an evolution in e.g. real time by an operator $\rme^{-\rmi H t}$, c.f. also \Sec{s:qtm}), the $\varepsilon_{j}$ are given in units of inverse length and represent inverse correlation lengths as established in \Sec{ss:oz}. As energies are given in units of inverse time, both quantities can thus be related by a velocity.}  and can serve as a first approximation for this energy if the velocity can be estimated. It appears that these velocities, which determine the energy scale for each line $\alpha$, can also vary between different excitation minima $E_{\rm min}^{\alpha}$ within each respective model in the shown examples.

Indeed, for the XY model, the momenta $k_{\rm min}$ of the minima of the elementary excitations and the three particle continuum are well reproduced by the eigenvalues of the mixed MPS-TM with largest magnitude with deviations of $\Or(10^{-3})$. Consequently the same holds for the minima of the two particle continuum and the regular MPS-TM. For the elementary excitations we estimate the characteristic velocity $v_{1}$ relating the lowest excitation energy $E_{\rm min}$ and $\varepsilon_{1}=-\log|\lambda_1|$, where $\lambda_1$ is the eigenvalue with second largest magnitude, as $v_{1}=E_{\rm min}/\varepsilon_{1}\approx0.9409$ (c.f. \Fig{fig:TMmixedvsE} inset), where we have extrapolated the value of $\varepsilon_{1}$ for $D\to\infty$. Towards the end of \Sec{ss:kl} we show how the value of this velocity can be estimated from assuming a Lorentz-invariant low energy behavior. There we obtain an estimate for $v_{1}$ which agrees with the value obtained above within $1\%$ accuracy. 

For the XXZ model, the momenta of the gapless excitations at multiples of $k_{\rm F}$ are even more precisely reproduced by the arguments $\phi_{j}$ of the eigenvalues of the regular and mixed MPS-TM with magnitude close to one (i.e. $\varepsilon_{j}$ close to zero) with relative deviations of $\Or(10^{-6})$. Notice that in the limit $D\to\infty$ we expect $\varepsilon_{j}\to 0$, i.e. the spectral radius of the mixed MPS-TM also becomes one and the rotational symmetry in the XY plane is restored (c.f. \Sec{ss:syms_mixedtm}).

For the extended Hubbard model, the star like structure of the eigenvalue spectrum in \Fig{fig:TMvsE2} is very pronounced. In the right plot the ratios of the variational dispersion minima and the lowest $\varepsilon_{j}$ at $k=\phi_{j}\approx0.4\pi$ and $k=\phi_{j}\approx\pi$ might suggest a characteristic velocity $v_{\alpha}>1$. However, one would expect to have gapless excitations at these momenta, as well as at $k\approx0.2\pi$, too, suggesting that the corresponding variational energies are not converged. It is instructive to either use larger bond dimension or enhance by using an ansatz including scattering, which we have not performed here. It is however interesting to note for these minima that the accuracy of the variational energies and the low lying $\varepsilon_{j}$ appears to be roughly on the same level.

At last, for the KLM we obtain a variational low energy dispersion that exhibits overall low variation in $k$, typical for the heavy fermion regime. We also observe dents in the slowly varying part coming from gapless excitations at multiples of the Fermi-momentum, for which we obtain an estimate of $k_{\rm F}\approx0.41541(3)\pi$ by extrapolating $D\to\infty$. While in \cite{KLM_kfermi} $k_{\rm F}$ is estimated by approximating the finite temperature QTM of the quantum system using TMRG and subsequently extrapolating $T\to0$  -- which is inherently prone to technical difficulties -- we directly work at $T=0$ and construct an approximation of the zero temperature QTM from an MPS ground state approximation. Indeed, our estimate of $k_{\rm F}$ at $T=0$ is of the same order as the value in \cite{KLM_kfermi} for lowest $T$; however there no claims about the value at $T=0$ were possible due to technical difficulties.

The above observations are truly remarkable. The mere knowledge of the ground state MPS-TM already yields important information about the excitation spectrum of the underlying Hamiltonian. More specifically, the momenta $k_{\rm min}$ of the excitation energy minima in the respective particle sectors can be determined accurately and the corresponding energies can be estimated in a first approximation just from static ground state properties. An advantage of this transfer matrix based approach over just considering static correlation functions is discussed in \Sec{ss:sma}.

\subsection{(1+1)-dimensional Field Theories}
\label{ss:1dfield}

\begin{figure}[t]
 \centering
 \includegraphics[width=\TMscale\linewidth,keepaspectratio=true]{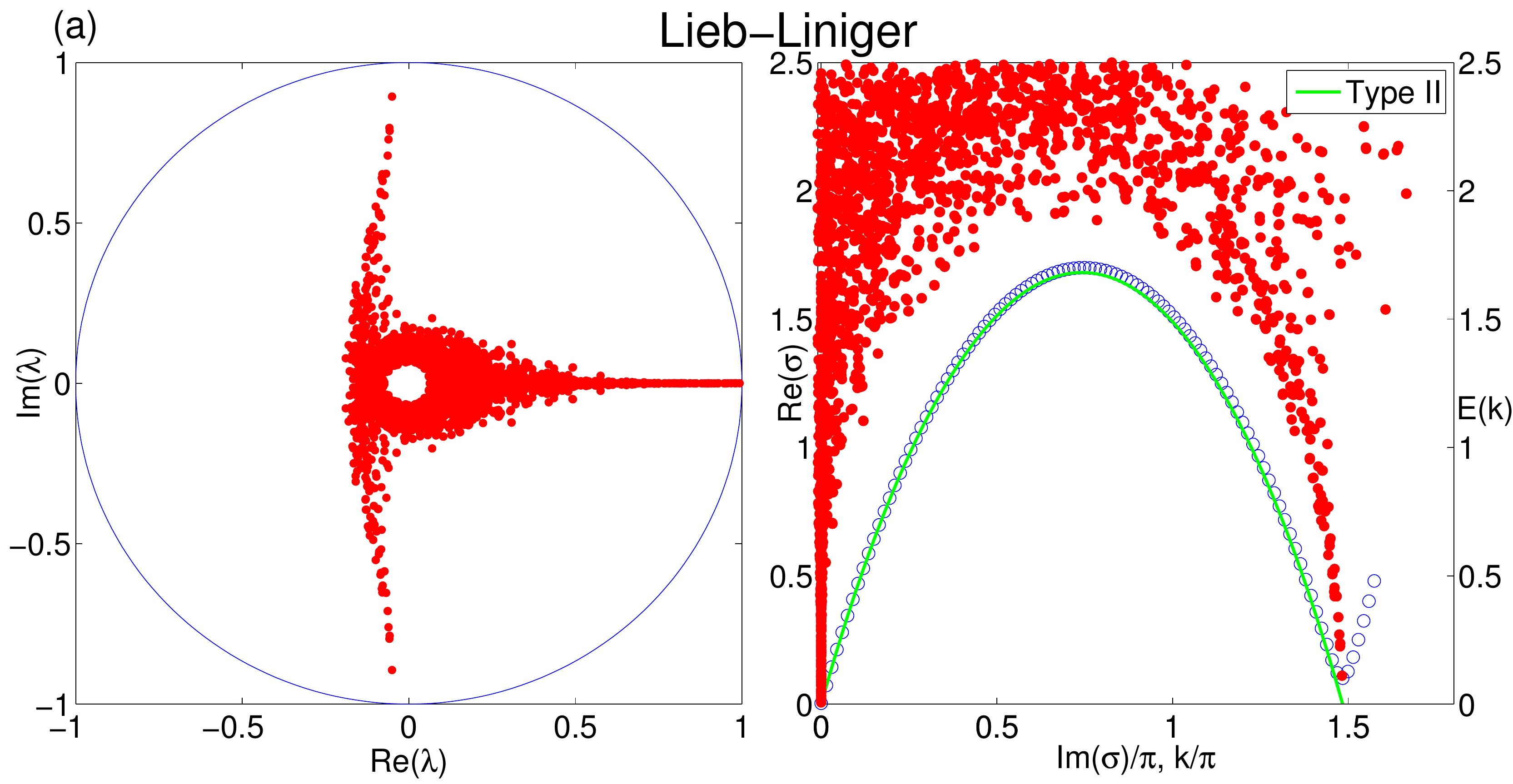}
 \includegraphics[width=\TMscale\linewidth,keepaspectratio=true]{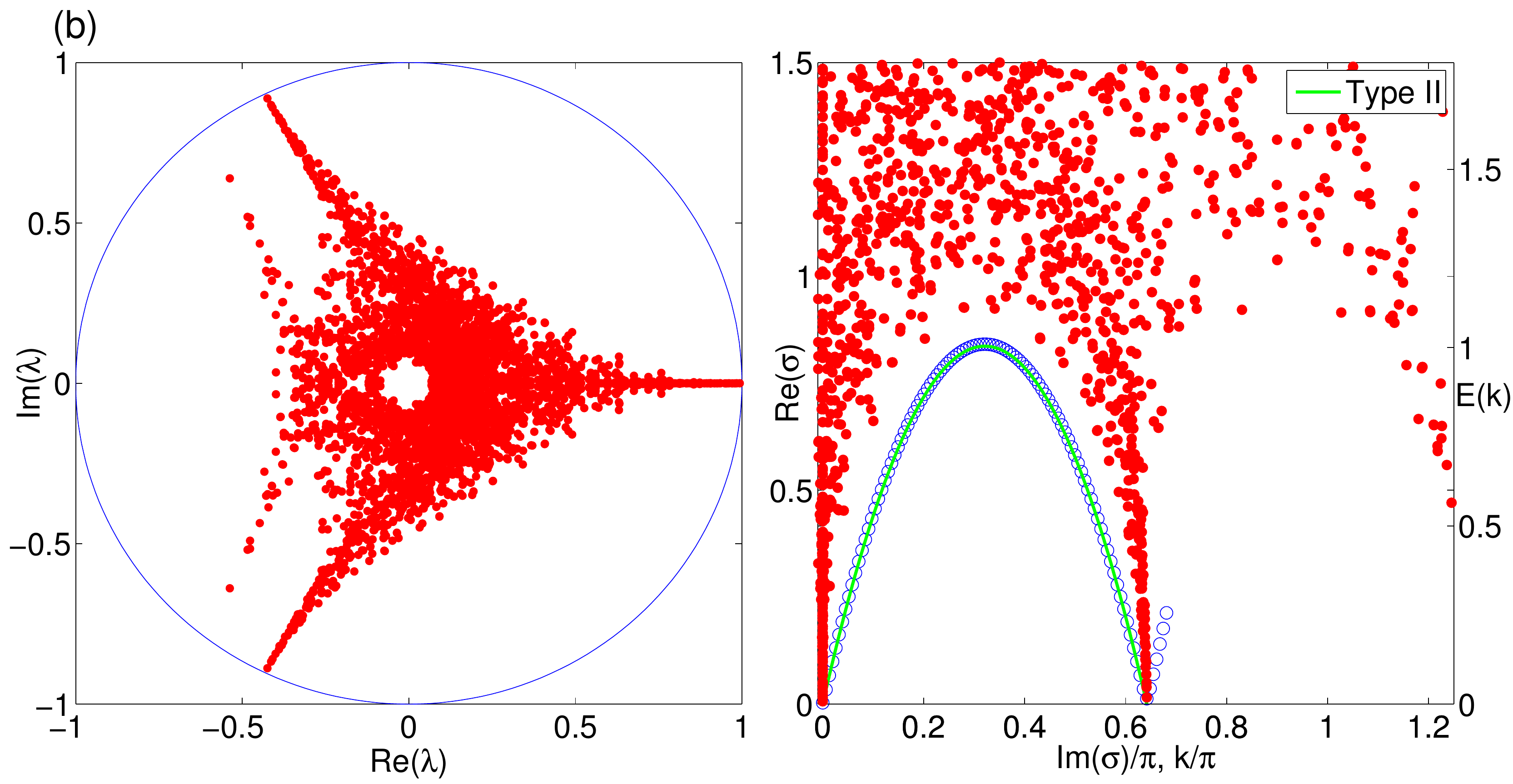}
 \caption{ 
 Eigenvalues of the (generator of the) \textit{regular} cMPS-TM and topologically trivial excitations for the Lieb-Liniger model at $D=64$ for (a) $\gamma\approx1.35$ and (b)
$\gamma\approx311.5$. \textit{Left column}: eigenvalues $\lambda_{j}$ of the cMPS-TM $\mpstm{}{}=\exp(\cmpstm{}{})$ on the complex plane within the unit circle. \textit{Right column}: Real versus imaginary part of the eigenvalues $-\sigma_{j}$ of the generator $\cmpstm{}{}$ (red symbols, left vertical axis) together with the lowest excitation energy obtained from a topologically trivial cMPS ansatz (blue symbols, right vertical axis). We also plot Lieb's Type \MakeUppercase{\romannumeral 2} excitations (lower boundary of hole-hole continuum, solid line, right vertical axis) for reference.}
 \label{fig:TMvsE-LL}
\end{figure}

\begin{figure}[t]
 \centering
 \includegraphics[width=\TMscale\linewidth,keepaspectratio=true]{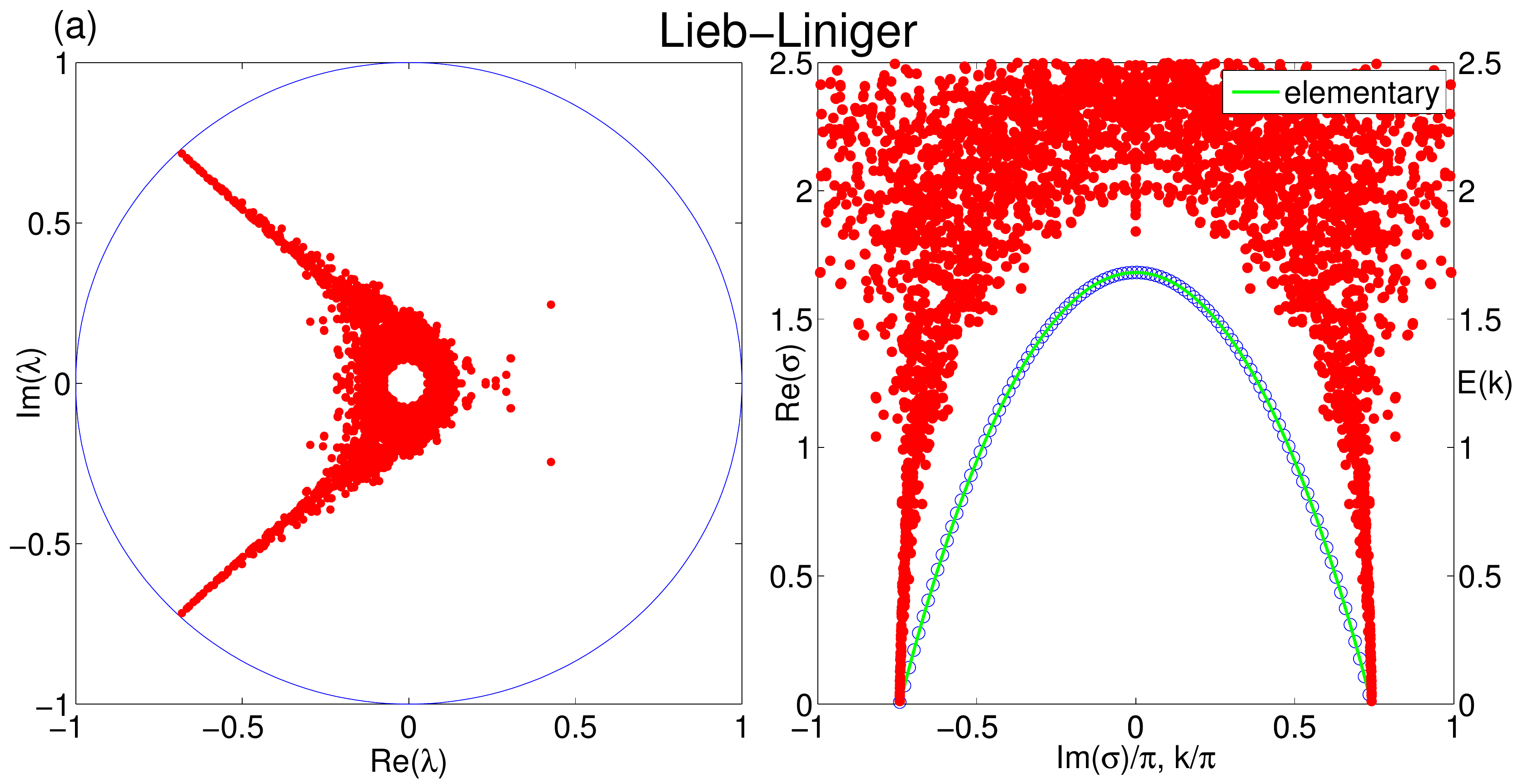}
 \includegraphics[width=\TMscale\linewidth,keepaspectratio=true]{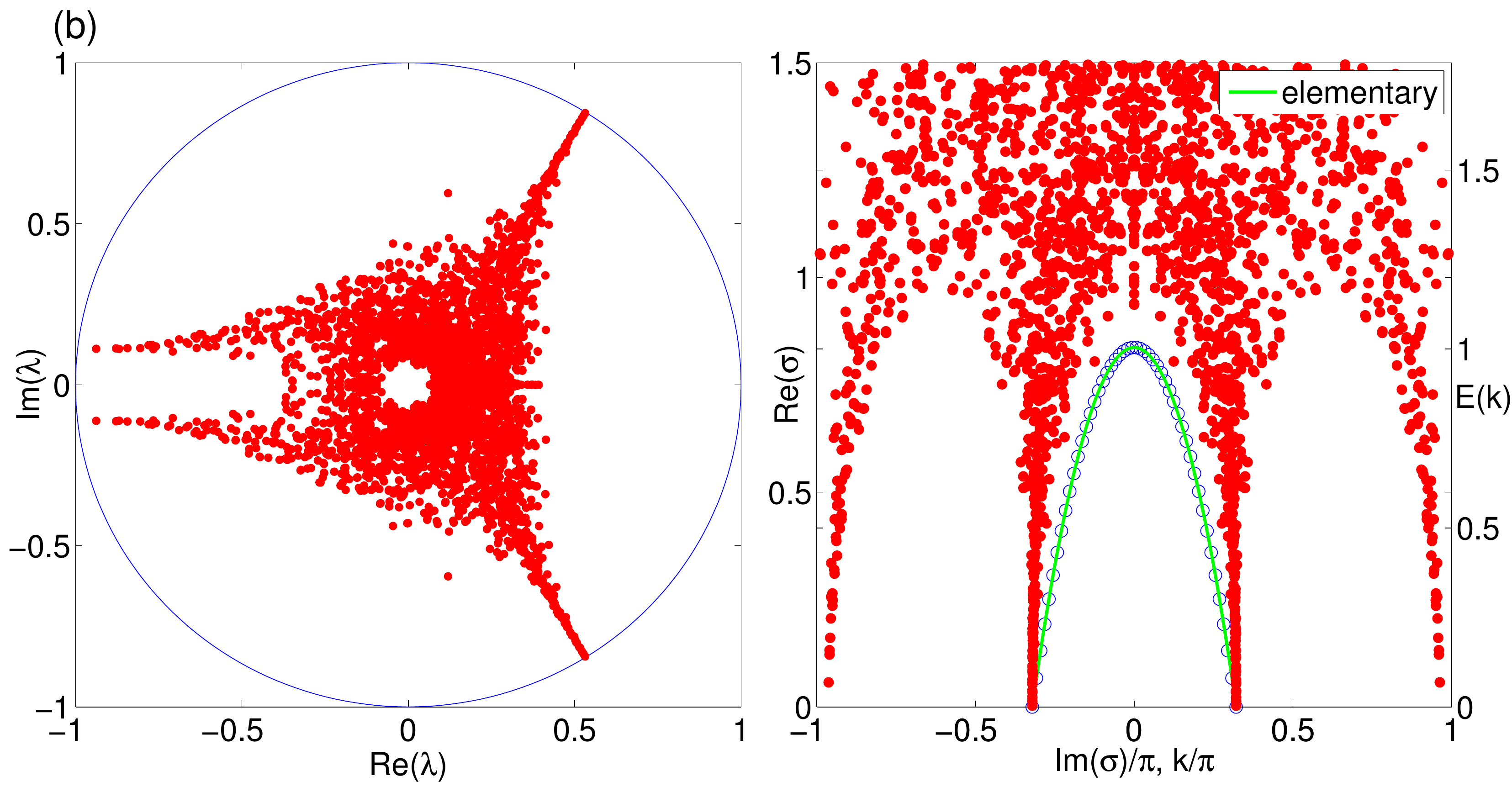}
 \caption{ 
 The same quantities as in \Fig{fig:TMvsE-LL} for the (generator of the) \textit{mixed} cMPS-TM and topologically non-trivial excitations. For reference we also plot the branch of
elementary hole excitations (solid line, right vertical axis).}
 \label{fig:TMmixedvsE-LL}
\end{figure}

We will now turn to continuous (1+1)-dimensional field theories and study the Lieb-Liniger model \cite{LiebLiniger} using cMPS methods. The Hamiltonian is given by 
\begin{equation}
H_{\rm{LL}} = \int_{-\infty}^{\infty}\rmd x\, \partial_{x}\psi^{\dagger}\partial_{x}\psi + c\psi^{\dagger}\psi^{\dagger}\psi\psi -\mu\psi^{\dagger}\psi,
\end{equation}
with repulsive interaction strength $c>0$ and chemical potential $\mu>0$, where $\psi$ and $\psi^{\dagger}$ are bosonic field operators. The model depends only on a single parameter $\gamma=\frac{c}{\rho}$, with $\rho=\braket{\psi^{\dagger}\psi}$ the ground state particle density and it is critical for all values of $\gamma$. 

In \Fig{fig:TMvsE-LL} and \Fig{fig:TMmixedvsE-LL} we show results for the eigenvalues of the regular and mixed cMPS-TM similar to the lattice case for (a) $\gamma\approx 1.35$ and (b) $\gamma\approx311.5$ and $D=64$. Given the relation in \eq{eq:relationmpscmpstm}, the right column now plots the real part of the eigenvalues $-\sigma_{j}$ of the generator $\cmpstm{}{}$ versus their imaginary part, which is now interpreted as momentum. In the left column the eigenvalues $\lambda_{j}$ of $\mpstm{}{}=\exp( \cmpstm{}{})$ are plotted on the complex plane within the unit circle as in the lattice case. In the continuum setting, momentum space is no longer $2\pi$-periodic and the definition of 
$\mpstm{}{}$ is not fully justified, as it can come with any real power $x\geq 0$ in continuum correlation functions, where only integer powers appear in lattice correlation functions. Nevertheless, it helps in illustrating that the spectrum of eigenvalues of the transfer matrix exhibits a similar structure. The fact that this structure in \Fig{fig:TMvsE-LL} and \Fig{fig:TMmixedvsE-LL} is less outspoken than for some of the lattice models indicates a larger contribution of microscopic effects for this specific case. 


Next we will study the simplest Lorentz-invariant theory available, the free (1+1)-dimensional Klein-Gordon boson described by the Hamiltonian:
\begin{equation}
\label{eq:KG}
H_{\rm{KG}} = \frac{1}{2}  \int_{-\infty}^{\infty} \rmd x \left[ \pi^2 +  \left( \partial_x \phi \right)^2 + m^2 \phi^2  \right]  \ ,
\end{equation}
where we have taken the speed of light to be unity. The field  operators $\phi$ and $\pi$ can be written in terms of the cMPS Fock space operators $\psi$ and $\psi^{\dagger}$ as:
\begin{equation}
\phi = \frac{1}{\sqrt{2 \nu } } ( \psi + \psi^{\dagger} ) \qquad  \pi = -\frac{\rmi}{2} \sqrt{2 \nu}  ( \psi -  \psi^{\dagger} ) \ ,
\end{equation}
where an arbitrary scale $\nu$ is introduced   \cite{VidFES}. The Hamiltonian (\ref{eq:KG}) needs to be regularised, and this is achieved by adding the term
$\left( \partial_x  \pi \right)^2\nu^{-2}$
to the Hamiltonian.

\begin{figure}[tb]
 \centering
 \includegraphics[width=\TMscale\linewidth,keepaspectratio=true]{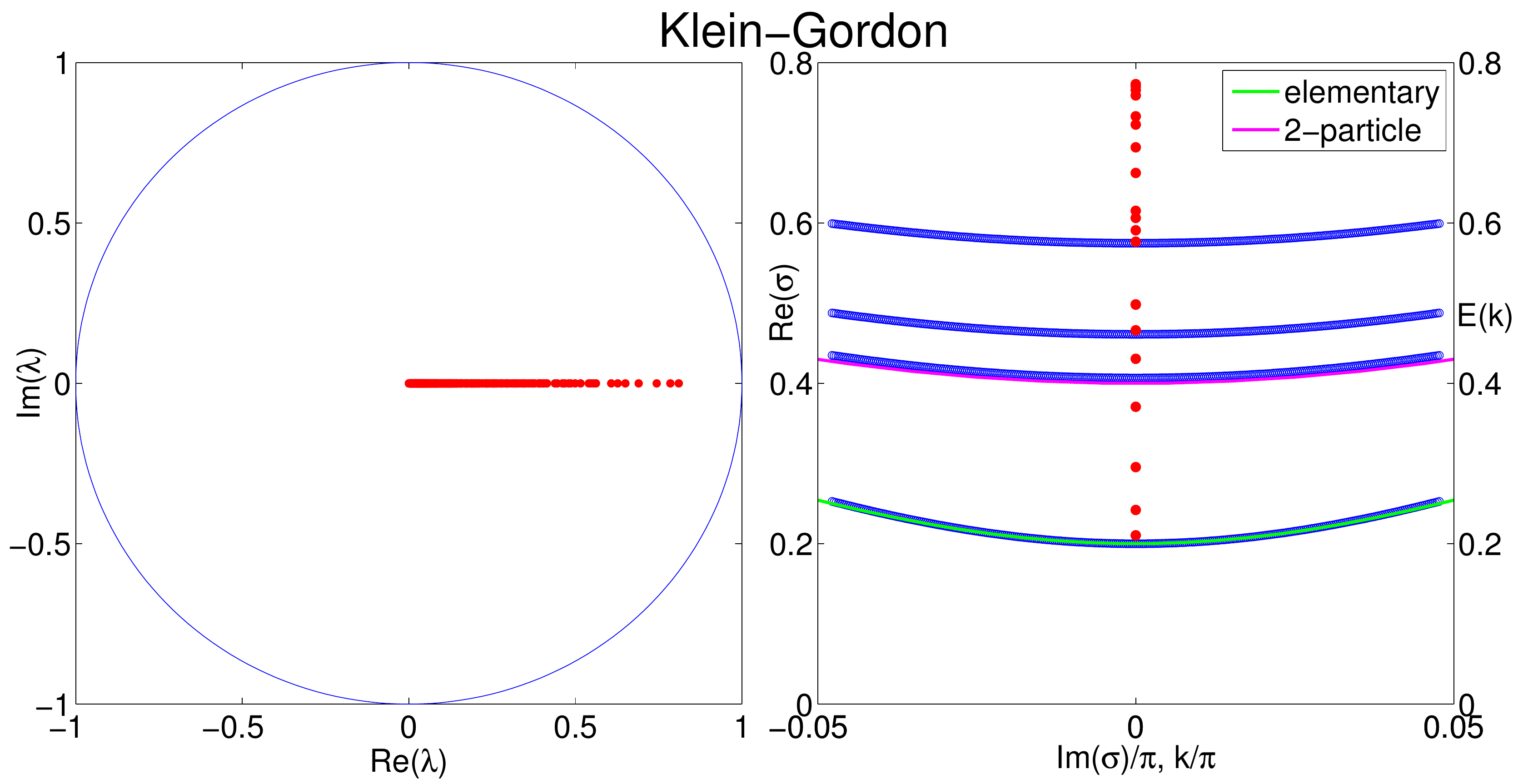}
 \label{fig:TMvsE-KG}
 \caption{ 
 The same quantities as in \Fig{fig:TMvsE-LL} for the Klein-Gordon boson for $m=0.2$ at $D=36$.
 The eigenvalues of $\mpstm{}{}$ and $\cmpstm{}{}$ are all real, resulting in the arrangement of all eigenvalues on a single line in both plots. We also plot the exact relativistic dispersion of elementary excitations (green solid line) and the lower boundary
 of the two-particle continuum (purple solid line) for reference.}
\label{fig:kleingordon}
\end{figure}

We make the following observations regarding the eigenspectrum of the generator of the cMPS transfer matrix  $\cmpstm{}{}$ corresponding to $H_{\rm{KG}}$, as plotted in \Fig{fig:kleingordon}. The eigenvalues $\sigma_j$ of $\cmpstm{}{}$ are all real and negative, for all values of $m$ and $D$. This reflects the fact that the relativistic dispersion relation has a single minimum at momentum zero. The eigenvalue with largest real part of $\cmpstm{}{}$ converges to $-m$ as $D \rightarrow \infty$. As is argued in \Sec{ss:imtime}, it follows from the Euclidean invariance of the quantum transfer matrix of a relativistic theory that in the limit $D \rightarrow \infty$ the eigenspectrum of $\cmpstm{}{}$ should be the same as that of $H_{\rm KG}$ (up to the sign), corresponding to a characteristic speed $v=c=1$. The above observation provides numerical support for this using and extrapolating from finite $D$ data.

We can also study the distribution of the eigenvalues of $\cmpstm{}{}$ as a function of the bond dimension. For any value of $m$, the (negative) eigenvalues $\sigma_j$ become dense in the region $[m,+\infty)$. In the gapped phase ($m>0$), the density of eigenvalues diverges at $m$, i.e. the ratio of the $n$th largest and second largest eigenvalue of $\cmpstm{}{}$ converges to unity for low lying $n$. This is similar to the density of states in a gapped single particle excitation branch near the minimum of the dispersion relation. As the mass $m$ is taken to zero the theory becomes critical, and the cMPS approximation enters the so-called "finite entanglement regime" \cite{PirvuFES}. The eigenvalues of  $\cmpstm{}{}$ are still all real and converge to zero as $D \rightarrow \infty$, but their ratios now converge to values larger than one, and are expected to encode universal data \cite{VidFES}. This is tantamount to the statement that the effect of finite bond dimension is to only introduce a single scale 
into the underlying conformal field theory, and implies that universal quantities can be extracted straightforwardly from cMPS data \cite{VidFES}.

Let us exemplify this by spelling out the results obtained by taking $m=0.2$ in (\ref{eq:KG}) and using the modest range of bond dimensions up to $D=36$. Scaling with $1/D$ and extrapolating to $D\rightarrow \infty$ yields that the second largest eigenvalue of $\cmpstm{}{}$ tends to $0.201$, thus reproducing the mass accurately. The ratio of the third and second eigenvalue of $\cmpstm{}{}$ is estimated to converge to $1.040$ as $D\rightarrow \infty$, and the ratio of the fourth and second to $1.086$. We note that, since the theory is free, the deviation from unity can not be due to convergence to some bound state just above the lowest branch, and can only be an effect of numerical accuracy. The same ratios for $m=0$ converge to approximately $2.51$ and $3.1$, respectively, and are related to properties of the excitation spectrum of the underlying conformal field theory \cite{VidFES}.

\subsection{Two-Dimensional Lattice Models}
\label{ss:2dlattice}
The observed connection between the eigenvalues of the transfer matrix and the minimum of the dispersion relation opens up a way to infer properties of the dispersion relation of two-dimensional systems, which are notoriously difficult to deal with.  To this end, given a two-dimensional translation invariant Projected Entangled Pair State (PEPS)~\cite{PEPS} on a square lattice cylinder with periodic boundary conditions in $y$ direction, we block all PEPS tensors in a ring around the cylinder (i.e., with the same $x$ coordinate along the cylinder). We then consider the quasi-one-dimensional system along the cylinder obtained that way, described by blocked tensors $A^{s}_{lr}$, and its transfer
matrix $\mpstm{}{}$.

As the original state was also translational invariant in $y$ direction, we can label the eigenvectors of $\mpstm{}{}$ by eigenvalues $\rme^{\rmi k_y}$ of the action of the translation operator on the auxiliary degrees of freedom, as given by \eq{eq:mpssym}. On a hexagonal lattice cylinder, we additionally block two neighboring PEPS tensors in order to obtain a quasi-square lattice before further blocking all obtained tensors in a ring around the cylinder. Assuming that the observed connection between the leading eigenvalues of the transfer matrix and the minimum of the dispersion relation holds for each $k_y$ independently, we obtain the location and relative energy of the minima of the dispersion relation for each possible value of $k_y$, which yields a cut through the dispersion relation.  By closing the periodic boundaries in different ways, we can obtain this information along different symmetry axes, allowing us to reconstruct the overall form of the dispersion relation.

We now apply this strategy to the AKLT model~\cite{AKLT} on the square and hexagonal lattice. Its ground state is constructed by placing spin-$\frac{1}{2}$ singlets on the edges of the lattice and projecting the spin-$\frac{1}{2}$'s at each vertex onto the sector with maximal spin ($S_\mathrm{phys}=2$ for the square lattice and $S_\mathrm{phys}=\frac{3}{2}$ for the hexagonal lattice); this construction yields the unique ground state of the $\mathrm{SU}(2)$ invariant Hamiltonian $H = \sum_{\langle i,j\rangle} \Pi_{i,j}$, where $\Pi_{i,j}$ is the projector onto the subspace with spin $S=S_\mathrm{phys}$ on neighboring sites $i$ and $j$. This construction corresponds exactly to a PEPS with bond dimension $D=2$. Even though the exact ground state is known, little is known about the excited states, although recently an indirect method was proposed to estimate the gap by means of a Tensor Network Renormalization Group method~\cite{TNRG}.

\begin{figure}[ht]
\centering
$\begin{array}{cc}
\includegraphics[width=0.56\columnwidth]{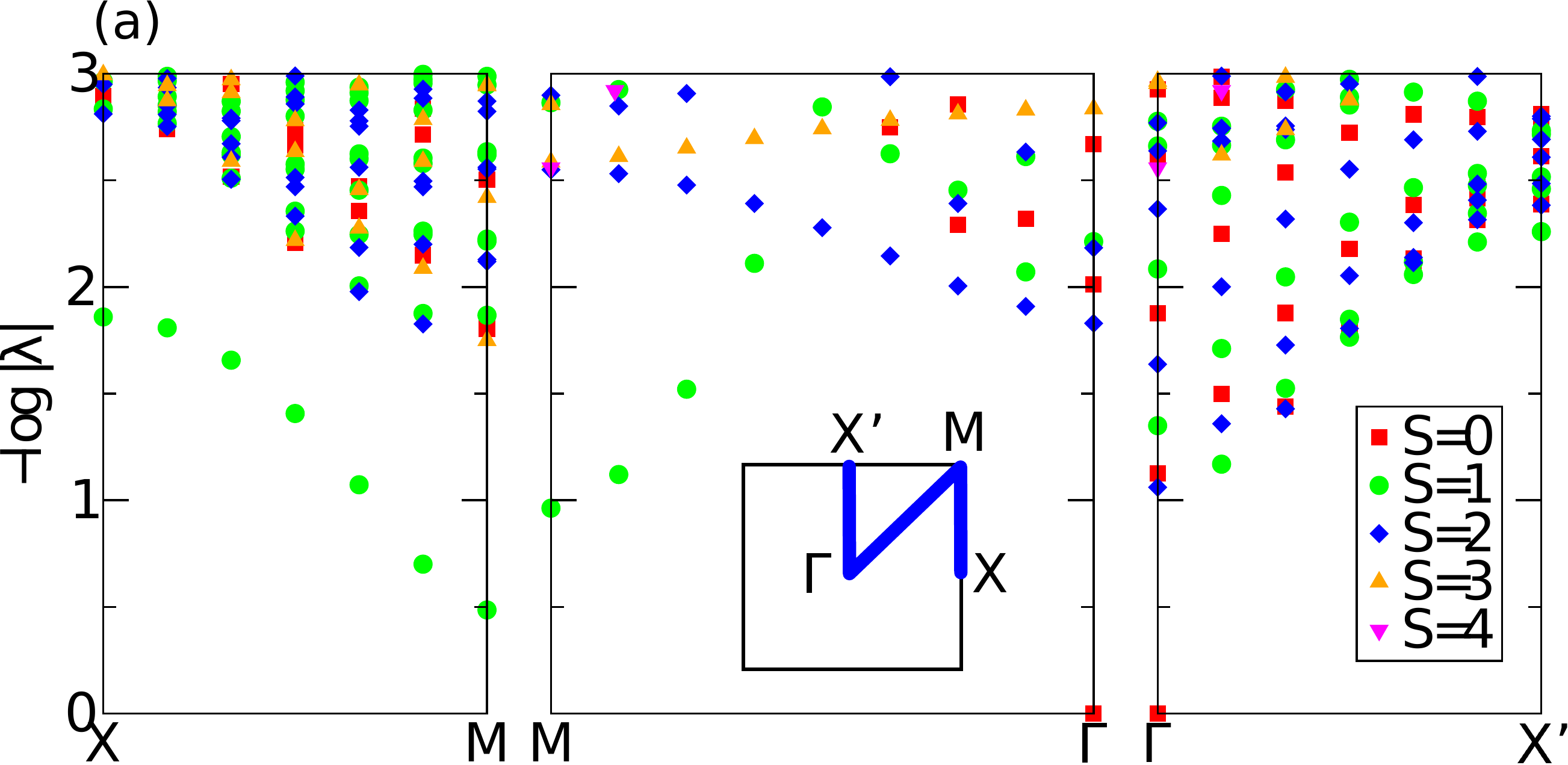}
\includegraphics[width=0.44\columnwidth]{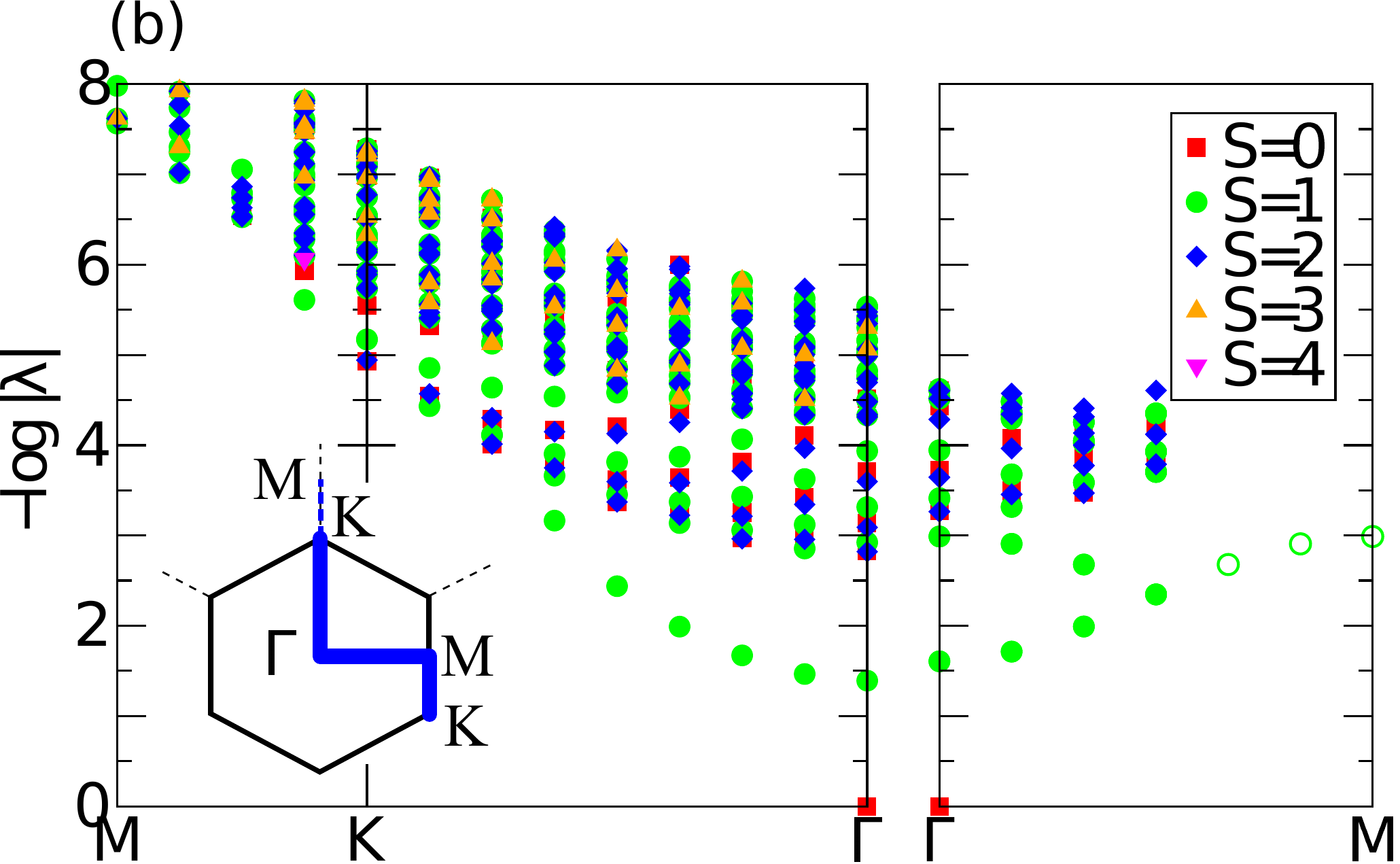}
\end{array}$
\caption{ 
Leading eigenvalues of the TM of the 2D AKLT model along the symmetry lines indicated in the insets. \textit{(a) Square lattice:} One can clearly see the minimum of the dispersion around the $M$ point with momentum $(\pi,\pi)$ and spin $S=1$, in accordance with the one-dimensional AKLT model. One can further recognize the continuum of two-particle states which sets in at $\Gamma$ at about twice the lowest quasi-energy of the single-particle band at $\mathrm{M}$, and can have spin $S=0,1,2$. 
\textit{(b) Hexagonal lattice:}
Again we find a branch of spin-$1$ excitations, whose minimum is now around the $\Gamma$ point with momentum $(0,0)$, as the unit cell contains two spins and there is a continuum of two-particle excitations at about twice the quasi-energy at $\Gamma$. For both lattices the $S=0$ point at $\Gamma$ corresponds to the ground state. 
Data has been obtained from cylinders with different circumferences ($N_y=12$ for $\Gamma$--$\mathrm{X'}$ and $\mathrm{X}$--$\mathrm{M}$, and $N_y=8\sqrt{2}$ for $\mathrm{M}$--$\Gamma$ for the square lattice and $N_y=6\sqrt{3}$ for $\Gamma$--$\mathrm{M}$, and $N_y=12$ for $\mathrm{M}$--$\mathrm{K}$ and $\mathrm{K}$--$\Gamma$ for the hexagonal lattice in units of the lattice constant) and different boundary conditions, resulting in different scales for the data in the individual panels.}

\label{fig:aklt}
\end{figure}


We can use an iterative eigensolver to exactly determine the low-lying spectrum of $\mpstm{}{}$ with very high precision on cylinders with sufficiently large circumference. Since the model possesses $\mathrm{SU}(2)$ symmetry, we can additionally label the eigenvectors of the transfer matrix by their spin (which corresponds to the spin of the excitation), thereby aiding the identification of different excitation branches. As the transfer matrix of the AKLT model is hermitian (up to a gauge transformation), its eigenvalues are real, and thus $k_x=0,\pi$. The pairs $(k_x,k_y)$ for all eigenvalues of $\mpstm{}{}$ are therefore arranged along lines in the Brillouin zone; by properly closing the periodic boundaries, we can thus obtain data points along different symmetry axes. The results for the square and hexagonal lattice are shown in \Fig{fig:aklt}.
In both cases, we find an isolated branch of antiferromagnetic spin-$1$ excitations, with a two-particle 
continuum starting at about twice the elementary quasi-energy gap, in agreement with known results for one-dimensional systems. In particular, for the square lattice we find the \mbox{minimum} of the dispersion at momentum $(k_{x},k_{y})=(\pi,\pi)$, whereas for the hexagonal lattice the minimum is found at $(k_{x},k_{y})=(0,0)$. For both lattices the minima appear on the isolated $S=1$ branches.

This approach has further been used by some of the authors to study anyon condensation in the Toric Code Model with string tension \cite{ShadowsAnyons}. 
\section{Static Correlation Functions and Excitations}
\label{s:statcorr}
In this section, we elaborate on the relation between the eigenvalues of the transfer matrix and static connected correlation functions and use this to provide several arguments for understanding the peculiar structure of the transfer matrix spectrum. Without loss of generality we consider the case of one-dimensional lattice systems and write the static connected correlation function for operators $A_{0}$ and $B_{n}$ acting on single sites $0$ and $n$ as
\begin{equation}
 C_{AB}(n)=\braket{A_{0} B_{n}} - \braket{A_{0}}\braket{B_{n}},
 \label{eq:corr}
\end{equation} 
where $\braket{\ldots}$ denotes the expectation value with respect to the ground state. These arguments can readily be extended to operators acting on multiple sites, as well as higher dimensional systems and continuum systems. 

In \Sec{ss:oz} we explain how the clustering of the eigenvalues of the transfer matrix in branches allows to recover the typical Ornstein-Zernike form of correlations (to be defined below) in the limit $D\to\infty$. \Sec{ss:sma} uses the single-mode approximation to relate these branches to minima in the dispersion relation and also discusses why generically the full spectrum of the transfer matrix can provide more information than selected typical static correlation functions. 

We also investigate this connection in the other direction by showing how the low-energy excitations of the Hamiltonian affect the static correlation functions in the ground state. \Sec{ss:kl} assumes a Lorentz-invariant low energy description to recognize the structure of the spectrum of the transfer matrix as the finite $D$ manifestation of the K\"{a}ll\'{e}n-Lehmann representation of correlation functions. Finally, \Sec{ss:corr_momentum} uses momentum filtering to formulate a momentum resolved version of the proof of \cite{HastingsLR} for the relation between the correlation length and the energy gap of the system.

\subsection{Recovering the Ornstein-Zernike Form}
\label{ss:oz}

Let us first recall how the regular MPS-TM gives access to all static correlation functions in the corresponding MPS. For this we assume a complete eigendecomposition of the transfer matrix
\begin{equation}
\mpstm{}{}=\sum_{j=0}^{D^{2}}\lambda_{j}\rket{j}\rbra{j},
\label{eq:TM_eigen}
\end{equation} 
where we have dropped the subscript $A$ for notational simplicity. Here $\rket{j}$ and $\rbra{j}$ are the right and left eigenvectors of $\mpstm{}{}$ respectively with $\left(i|j \right)=\delta_{ij}$ and we again write $\lambda_{j}=\rme^{-\varepsilon_{j} + \rmi\phi_{j}}$ for the eigenvalues. We demand the ground state MPS representation to be injective, such that there is a unique dominant eigenvalue $\lambda_{0}=1$ and $|\lambda_{j>0}|<1$ for all other eigenvalues. Furthermore we define the \textit{operator transfer matrix} (OTM) for operators $O$ acting on $n$ sites as 
\begin{equation}
\otm{O}=\sum_{i_{1}\ldots i_{n} \atop j_{1}\ldots j_{n}} O^{i_{1}\ldots i_{n}}_{j_{1}\ldots j_{n}}
\left(\bar{A}^{i_{1}}\ldots\bar{A}^{i_{n}}\right) \otimes \left({A}^{j_{1}}\ldots{A}^{j_{n}}\right),
\label{eq:OpTM}
\end{equation} 
where $O^{i_{1}\ldots i_{n}}_{j_{1}\ldots j_{n}}=\braket{i_{1}\ldots i_{n}|O|j_{1}\ldots j_{n}}$.

With the above definitions it is well known that \eq{eq:corr} can be written as
\begin{equation}
C_{AB}(n+1)=\sum_{j>0}f^{\rm AB}_{j}\;\lambda_{j}^{n}=\sum_{j>0}f^{\rm AB}_{j}\;\rme^{-\varepsilon_{j}n}\rme^{\rmi\phi_{j}n}
\label{eq:corr_MPS}
\end{equation} 
where we have defined the form factors
\begin{equation}
 f^{\rm AB}_{j}=\rbra{0}\otm{A}\rket{j}\rbra{j}\otm{B}\rket{0}.
 \label{eq:formfactors}
\end{equation} 
As for a finite bond dimension \eq{eq:corr_MPS} is a finite sum of exponentials, connected correlation functions for sufficiently large distances $n$ must decay as a pure exponential $\nrm{C_{AB}(n)}\sim \exp(-n/\xi)$ where the correlation length $\xi$ corresponds to the inverse of the smallest $\varepsilon_j$ with non-zero form factor.

In contrast, typical correlation functions in gapped phases are expected to decay at large distances $n$ as
\begin{equation}
\nrm{C_{AB}(n)}\sim n^{-\eta}\rme^{-\frac{n}{\xi}},
 \label{eq:ornzern-form}
\end{equation} 
where there is an additional power law contribution to the decay with an exponent $\eta$, which in principle depends on the operators $A$ and $B$. Close to a critical point, this form can be motivated from conformal field theory or from a general renormalization group argument. Approaching the critical point takes the correlation length $\xi\to\infty$, and a pure power law decay of correlations remains, where the scaling exponent $\eta$ can depend on the choice of operators $A$ and $B$. Sufficiently deep in a gapped phase, on the other hand, \eq{eq:ornzern-form} is known as the \emph{Ornstein-Zernike} form and $\eta$ \textit{typically} depends on the number of spatial dimensions $d$ as $\eta=d/2$ \cite{OrnsteinZernike,Kennedy}, i.e. for a one-dimensional quantum system
\begin{equation}
 \nrm{C_{AB}(n)}\sim \frac{1}{\sqrt{n}}\rme^{-n/\xi}
 \label{eq:ornzern-1d}
\end{equation} 
for large distances $n$.


\begin{figure}[tb]
 \centering
 \includegraphics[width=0.5\linewidth,keepaspectratio=true]{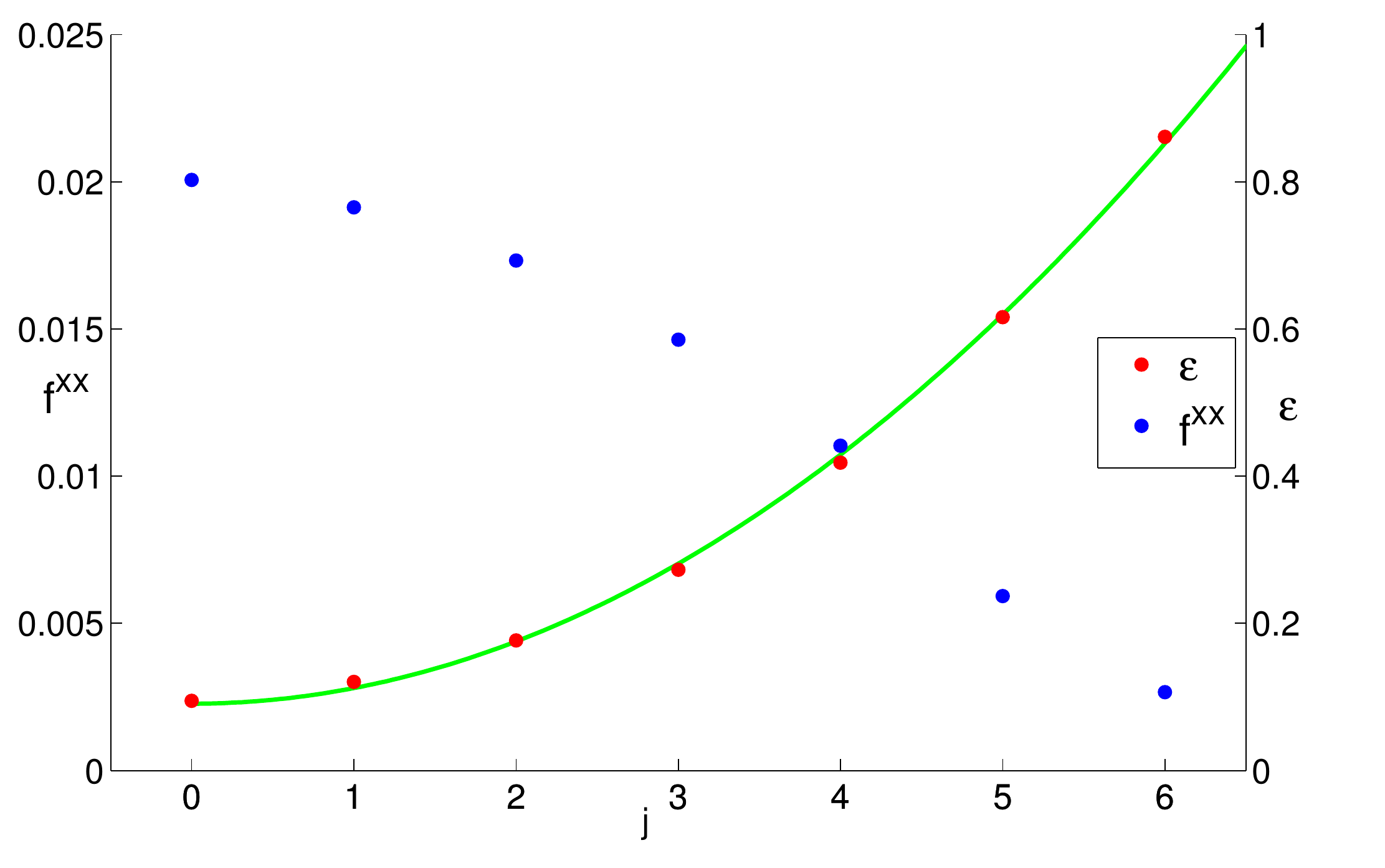}
 \caption{ 
 Plot of the form factors $f^{\rm XX}_{j}$ (blue symbols, left vertical axis, $X$ stands for $S^{x}$) as well as $\varepsilon_{j}$ along with a quadratic fit (red symbols and green line, right vertical axis) for the XY model at $\gamma=0.5$ and
$g=1.05$ for $j<7$.
Results have been obtained from an MPS ground state approximation with $D=32$ and confirm the expected values of exponents as $\kappa=2$ and $\rho=0$.}
 \label{fig:XY_formfac}
\end{figure}

In the limit $n \to \infty$, an MPS with finite bond dimension would correspond to $\eta=0$. Nevertheless, for $D\to\infty$ the correct form of the correlation functions should be restored.
We proceed by noting that the scaling form in \eq{eq:ornzern-form} can be obtained from \eq{eq:corr_MPS} under the following assumptions:
\begin{enumerate}
\item The leading eigenvalues $\lambda_{j}$ arrange themselves on a finite number of lines in the complex plane with constant complex argument $\phi_{\alpha}$, where we label these lines with index $\alpha$. Consider also the form factors $f^{\rm AB}_{j}$ as defined in \eq{eq:formfactors} along such a line. We define the subset of indices $\left\{j_{\alpha}\big\vert\phi_{j}=\phi_{\alpha}\;{\rm and}\; f^{\rm AB}_{j}\neq0\right\}$, i.e. indices $j_{\alpha}$ label all eigenvalues with constant complex argument $\phi_{\alpha}$ for which the form factors $f^{\rm AB}_{j}$ are non-zero.
\item On each of these lines the $\lambda_{j_{\alpha}}$ become sufficiently dense for $D \gg 1$ and the corresponding $\varepsilon_{j_{\alpha}}$ will follow some dispersion. We will then reorder the indices $j_{\alpha}$, such that the $\varepsilon_{j_{\alpha}}$ are in ascending order and $\Delta_{\alpha}=\varepsilon_{j_\alpha=0}$ is the minimum.  In the complex plane this parameterization corresponds to going from the $\lambda_{j_{\alpha}}$ closest to the unit circle along 
line $\alpha$
towards the center. For small $j_\alpha$,  $\varepsilon_{j_\alpha}$ changes smoothly as $\varepsilon_{j_\alpha} = \Delta_{\alpha} + g\, j_\alpha^{\kappa}$ to leading order with some constants  $g,\kappa>0$, possibly depending on $\alpha$. 
\item The form factors $f^{\rm AB}_{j_{\alpha}}$ also vary smoothly and follow, to leading order, some power law $f^{\rm AB}_{j_{\alpha}} \sim j_{\alpha}^\rho$ where the exponent $\rho$ depends on the operators $A$ and $B$ and possibly $\alpha$. For $\rho=0$ the leading order corresponds to a non-zero constant. 
\end{enumerate}

With the above assumptions and using the Euler-McLaurin formula to approximate the discrete sum with an integral, the contribution of one line $\alpha$ of eigenvalues in \eq{eq:corr_MPS} becomes approximately
\begin{equation}
\rme^{\rmi\phi_{\alpha}n}\rme^{-\Delta_{\alpha}n}\int_{0}^{\infty}\rmd z\, f^{\rm AB}_{\alpha}(z)\,\rme^{-ngz^{\kappa}}
\end{equation} 
where we have replaced $j_\alpha$ with a continuous parameter $z$ and integrate to $\infty$ for convenience, as $\rme^{-ngz^{\kappa}}$ decays sufficiently fast with increasing $z$, even for moderate $n$. 

Using the saddle point approximation allows to replace $f^{\rm AB}_{\alpha}(z)$ by its dominant behavior near $z=0$, and we obtain for $n$ sufficiently large
\begin{equation}
\rme^{\rmi\phi_{\alpha}n}\rme^{-\Delta_{\alpha}n}\int_{0}^{\infty}\rmd z\, z^\rho\,\rme^{-ngz^{\kappa}} \sim \rme^{\rmi\phi_{\alpha}n} n^{-\frac{1+\rho}{\kappa}} \rme^{-\Delta_{\alpha}n}
\end{equation} 
and we have recovered \eq{eq:ornzern-form} with $\eta=\frac{1+\rho}{\kappa}$. Deep in a gapped phase, we can reasonably expect that $\kappa=2$ and with $\rho=0$ we then recover the typical Ornstein-Zernike form for one-dimensional quantum systems with $\eta=1/2$. The correlation length $\xi$ is then given by $\frac{1}{\xi}=\varepsilon_{1}=\Delta_{1}$ with $\varepsilon_{1}$ the smallest non-zero $\varepsilon_{j}$.

In \Fig{fig:XY_formfac} we show corresponding numerical evidence for the XY model, defined in \eq{eq:XY}, at $\gamma=0.5$ and $g=1.05$, i.e. in the gapped paramagnetic phase, with bond dimension $D=32$. In this phase all the eigenvalues of the transfer matrix are real, i.e. $\phi_{j}=0$. We plot $\varepsilon_{j}$ along with a quadratic fit $\varepsilon(j)=\Delta + a\,j^{2}$, as well as the form factors $f^{\rm XX}_{j}$, where $X$ stands for $S^{x}$, vs. index $j$ for $j<7$. Again, $j$ labels eigenvalues for which the form factors $f^{\rm XX}_{j}$ are non-zero, in ascending order. The data confirms the expected values of the exponents as $\kappa=2$ and $\rho=0$, yielding the Ornstein-Zernike behavior of the correlation function 
\begin{equation}
 C_{\rm XX}(n)\sim \frac{\rme^{-\Delta n}}{\sqrt{n}}
\end{equation} 
expected for this model in this parameter regime \cite{XY_Barouch}.

For a more rigorous analysis in the framework of an exact MPS formulation for the ground state of the XY model, see \cite{MarekXY}.


\subsection{Static Structure Factor and the Single Mode Approximation (SMA)}
\label{ss:sma}

Whereas the previous subsection indicates why the peculiar structure of the eigenvalue spectrum of the MPS-TM allows to recover the typical form of static correlation functions for $D\to\infty$, it makes no connection between the branches of eigenvalues appearing in this spectrum and the dispersion relations of the elementary excitations of the model. To make this connection more explicit, we now reiterate the well-known result that the Single Mode Approximation (SMA) produces dispersion relations which are strongly dependent on the static structure factor \cite{SMA1,SMA2,SMA3,SMA4,SMA5}. In particular, we will illustrate why the energies $E(k)$ of the lowest energy-momentum eigenstates $\ket{E^{\alpha}_{k}}$ of a local translation invariant Hamiltonian $H=\sum_{n}h_{n}$ become very small at momenta $k$ where the TM has eigenvalues $\lambda_{j}=\rme^{-\varepsilon_{j} + \rmi\phi_{j}}$ with $\varepsilon_{j}$ approaching zero and $\pm k=\phi_{j}$.

The generality of this subsection is based on the recent proof that elementary excitations on top of a gapped strongly correlated ground state tend to be localized or particle-like \cite{JuthoLocalized}. This means that there exists a representation of all lowest lying excited states by acting with the Fourier transform of a quasi-local operator $O^{(\ell)}$ centered around site $n=0$ and having support on sites $n\in\left[-\ell,\ell\right]$, on the ground state:
\begin{equation}
 \ket{\phi_{k}(O^{(\ell)})} = \frac{1}{\sqrt{V}}\sum_{n}\rme^{\rmi k n}O^{(\ell)}_{n}\ket{\psi_{0}},
 \label{eq:lowEvarstate}
\end{equation}
where $O^{(\ell)}_{n}=U_{n}O^{(\ell)}U_{n}^{\dagger}$ with $U_{n}$ the lattice translation operator over $n$ sites and $V$ is the volume of the system. For the remainder of this subsection, we work in a finite system with periodic boundary conditions in order to be able to define normalizable energy-momentum states. The unnormalized state in \eq{eq:lowEvarstate} becomes exponentially close to a true isolated energy-momentum eigenstate $\ket{E^{\alpha}_{k}}$ with increasing linear size $\ell$ of the support of $O^{(\ell)}$ \cite{JuthoLocalized}. 

We can readily use this set of states in a variational ansatz for low-energy excited states and we will consider their energy expectation value
\begin{equation}
 E(k)= \frac{\braket{\phi_{k}(O^{(\ell)})|H|\phi_{k}(O^{(\ell)})}}{\braket{\phi_{k}(O^{(\ell)})|\phi_{k}(O^{(\ell)})}}
 \label{eq:Eexp}
\end{equation} 
which is a good approximation for the lowest excitation energies of $H$ at momentum $k$. The operator $O^{(\ell)}$ should have zero vacuum expectation value, and can be chosen hermitian \footnote{The proof of \cite{JuthoLocalized} could have been formulated equally well using a symmetric energy filtering operation, by which a hermitian operator would be mapped to a hermitian operator under energy filtering.}. Together with the assumption of parity invariance for both $H$ and $O^{(\ell)}$, it is well known \cite{SMA4} that \eq{eq:Eexp} can be rewritten as
\begin{equation}
 E(k)= \frac{1}{2} \frac{\braket{\psi_{0}|[O_{-k},[H,O_{k}]]|\psi_{0}}}{\braket{\psi_0|O_{-k} O_{k}|\psi_0}} = \frac{1}{2}\frac{F(k)}{S(k)},
\label{eq:Eexp2}
\end{equation}
where $O_{k}=\frac{1}{\sqrt{V}}\sum_{n} \rme^{\rmi k n} O_{n}$ and we have omitted the superscript $(\ell)$ for notational simplicity. Here $F(k)$ is the double commutator expectation value known as the oscillator strength and $S(k)=S_{OO}(k)$ is the static structure factor, which is related to the static correlation function by a Fourier transform, i.e.
\begin{equation}
C_{AB}(n)=\int_{0}^{2\pi} \frac{\rmd k}{2\pi} \,S_{AB}(k)\, \rme^{\rmi k n}.
\end{equation}
In \eq{eq:Eexp2}, we have again discarded subscripts denoting the operators $O$ for notational simplicity.

For a Hamiltonian consisting of strictly local terms $h^{(m)}_{j}$ acting on sites $[j,j+m]$, i.e. with support on $m+1$ sites, $F(k)$ is a finite polynomial in powers of $\rme^{\rmi k}$ and can thus be bounded as
\begin{eqnarray}
 |F(k)|&=\Big|\sum_{nj}\rme^{\rmi kn}\braket{[O_{0},[h_{j},O_{n}]]}\Big|\nonumber\\
 &\leq4(4\ell+2m+1)(2\ell+m+1)\| O\|^{2}\| h\|.
\end{eqnarray}
The static structure factor $S(k)$ can however become very large. This will happen when the momentum $k$ corresponds to the period of an oscillating static correlation function $C(n)$ with very large correlation length. For these momenta one can therefore optimize over possible operators $O$ such that the resulting energy expectation value is very small and thus, by virtue of the variational principle, there exist excitations with small energies.

Generically, there is a one-to-one correspondence between the momenta $k$ where $S(k)$ peaks and the complex arguments $\phi_{j}$ of the transfer matrix eigenvalues. This can easily be shown in the framework of matrix product states where we use the same notation as in subsection \ref{ss:oz}. Due to translation invariance the static structure factor can be written as
 \begin{equation}
S(k)=\sum_{n}\rme^{\rmi kn}\braket{{O_{0}}O_{n}}.
\label{eq:Sk_start}
 \end{equation}
This sum can be split into a part $|x|\leq2l$ where operators ${O_{0}}$ and $O_{n}$ overlap and a part $|x|>2l$ where they commute. The first part can be bounded as 
\begin{equation}
 \sum_{n=-2\ell}^{2\ell}\rme^{\rmi kn}\braket{{O_{0}}O_{n}}=D_{\ell}(O)\leq(4\ell+1)\nrm{O}^{2}.
\end{equation} 
The remaining part can be written as
\begin{eqnarray}
 &2\Re\sum_{n>2\ell}\rme^{\rmi kn}\braket{{O_{0}}O_{n}} =\nonumber\\
 &2\Re\sum_{n>2\ell}\rme^{\rmi kn}\rbra{0}\otm{O} \mpstm{}{n-(2l+1)} \otm{O}\rket{0}=\nonumber\\
 &2\Re\,\rme^{\rmi k(2\ell+1)}\sum_{n=0}^{\infty}\rbra{0}\otm{O}\left[\rme^{\rmi k} \mpstm{}{}\right]^{n} \otm{O}\rket{0},
\end{eqnarray}
where $\otm{O}$ is the operator transfer matrix defined in \eq{eq:OpTM}.
To perform the geometric sum in the last line, we define the projector $Q = \unity - \rket{0}\rbra{0}$ that projects out the dominant eigenvector with eigenvalue one, i.e. $\mathcal{E} = Q\mpstm{}{}=\mpstm{}{}Q=Q\mpstm{}{}Q = \mpstm{}{}-\rket{0}\rbra{0}$. We can then write $\mpstm{}{n}=Q\mathcal{E}^{n}+\rket{0}\rbra{0}=\mathcal{E}^{n}Q+\rket{0}\rbra{0}$, where the second term will not contribute to $S(k)$ due to the zero vacuum expectation value of $O$. Assuming that there are no other eigenvalues with magnitude one -- which is guaranteed if the MPS is injective -- we can now safely perform the geometric sum and obtain
\begin{eqnarray}
S(k)&=D_{\ell}(O) + 2\Re \rbra{0}\otm{O}\frac{\rme^{\rmi k(2\ell+1)}}{\unity-\rme^{\rmi k}\mathcal{E}}\;\otm{O}\rket{0}\\
&=D_{\ell}(O) + 2\Re\sum_{j>0}f^{OO}_{j}\frac{\rme^{\rmi k(2\ell+1)}}{1-\rme^{-\varepsilon_{j}}e^{\rmi(k+\phi_{j})}},
\label{eq:Sk_end}
\end{eqnarray} 
where the form factors $f^{OO}_{j}$ are given by \eq{eq:formfactors}.
It is apparent that the fraction can become very large for an eigenvalue of $\mpstm{}{}$ with $\varepsilon_{j}$ close to zero (i.e. magnitude close to one) with $k$ approaching the argument $\pm\phi_{j}$ \footnote{Complex eigenvalues of $\mpstm{}{}$ come in conjugate pairs. Assuming they are arranged successively we therefore have $\varepsilon_{j}=\varepsilon_{j+1}$ and $\phi_{j}=-\phi_{j+1}$ within such a pair and \eqref{eq:Sk_end} will become large for $k$ close to $\pm \phi_{j}$ if $\varepsilon_{j}$ small.}. By the above arguments we can therefore expect low energy excited states at these momenta, assuming that the form factors $f^{OO}_{j}$ do not vanish.

In the case of symmetry breaking and topologically non-trivial excitations, similar arguments lead to the same form of $S(k)$ as in \eq{eq:Sk_end}, where $\mathcal{E}$ is replaced with the mixed MPS-TM \footnote{In this case, there is no need for the projector $Q$ as here the mixed MPS-TM has a spectral radius strictly smaller than one.}.

As a final point, we elaborate further on the relation between the static structure factor and the spectrum of the transfer matrix. For an injective MPS, it will always be possible to find an operator, whose OTM exactly excites one or more of the eigenvalues of the TM on a branch with fixed $\phi_\alpha$, i.e. $\rbra{j_{\alpha}}\otm{O}\rket{0}\neq0$. This, however, could be an operator with very large support. Conversely, it is possible that for operators with small support, several eigenvalues are excited on branches with different arguments $\phi_{\alpha}$ which are close together. For these operators, the static structure factor might then have a maximum at a momentum $k$ which does not exactly correspond to one of the arguments $\phi_{\alpha}$. 

\begin{figure}
 \centering
 \includegraphics[width=0.75\linewidth,keepaspectratio=true]{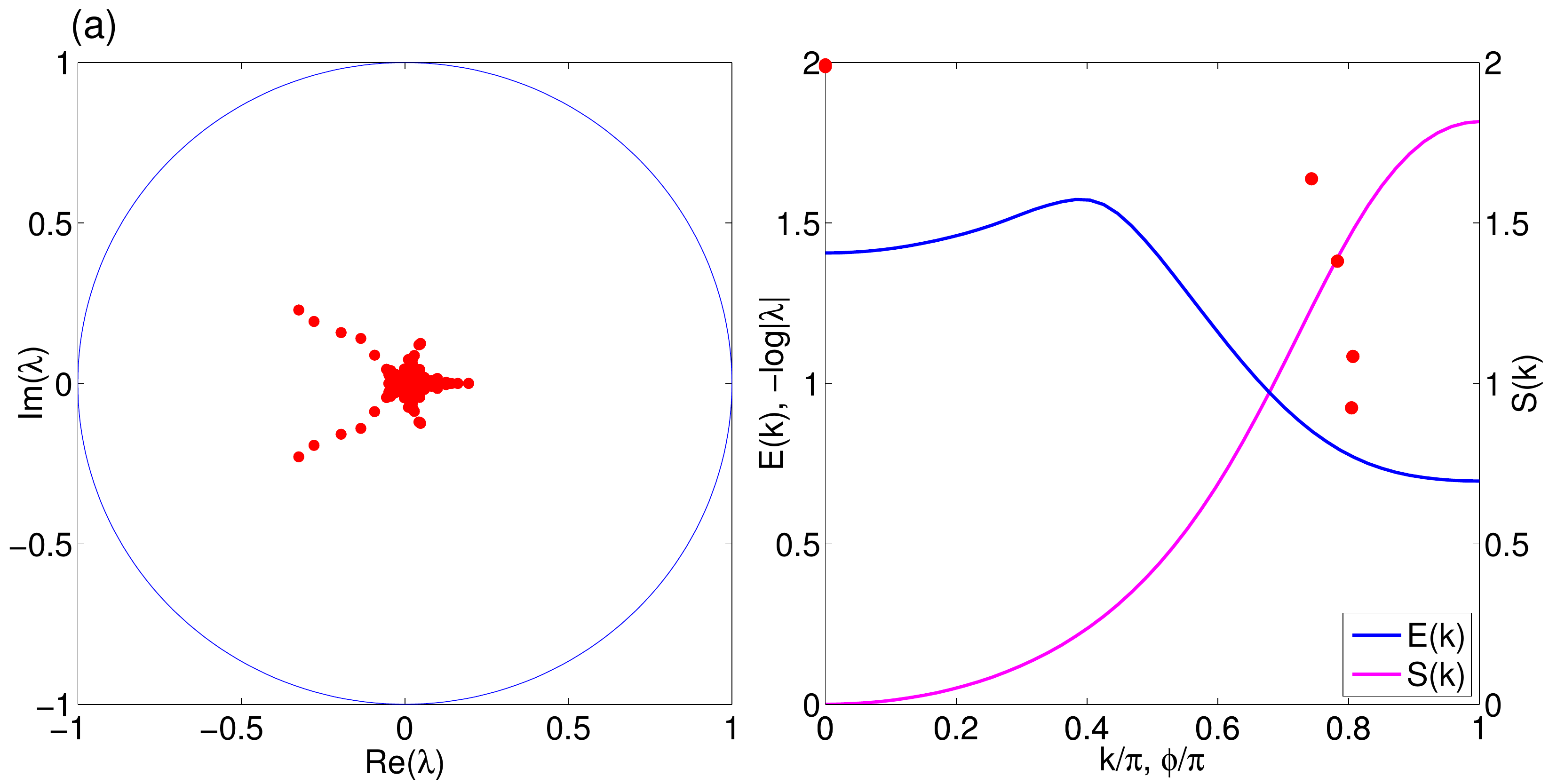}
 \includegraphics[width=0.75\linewidth,keepaspectratio=true]{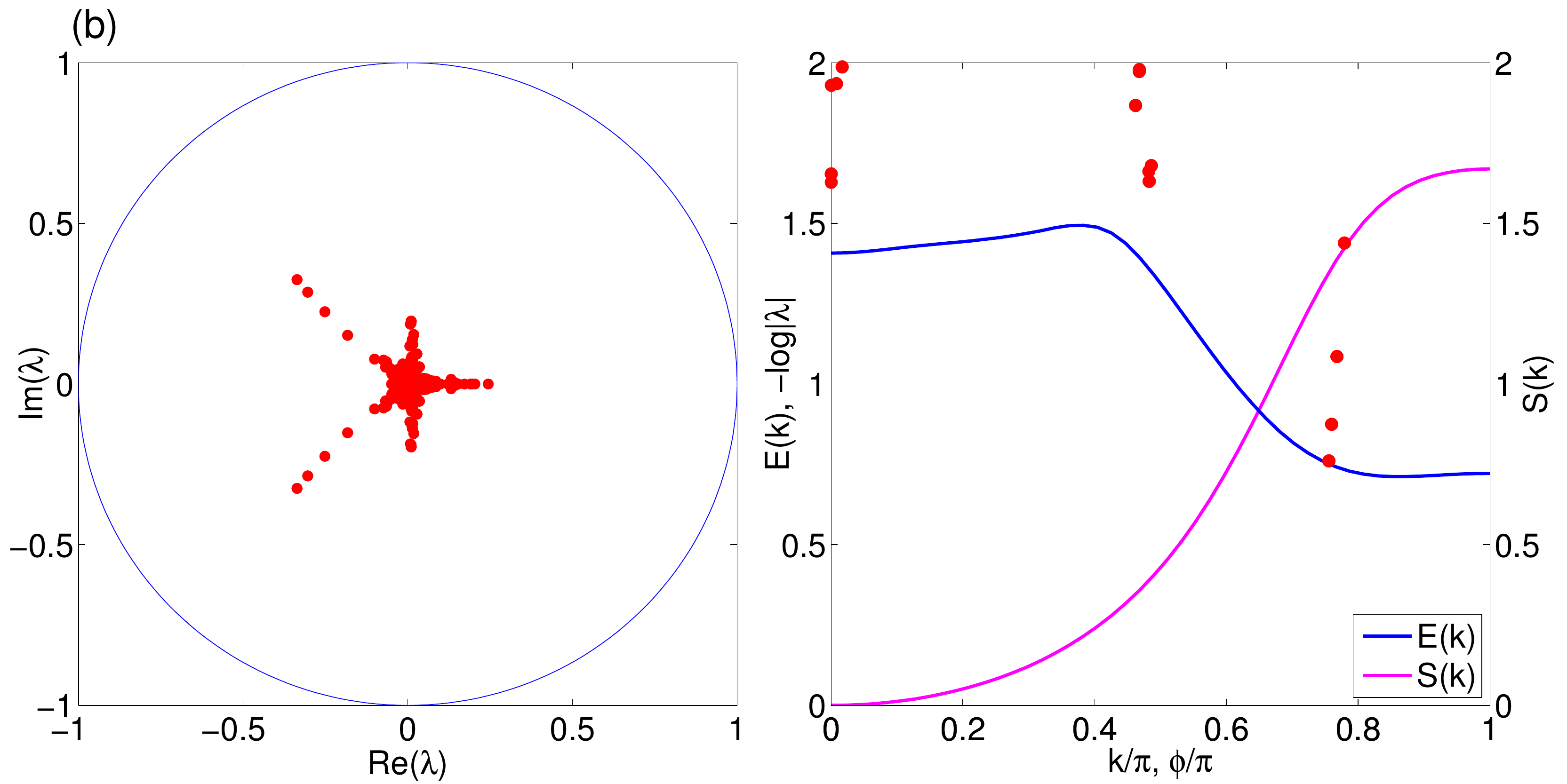}
 \includegraphics[width=0.75\linewidth,keepaspectratio=true]{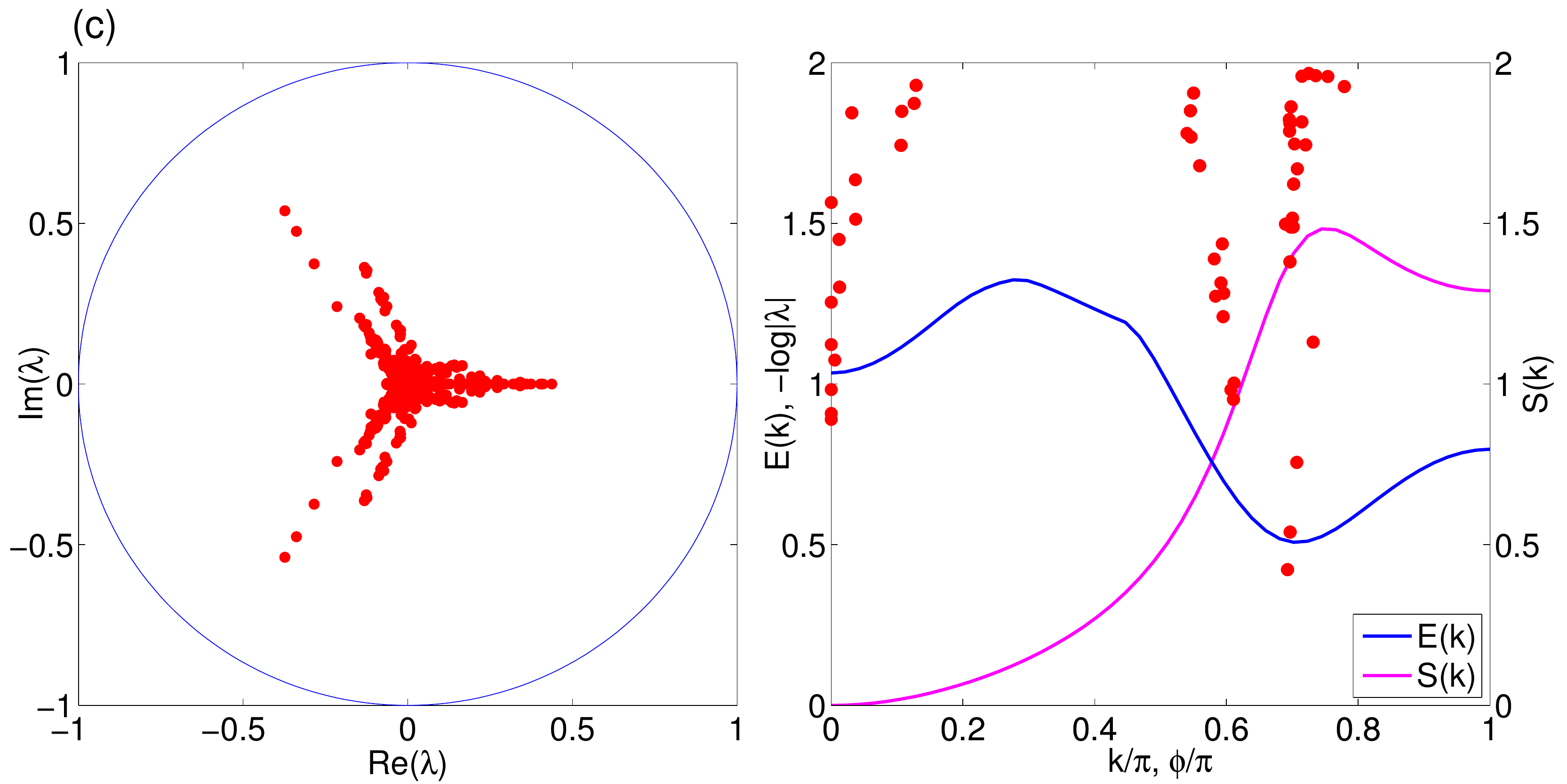}
 \caption{ 
 \textit{left column}: Eigenvalues $\lambda_{j}=\rme^{-\varepsilon_{j} + \rmi\phi}$ of the transfer matrix on the complex plane within the unit circle for the bilinear-biquadratic $S=1$ Heisenberg chain \eq{blbq} with (a) $\theta=0.11478\pi$, (b) $\theta=0.12522\pi$, (c) $\theta=0.15652\pi$ and $D=64$. \textit{right column}: static structure factor $S(k)=S_{ZZ}(k)$, where $Z$ stands for $S^{z}$ (purple line) and variationally obtained dispersion $E(k)$ (blue line) vs. momentum $k$ along with $\varepsilon_{j}$ vs. $\phi_{j}$ (red symbols). Whereas $S(k)$ still has its maximum at momentum $\pi$ in (b), the minimum $k_{\rm min}$ of the dispersion has already started shifting away from $k_{\rm min}=\pi$. In the spectrum of the transfer matrix this is reflected by the fact that the eigenvalues with largest magnitude have finished aligning along $\phi_{\alpha}\approx 0.755\pi$. The maximum of $S(k)$ doesn't start shifting until around $\theta\approx0.1314\pi$.}
\label{fig:blbq}
\end{figure}

An example for this is the transition from commensurate to incommensurate order in the bilinear-biquadratic $S=1$ Heisenberg chain \cite{SchollIncom,Nomura1,Nomura2}
\begin{equation} \label{blbq}
H_{\rm BLBQ}=\sum_{n}\cos(\theta)\, \bi{S}_{n} \cdot \bi{S}_{n+1}+\sin(\theta)\, (\bi{S}_{n} \cdot \bi{S}_{n+1})^2.
\end{equation}
We consider the regime $0<\theta<0.25\pi$, where the oscillation period of static correlation functions changes from $\pi$ to some incommensurate period exactly at the AKLT-point $\theta_{\rm VBS}=\arctan(1/3)\approx 0.1024\pi$. However by looking at the static structure factor of a simple one-site operator such as $S^z$, the peak stays at $k=\pi$ until some significantly larger value $\tilde{\theta}\approx 0.1314\pi$. Based on the relation between the structure factor and the minima of the dispersion relation, we have reason to expect that the dispersion relation starts shifting away from $\pi$ at this latter value of $\theta$. It appears, however, that the minimum of the dispersion relation starts shifting at $\theta\approx 0.12\pi$, a value in between $\theta_{\rm VBS}$ and $\tilde{\theta}$. In \Fig{fig:blbq} we observe from the full spectrum of the transfer matrix, that this happens when the eigenvalues with largest magnitude have finished aligning along the line of constant 
phase $\phi_{\alpha}\approx0.755\pi$. At this point it appears that the support of the operator $O$ generating the excitation, has shifted from the -- by now very small -- branch with $\phi_{\alpha}=\pi$ to the now fully aligned branch with $\phi_{\alpha}\approx0.755\pi$. This value of $\phi_{\alpha}$ further shifts towards $2\pi/3$ with $\theta\to0.25\pi$, where the gap then closes at momenta $k=0,\pm 2\pi/3$ \cite{LaiSutherland1,LaiSutherland2,LaiSutherland3}. 

From this example it is apparent that especially in the vicinity of such peculiar crossover points, it is worthwhile to look at the full spectrum of the transfer matrix. It generically contains more information than a simple static structure factor, as the transfer matrix is completely independent of the choice of operator. It illustrates that there is a crossover regime where the MPS-TM has to react by developing additional branches with constant $\phi_{\alpha}$ and the support of the operator $O$ generating the excitation has to shift to this newly developed branch. It appears that this process is completed after the eigenvalues have fully aligned along the newly developed branch. It would be interesting to investigate this process further with the precise knowledge about the operator $O$ generating the excitation and how the form factors $\rbra{j_{\alpha}}\otm{O}\rket{0}$ develop with $\theta$. 

For now we conclude with the observation that there are indeed also situations where the locations of the lowest lying TM eigenvalues do not precisely coincide with the minima of the dispersion. These are however very special cases like the one discussed above,
where the TM has to adapt to changing conditions within some crossover regime. There the peculiar structure of the TM eigenvalue spectrum is however also different from normal situations as investigated in \Sec{s:numerics}, which can serve as an indicator for such exceptional situations.


\subsection{K\"{a}ll\'{e}n-Lehmann representation}
\label{ss:kl}
In the previous two subsections, we have discussed two effects of the peculiar distribution of the eigenvalues of the MPS-TM. Firstly, the clustering of eigenvalues onto lines starting from the origin allows to recover the typical form of static correlation functions in gapped quantum ground states in the limit $D\to\infty$. Secondly, this distribution causes peaks in the static structure factor, which can be related to minima in the dispersion relation of excitations using the single mode approximation. In retrospect, the first effect only requires a dense distribution of eigenvalues along a line, without any connection between the location of these lines and the dispersion relation of the excitations of the system. The second argument only requires the existence of a single dominant eigenvalue with a phase corresponding to the momentum of the minimum of the dispersion relation, and does not explain why there needs to be a dense distribution of eigenvalues. This leaves open the question whether a different 
structure of the eigenvalue distribution could give rise to similar effects. Put differently, we would like to answer the reverse question, i.e. to what extent are static correlation functions and the clustering of the eigenvalues of the MPS-TM determined by the excited states of the Hamiltonian.

For Lorentz-invariant field theories, the K\"{a}ll\'{e}n-Lehmann representation of two-point correlation functions provides such a direct connection to the excitation spectrum of the Hamiltonian. Whereas the K\"{a}ll\'{e}n-Lehmann representation exists for arbitrary dynamical correlation functions, we here specialize to the case of static correlation functions between two scalar operators $A(x)$ and $B(y)$, where it is given by
\begin{eqnarray}
\fl
\braket{\Psi|A(x)B(y)|\Psi}=\nonumber\\
\int \rmd M^2 \rho(M^2) \int \frac{\rmd^d k}{(2\pi)^d} \frac{\rme^{\rmi k (x-y)}}{2 \sqrt{M^2+k^2}}
\braket{\Psi|A(0)|M^2,0}\braket{M^2,0|B(0)|\Psi},
\label{eq:kallenlehmann}
\end{eqnarray}
where -- contrary to standard field theory notation -- $x$ and $y$ denote spatial vectors, $k$ denotes momentum and $d$ represents the number of spatial dimensions. The first integral is over all possible masses in the theory, and $\rho(M^2)$ corresponds to the density of states. If the lowest lying excitations correspond to single particle excitations with discrete masses $M_\alpha$, then $\rho(M^2)$ will contain a contribution $\sum_{\alpha} \delta(M^2-M_{\alpha}^2)$. The state $\ket{M_{\alpha}^2,0}$ corresponds to the presence of such an excitation with mass $M_{\alpha}$ and momentum zero. 

Let us again restrict to the case of $d=1$. We are interested in the long-range behavior of correlation functions in a lattice model, which is clearly dictated by the low-energy behavior of the model. If this low-energy behavior can be captured by a Lorentz-invariant theory with masses $M_{\alpha}$, a remnant of the K\"{a}ll\'{e}n-Lehmann representation of correlation functions should exist in the lattice model. For the field theory, the minimum of all dispersion relations is at momentum zero. However, in many cases taking the continuum limit of a lattice theory requires that e.g. $N$ sites are blocked, and a single lattice dispersion relation with several minima for momenta $\phi_{\alpha}=2\pi\alpha/N$ with $\alpha=0,\ldots,N-1$ gives rise to $N$ independent dispersion relations of the field theory. A prototypical example for $N=2$ is the XX-model, which corresponds to the staggered fermion discretization of relativistic Dirac fermions, where the two components of the Dirac spinor are put on even 
and odd sites respectively \cite{Susskind}.

Equation \eq{eq:kallenlehmann} represents the static correlation function in real space as an integral over momentum space. The corresponding representation for the static structure factor in momentum space can therefore easily be identified. In the Lorentz-invariant case, the single particle excitations add a contribution to the static correlation function that is given by a constant times the inverse of the dispersion relation of that excitation. For typical operators, these will be the dominant contributions. If the low-energy behavior of a lattice model is Lorentz-invariant, we can thus expect that $S(k)$ should also receive contributions $S_{\alpha}(k)$ of the form
\begin{equation}
S_{\alpha}(k)=c_{\alpha}\,[m_\alpha^2+v_{\alpha}^2 (k-\phi_{\alpha})^2]^{-1/2}.
\label{eq:rel_disp}
\end{equation}
Here, $c_{\alpha}$ is a constant depending on the choice of operators $A$ and $B$, $m_{\alpha}$ is the mass of the excitations in units of the inverse lattice spacing and $v_{\alpha}$ is the characteristic velocity, which for a proper Lorentz-invariant low-energy behavior should be the same (and thus independent of $\alpha$) for all excitations.

\begin{figure}
 \centering
 \includegraphics[width=0.3\linewidth,keepaspectratio=true]{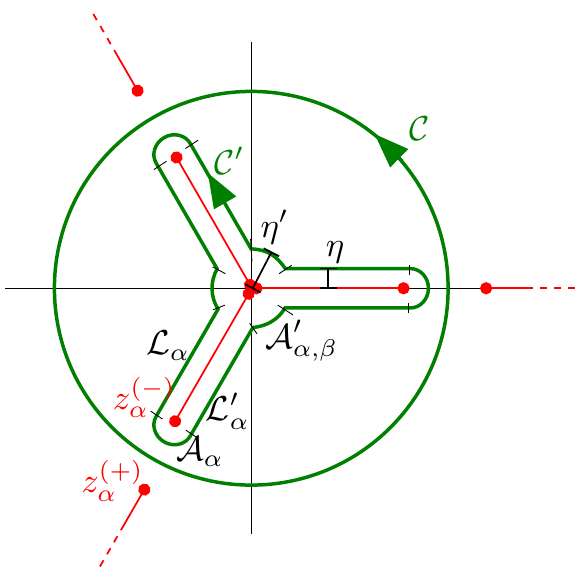}
 \caption{ 
 Change of integration contour: the original contour $\mathcal{C}$ corresponds to the unit circle, and is mapped to a contour $\mathcal{C}'$ consisting of line sigments $\mathcal{L}_{\alpha}$ and $\mathcal{L}'_{\alpha}$ at distance $\eta$ from every branch cut between $z_{\alpha}^{(-)}$ and $0$, as well as arc segments $\mathcal{A}_{\alpha}$ rotating by $\pi$ around $z_{\alpha}^{(-)}$ with radius $\eta$, and arc segments $\mathcal{A}'_{\alpha,\beta}$ rotating by $2\pi/N$ around the origin at radius $\eta'$, between the branch cuts corresponding to $z_{\alpha}^{(-)}$ and $z_{\beta}^{(-)}$. }
 \label{fig:contour}
\end{figure}

Since this form is only expected to hold for $k\approx \phi_{\alpha}$ we can instead choose a $2\pi$-periodic version of the dispersion relation to get contributions of the form
\begin{equation}
S_{\alpha}(k)=c_{\alpha}\,\{m_\alpha^2+v_{\alpha}^2 [2-2\cos(k-\phi_{\alpha})]\}^{-1/2}.
\label{eq:Scontributions}
\end{equation} 
We can now transform from the $2\pi$-periodic variable $k$ to the complex variable $z=\rme^{\rmi k}$ and we change the notation of the static structure factor as $S(k)\to S(z=\rme^{\rmi k})$ to write
\begin{equation}
C(n)=\oint_{\mathcal{C}} \frac{\rmd z}{2\pi \rmi}\, S(z)\, z^{n-1},
\label{eq:correlation_contour}
\end{equation}
where the contour integral is over the unit circle $\mathcal{C}$, which is where $S(z)$ is originally defined (see \Fig{fig:contour}). 

If we now were to construct an analytic continuation of $S(z)$, we expect that every contribution of the form of \eq{eq:Scontributions} produces a square root singularity at $z=0$ and inverse square root singularities at the points
\begin{eqnarray}
z^{(\pm)}_{\alpha} &= \rme^{\rmi\phi_{\alpha}} \left[1 \pm \Delta_{\alpha}\left(\sqrt{1+\Delta_{\alpha}^2/4}\pm \Delta_{\alpha}/2\right)\right] \nonumber\\
&\approx \rme^{\rmi \phi_{\alpha}\pm \Delta_{\alpha}+\Or(\Delta_{\alpha}^3)}
\label{eq:branch_roots}
\end{eqnarray}
with $\Delta_{\alpha}=m_{\alpha}/v_{\alpha}$. We can thus choose the branch cuts to go from $z_{\alpha}^{(-)}$ to $0$ and from $z_{\alpha}^{+}$ to $+\infty$. Assuming that there
are no other singularities, we can then deform the integration contour as in \Fig{fig:contour}. In the limit $\eta,\eta'\to 0$ the arc segments $\mathcal{A}_{\alpha}$ and
$\mathcal{A}'_{\alpha,\beta}$ do not contribute, as they correspond to square root and inverse square root singularities, whereas line segments $\mathcal{L}_{\alpha}$ and
$\mathcal{L}_{\alpha}'$ produce equivalent contributions
\begin{equation}
\rme^{\rmi\phi_{\alpha}n} \int_{0}^{\rme^{-\Delta_{\alpha}}}  \rmd y\,y^{n-1/2}\,[(y-\rme^{-\Delta_{\alpha}})(y-\rme^{+\Delta_{\alpha}})]^{-1/2}.
\label{eq:branch_contrib}
\end{equation}
where we have transformed to $y=z\rme^{-\rmi\phi_{\alpha}}$ and the contributions to the correlation function $C(n)$ are now written as 
integrals over $0\leq y\leq \rme^{-\Delta_{\alpha}}$ along the branch cuts.

It is apparent that a finite $D$ MPS-approximation of the static correlation function tries to reproduce this continuum form with a discrete sum over the eigenvalues of the transfer matrix. These eigenvalues cluster on the branch cuts of the static structure factor, which we have related to the single particle excitation spectrum by assuming a Lorentz-invariant low-energy behavior. 

For large $n$, where the low-energy contributions are dominating and the approximations are valid, we can again use the saddle point approximation around the point $y=\rme^{-\Delta_{\alpha}}$ and change variables to $y=\rme^{-(\Delta_{\alpha}+x)}$, to obtain for the contributions $C_{\alpha}(n)$ to $C(n)$ up to some factor
\begin{equation}
C_{\alpha}(n)\sim \rme^{\rmi\phi_{\alpha} n}\rme^{-\Delta_{\alpha} n}\int_{0}^{\infty} \rmd x\, \frac{\rme^{-x (n-\frac{1}{2})}}{\sqrt{x}}\,\Big(1+\Or(x)\Big)\sim \frac{\rme^{\rmi\phi_{\alpha} n}\rme^{-\Delta_{\alpha} n}}{\sqrt{n-1/2}},
\label{eq:OZlorentz}
\end{equation}
which reproduces correlations of the Ornstein-Zernike form, with correlation lengths $\xi_{\alpha}^{-1}=\Delta_{\alpha}$. The dominant contribution for $n$ large therefore stems from the branch with smallest $\Delta_{\alpha}=m_{\alpha}/v_{\alpha}$.

It is interesting to compare this to the discussion of the previous section. The single mode approximation for the dispersion relation $E(k)$ was given in \eq{eq:Eexp2} as $E(k)=F(k)/[2 S(k)]$, which allowed to conclude that peaks in the static structure factor $S(k)$ produce minima in the corresponding dispersion relation. Assuming Lorentz-invariance, the K\"{a}ll\'{e}n-Lehmann representation produces contributions to $S(k)$ of the form $S_{\alpha}(k)\sim c_{\alpha}/E(k)$ with $c_{\alpha}$ a $k$-independent constant, which looks like the reverse relation: minima in the dispersion relation give rise to peaks in the static structure factor. However, the single mode approximation also allows for minima which are not caused by $S(k)$ but rather by a small or vanishing value for the oscillator strength $F(k)$. It would be interesting if one could show that such excitations are necessarily related to low-energy features which have an intrinsically non-relativistic description.

As a final justification for the argumentation in this subsection, we apply the above results to the case of the XY-model in the incommensurate gapped phase investigated in \Sec{ss:1dlattice}. There it is observed that for the elementary excitations the smallest excitation energy $E_{\rm min}$ (i.e. the energy gap) and the eigenvalue of the transfer matrix with second largest magnitude $\lambda_{1}$ are related by some characteristic velocity $v_{1}$ via $\varepsilon_{1}=E_{\rm min}/v_{1}$, where $\varepsilon_{1}=-\log|\lambda_{1}|$ and $\varepsilon_{1}=\xi^{-1}$ with $\xi$ the correlation length as established in \Sec{ss:oz}. If we assume a Lorentz-invariant low energy behavior -- with some characteristic velocity $v_{c}$ -- of the dispersion $E(k)$  around the minimum of the form
\begin{equation}
 E(k)\approx\sqrt{E_{\rm min}^{2} + v_{c}^{2}(k-k_{\rm min})^{2}},
 \label{eq:lorentz_disp}
\end{equation}
as also used in \eq{eq:rel_disp}, we can deduce by virtue of \eq{eq:branch_roots} and \eq{eq:OZlorentz} that $\xi^{-1}=\Delta_{1}=E_{\rm min}/v_{c}$, where we have interpreted the energy gap $E_{\rm min}$ as the lowest mass $m_{1}$. On the other hand with $\xi^{-1}=\varepsilon_{1}=E_{\rm min}/v_{1}$ we thus see that $v_{1}=v_{c}$ is exactly the characteristic velocity appearing in \eq{eq:lorentz_disp}.

With $E(k)=\sqrt{(g-\cos(k))^{2}+\gamma^{2}\sin^{2}(k)}$ known exactly for the XY-model \cite{XY_Katsura,XY_Barouch,XY_BunderMcKenzie} and assuming \eq{eq:lorentz_disp} to hold around $k=k_{\rm min}$ we can then estimate $v_{1}$ as
\begin{equation}
 v_{1}^{2}=\frac{1}{2}\frac{\rmd^{2}E(k)^{2}}{\rmd k^{2}}\Big|_{k=k_{\rm min}}=\frac{(1-\gamma^{2})^{2} - g^{2}}{1-\gamma^{2}}.
 \label{eq:v_from_lorentz_disp}
\end{equation}
For the parameters $\gamma=0.3$ and $g=0.2$ considered in \Sec{ss:1dlattice} this yields an estimate of $v_{1}=0.9306$, which differs from the value $v_{1}=0.9409$ obtained from $v_{1}=E_{\rm min}/\varepsilon_{1}$ by only $\approx 1\%$.

\subsection{Momentum Resolved Relation between Correlation Length and Gap}
\label{ss:corr_momentum}

In this section we derive more rigorous statements connecting the decay of static connected correlation functions to the dispersion $E(k)$ of low energy excitations of local translation invariant Hamiltonians in the thermodynamic limit. We generalize the seminal work of Hastings \cite{HastingsLR} in which it is proven that the inverse of the energy gap $\Delta$ of a local translation invariant Hamiltonian times a constant serves as an upper bound for the correlation length $\xi$ of connected static correlation functions. This implies that if the gap vanishes, these correlations may (and in most cases will) be long ranged with a diverging correlation length. 

The proof gives a statement relating the smallest overall excitation energy and the largest correlation length in the system, but does not take into account momentum information. In this work we extend the results in \cite{HastingsLR} to derive bounds on the decay of \textit{momentum-filtered} correlation functions and relate them to the dispersion $E(k)$ of low energy excitations. Specifically, we show that the inverse of the energy gap $E(k)$ at a specific momentum $k$ times a constant serves as an upper bound for a momentum-resolved correlation length $\xi_{k}$. Conversely, the existence of a finite correlation length $\xi_{k}$ thus implies an upper bound for the energy $E(k)$.

The detailed derivation of the bound in the most general setting is given in \ref{a:proof}; here we present the result for one-dimensional lattice systems and operators acting on single sites for the sake of simplicity. We start from the static connected correlation function of operators $A_{i=0}$ and $B_{j=\ell>0}$, for which we assume zero vacuum expectation value. We now attempt to extract momentum-space information while retaining real space information by replacing operator $B_{\ell}$ at site $\ell$ with a gaussian wave packet centered around site $\ell$, defined as
\begin{equation}
 \tilde{B}_{\ell}(k) = N_{r} \sum_{n} \rme^{-\frac{n^{2}}{2r}}\rme^{\rmi kn}B_{\ell+n}
\end{equation} 
where $N_{r}$ is a normalization constant. We define the momentum-filtered correlation function as
\begin{equation}
 C_{k}(\ell) = \braket{A_{0}\tilde{B}_{\ell}(k)},
 \label{eq:corr_kfilter}
\end{equation}
which corresponds to a Fourier transform of the product of the static correlation function and a gaussian wave packet centered around site $\ell$. In momentum space this yields the convolution of the static structure factor $S(k)$ and another gaussian wave packet in momentum space.

In \ref{a:proof} it is proven that by tuning $r$ as a fraction of $\ell$, the momentum-filtered correlation function $C_k(\ell)$ can -- for sufficiently large $\ell$ -- be bounded by
\begin{equation}
 |C_{k}(\ell)|\leq c_{1} \| A\| \| B\| \rme^{-c_{2}\ell},
 \label{eq:corrk_bound1}
\end{equation} 
where $c_{1}$ and $c_{2}$ are some constants and the inverse of $c_{2}$ is an upper bound for the momentum-filtered correlation length $\xi_{k}$, given by
\begin{equation}
\xi_{k}\leq \frac{1}{c_{2}}=\frac{1}{\delta} + \frac{v_{\rm LR}}{E^{\ast}(k,\delta)}.
\label{eq:corrk_bound2}
\end{equation} 
Here $v_{\rm LR}$ is the characteristic Lieb-Robinson velocity \cite{LR} and $E^{\ast}(k,\delta)=\min_{|k-k'|\leq\delta}E(k')$ is the minimum of the dispersion $E(k)$ in an interval around $k$ given by $\delta$.

The constant $\delta$ is introduced in the proof and and can be tuned to obtain the sharpest bound. From \eq{eq:corrk_bound2} it is clear that there is a tradeoff, as increasing $\delta$ generally leads to a decrease of $E^{*}(k,\delta)$. If $k$ however corresponds to a minimum in the dispersion $E(k)$ then the function $E^{*}(k,\delta)$ is largely insensitive to $\delta$ in some region around the minimum and we can choose $\delta$ as large as possible within this region. If $k$ on the other hand corresponds to a regular point where $\frac{\rmd E}{\rmd k}\neq 0$ then there is a direct effect from increasing $\delta$ to decreasing $E^{*}$. An optimal choice of $\delta$ is thus dependent on the form of $E(k)$.

Colloquially, the above result means that the decay of the momentum-filtered correlation functions is dictated by the corresponding low energy states around that given momentum. Equivalently, one can say that a large correlation length for a given momentum $k$ implies a small excitation energy $E(k)$. 

In practice, this means that one can deduce an upper bound for the low-energy spectrum of the Hamiltonian by looking at the momentum dependence of momentum-filtered correlation functions, where one would have to optimize over the parameter $\delta$ \footnote{To find the optimal bound one would have to optimize over all possible operators $A$ and $B$ with arbitrary finite support as well.}.

\section{The Quantum Transfer Matrix (QTM)}
\label{s:qtm}
This section uses renormalization group arguments to establish a close relationship between the MPS-TM and the exact Quantum Transfer Matrix of the model, thus providing a direct connection between the MPS-TM and the spectral properties of the Hamiltonian. 

\subsection{Imaginary Time Evolution as Tensor Network}
\label{ss:imtime}
Consider a one-dimensional local lattice Hamiltonian $H$ with translation invariance and a unique ground state $\ket{\psi_{0}}$ with ground state energy $E_{0}=0$. To avoid issues with the infrared orthogonality catastrophe, we work in a finite system of $N$ sites with periodic boundary conditions for this section. The ground state can be obtained by an evolution in imaginary time $\beta$
\begin{equation}
\ket{\psi_{0}}=\lim_{\beta\to\infty}\frac{\rme^{-\beta H}\ket{\phi_{\rm init}}}{\| \rme^{-\beta H}\ket{\phi_{\rm init}} \|},
\end{equation} 
with $\ket{\phi_{\rm init}}$ some initial state which is non-orthogonal to $\ket{\psi_{0}}$. Since $H$ is a sum of local terms $h_{n}$, we can break $\beta$ into small imaginary-time steps $\delta$ and use a Suzuki-Trotter decomposition of $\rme^{-\delta H}\approx\prod_{n}\rme^{-\delta h_{n}}$\cite{Trotter,Suzuki}. There are several strategies to write this (or alternative) decomposition(s) as a 2-dimensional tensor network with translation invariance in the spatial direction \cite{TITrotter1,Sirker,Huang}. Obviously, all the information of $H$ is thus encoded into this tensor network. 

\begin{figure}
 \centering
 \includegraphics[width=0.6\linewidth,keepaspectratio=true]{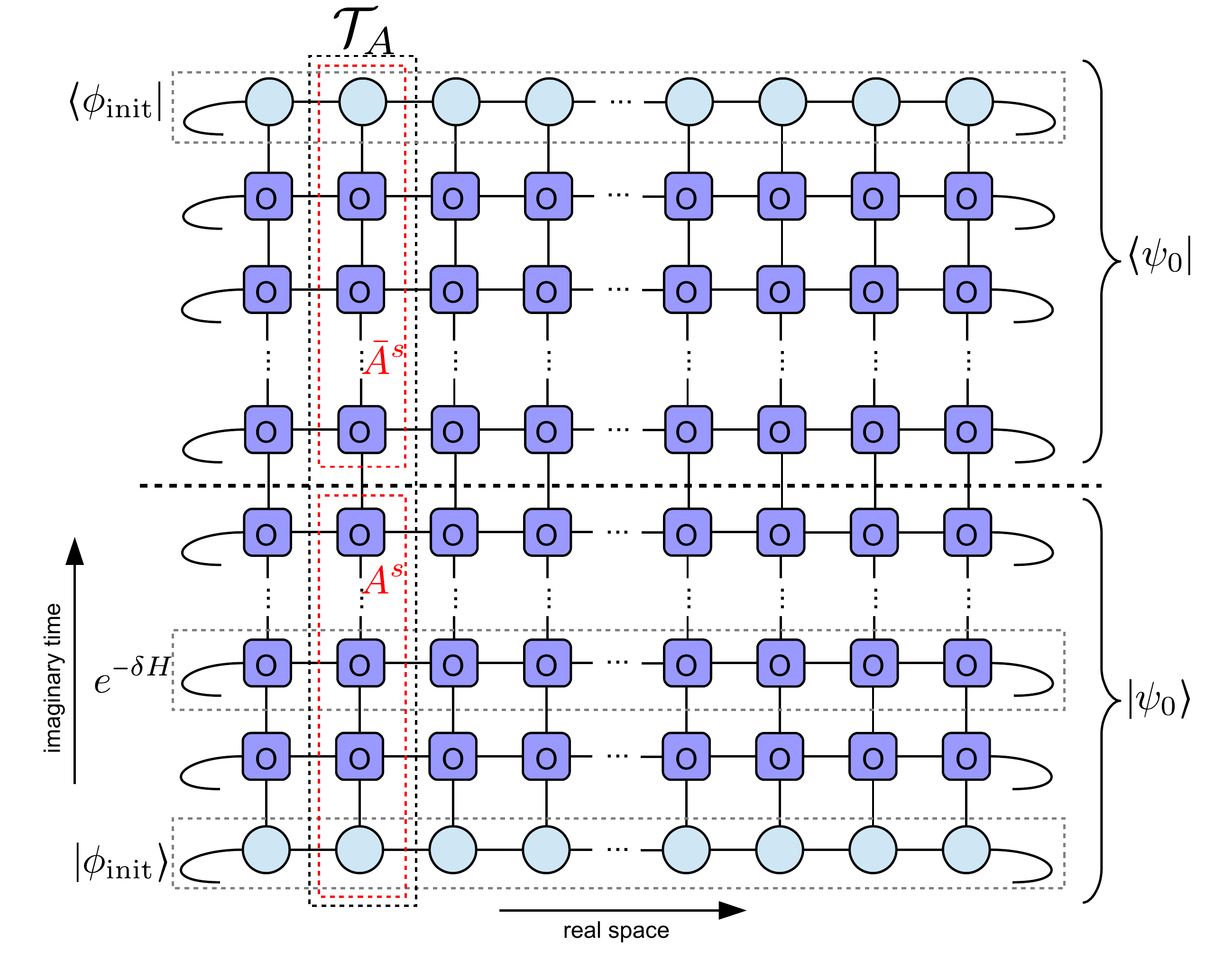}

 \caption{ 
 Two-dimensional tensor network representing the ground state $\ket{\psi_{0}}$ of a one-dimensional local lattice Hamiltonian $H$. The horizontal slices correspond to a
decomposition of $\rme^{-\delta H}$ into a translation invariant MPO with tensors $O$. The ground state is obtained by successively applying $\rme^{-\delta H}$ onto an initial state
$\ket{\phi_{\rm init}}$.
Grouping contractions along the vertical imaginary time axis, $A^{s}$ can be interpreted as a translation invariant MPS representation of the ground state.
The MPS-TM is then identified as a single column of the tensor network representing the partition function at zero temperature
$Z_{\beta\to\infty}=\braket{\psi_{0}|\psi_{0}}$. 
 }
 \label{fig:ImTimeTN}
\end{figure}

If the state $\ket{\phi_{\rm init}}$ is initially in the form of a translation invariant MPS, then we can also interpret $\ket{\psi_{0}}$ as a translation invariant MPS by grouping contractions along imaginary time. Each column of the tensor network is an MPS matrix $A^{s}$, itself being a half-infinite matrix product operator (MPO). $A^{s}$ then exactly represents the ground state up to a Trotter error \footnote{The limit of infinite Trotter number ($\delta\to 0$) is in general possible \cite{CT1,CT2,CT3}.}. A graphical representation of this construction is given in \Fig{fig:ImTimeTN}. 

We immediately see that $\braket{\psi_{0}|\psi_{0}} = \Tr\mpstm{A}{N}$ where the MPS transfer-matrix $\mpstm{A}{}$ is defined in \eq{eq:mpstm}. For a system with a unique ground state, the boundary conditions at $\beta=0$ are irrelevant and an equivalent network with periodic boundary conditions in the temporal direction would be obtained for the thermal partition function $Z_{\beta}=\Tr \rme^{-\beta H}$ in the limit $\beta\to\infty$. The exact MPS-TM for the ground state $\ket{\psi_{0}}$ thus corresponds to the quantum transfer matrix (QTM) at zero temperature, defined in Refs.~\cite{Suzuki,Betsuyaku}.

It is important to note that all the information about the QTM -- in particular its eigenvalues -- is thus contained within the ground state $\ket{\psi_{0}}$ and its exact MPS representation $A^{s}$. Note, however, that this exact representation with exponentially diverging bond dimension differs from a finite $D$ approximation $\tilde{A}^{s}$ that can be obtained for instance from some variational algorithm. For an example of an analytic derivation of such an exact ground state MPS representation $A^{s}$ and the effect of truncating to finite $D$ for the case of the $S=1/2$ XY model, see \cite{MarekXY}.

In \Sec{ss:RG} we present a construction how such a finite $D$ approximation -- which only retains degrees of freedom relevant for the physical degree of freedom $s$ -- can be obtained from this exact MPS representation $A^{s}$ and thus from the true QTM. To understand how this relates the MPS-TM to the Hamiltonian of the system, we first need to discuss how the latter relates to the exact QTM, which is the topic of the next subsection. 

\subsection{The QTM and the Hamiltonian}
\label{ss:qtm_ham}
To understand what information about the underlying Hamiltonian can be extracted from the knowledge about the QTM, we first consider the case of relativistic (1+1)-dimensional field-theories. The vacuum or ground state can be expressed in terms of a path integral formulation very similar to 
\Fig{fig:ImTimeTN}, but where both real space and imaginary time are continuous. We can then identify a factor $\rme^{-\delta H}$ for some infinitesimally small $\delta$ with a narrow horizontal slice of the entire network. Due to relativistic invariance, however, such slices are invariant under Euclidean rotations between real space and imaginary time. This means in particular that a vertical slice-- corresponding to a field theory analogue of $\mpstm{}{}$ -- is also equal to $\rme^{-\delta H}$. Hence, knowing the spectrum of $\mpstm{}{}$ immediately yields knowledge about the spectrum of $H$. This in turn implies that all the information about the eigenvalues of $H$ is already contained within the ground state. Of course, Lorentz-invariance also strongly restricts the dispersion relations of the theory, such that they are completely 
determined by a single parameter, being the mass of the excitation.

\begin{figure}[ht]
 \centering
 \includegraphics[width=\linewidth,keepaspectratio=true]{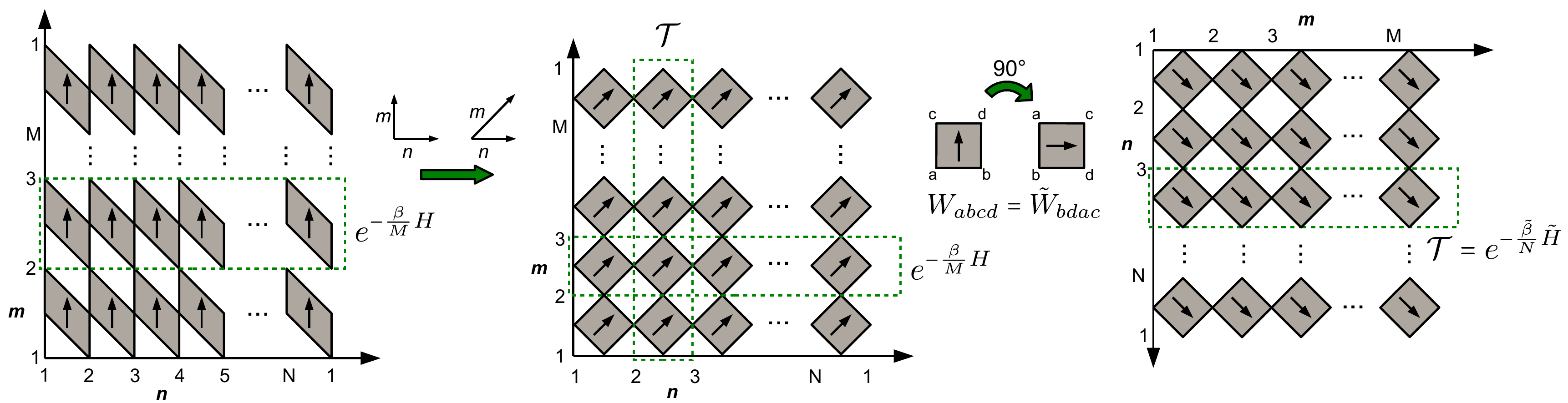}
 \caption{ 
 The system described by a Hamiltonian $H$ can be mapped onto a classical eight vertex model with statistical weigths $W$ determined by the model parameters by means of a Suzuki-Trotter decomposition. Here we choose a real space decomposition where the quantum transfer matrix $\mpstm{}{}$ is translation invariant on a slanted lattice. An effective model Hamiltonian $\tilde{H}$ is obtained by rotating the system by 90 degrees. The original QMT can then be written as $\mpstm{}{}=\exp(-\frac{\tilde{\beta}}{N}\tilde{H})$ and the effective model parameters can be extracted from the effective statistical weights $\tilde{W}$ which are related to the original statistical weigths $W$ through $\tilde{W}_{bdac}=W_{abcd}$.}
 \label{fig:qtm_rotation}
\end{figure}

The relevant question is thus how much of Euclidean invariance between real space and imaginary time remains in non-relativistic lattice systems. As a concrete example we consider the one-dimensional XYZ model on a chain with $N$ sites and periodic boundary conditions
\begin{equation}
H_{\rm XYZ}=-\sum_{j}J^{x}S_{j}^{x}S_{j+1}^{x} + J^{y}S_{j}^{y}S_{j+1}^{y} + J^{z}S_{j}^{z}S_{j+1}^{z}.
\label{eq:XYZ}
\end{equation}
By invoking a Suzuki-Trotter decomposition \cite{Trotter,Suzuki} the finite temperature partition function $Z_{\beta}$ of this model can be mapped onto a classical 2-dimensional eight-vertex model \cite{Takahashi,Baxter} of dimension $N\times M$, where the statistical weights $W$ of arrow configurations depend on the model parameters, inverse temperature $\beta$, the Trotter number $M$ and the type of chosen decomposition. Here we use a generalized Suzuki-Trotter decomposition $\rme^{-\delta H}=\rme^{-\delta\sum_{n}h_{n}}\approx \prod_{n}\rme^{-\delta h_{n}}$, with $\delta=\beta/M$. This leads to a real space decomposition introduced by Suzuki \cite{Suzuki}, where the quantum transfer matrix $\mpstm{}{}$ is translation invariant on a slanted lattice \cite{Betsuyaku} and we assume $M$ to be a multiple of $N$. The statistical weights are then given by $W_{abcd}=\braket{cd|\rme^{-\delta h_{n}}|ab}$\cite{Takahashi}. The same network is obtained after a rotation of the lattice by 90 degrees and considering an 
equivalent effective model on a chain of $M$ sites and Trotter number $N$ at an effective inverse temperature $\tilde{\beta}$. The QTM of the original lattice can therefore be written in terms of the effective model Hamiltonian $\tilde{H}$ as $\mpstm{}{}=\rme^{-\tilde{\delta}\tilde{H}}$ with $\tilde{\delta}=\tilde{\beta}/N$ and effective statistical weights $\tilde{W}$ up to a Trotter error in $N$. The effective model parameters and the effective inverse temperature $\tilde{\beta}$ can be obtained from the vertex weights after rotating, i.e. from $\tilde{W}_{bdac}=W_{abcd}$. For a graphical representation of this mapping see \Fig{fig:qtm_rotation}. From this relation the spectrum of the effective Hamiltonian can in principle be studied by looking at the QTM generated by the ground state of the original Hamiltonian.

Note that depending on the model and parameter regime, the effective parameters may also become complex, resulting in a non-hermitian effective Hamiltonian and thus accounting for non-hermitian transfer matrices. This particular fact was already discussed in the case of systems with Lorentz-invariance, as the continuum limit can often only be taken after blocking $N$ sites. In that case, all eigenvalues of the transfer matrix would have phases $\phi_{\alpha}=2\pi n/N$ with $n=0,\ldots,N-1$, and by defining $\tilde{H}=-\log(\mpstm{}{N})$ up to some energy scale, a hermitian effective Hamiltonian would be obtained. For systems with incommensurate order, where the eigenvalues can have arbitrary phases which are not fractions of $2\pi$, this is no longer possible.

\subsection{Truncation of the Virtual System}
\label{ss:RG}

In this subsection we show how to obtain an MPS approximation $\tilde{A}^{s}$ with finite bond dimension from the exact QTM constructed in the previous subsections. As was shown there, an exact MPS representation of the ground state can be constructed from imaginary time evolution, where the MPS matrices $A^s$ are given as a semi-infinite MPO with exponentially diverging virtual dimension (c.f. \Fig{fig:ImTimeTN}). This construction allows to identify the MPS-TM with the exact QTM at zero temperature. 

In the following we assume that the QTM can be written in terms of an effective local Hamiltonian $\tilde{H}=\sum_{n}\tilde{h}_{n}$ as $\mpstm{}{}=\rme^{-\tilde{H}}$, for instance via the construction of the previous subsection. In this representation, the MPS-TM appears to be completely translation invariant in the imaginary time direction, with no special role being played by the physical index s, where the matrices $A^{s}$ and $\bar{A}^s$ of the exact MPS representation of ket and bra are connected. The point is of course that, for expectation values of operators $O$ different from the identity, there would be an extra insertion at imaginary time $\tau=0$ corresponding to the operator transfer matrix $\mathcal{J}_{O}$ defined in \eq{eq:OpTM}. Here we have set the origin $\tau=0$ of the imaginary time axis at the point where $A^{s}$ and $\bar{A}^{s}$ are connected. 

We can then interchange the roles of real space and imaginary time and give two distinct new interpretations to the expectation value $\braket{\Psi|O(x)|\Psi}$. Firstly, we can interpret the new imaginary time direction as the evolution of a pure state of an infinite one-dimensional virtual system in the $x$ direction according to the MPS-TM $\mpstm{}{}$ or equivalently, the corresponding Hamiltonian $\tilde{H}$. At certain ``times'' $x$, there is an insertion of some impurity $O$ at the fixed coordinate $\tau=0$, which destroys the translation invariance of the virtual system. An alternative interpretation is given below while constructing an approximation with finite bond dimension of the exact ground state MPS $A^{s}$.

To arrive at a finite virtual dimension $D$, it is necessary to restrict the exponentially diverging amount of virtual degrees of freedom (DOF) of $A^{s}$ to a finite subset that is relevant for the physical DOF $s$ at $\tau=0$. We thus need to identify the relevant low-energy subspace of the virtual system to describe the evolution of the impurity at position $\tau=0$ as a function of $x$; the finite $D$ approximation $\tilde{A}^{s}$ is then obtained from the exact MPS $A^{s}$ by projecting onto this relevant subspace. 

The relevant low-energy subspace for an impurity problem can be obtained by applying real space RG transformations, as was first shown in the seminal work of Wilson using his numerical renormalization group (NRG) \cite{WilsonRG}. However, we here follow the more recent construction using the multiscale entanglement renormalization ansatz (MERA) \cite{MERA,MERA_rev,MERA_criticality,MERA_impurity1,MERA_impurity2}. This approach allows to identify the relevant degrees of freedom as those living at the causal cone of the impurity \cite{MERA_impurity1,MERA_impurity2}. For completeness, we repeat this argument for our specific case.

\begin{figure}[tb]
 \centering
 \includegraphics[width=0.6\linewidth,keepaspectratio=true]{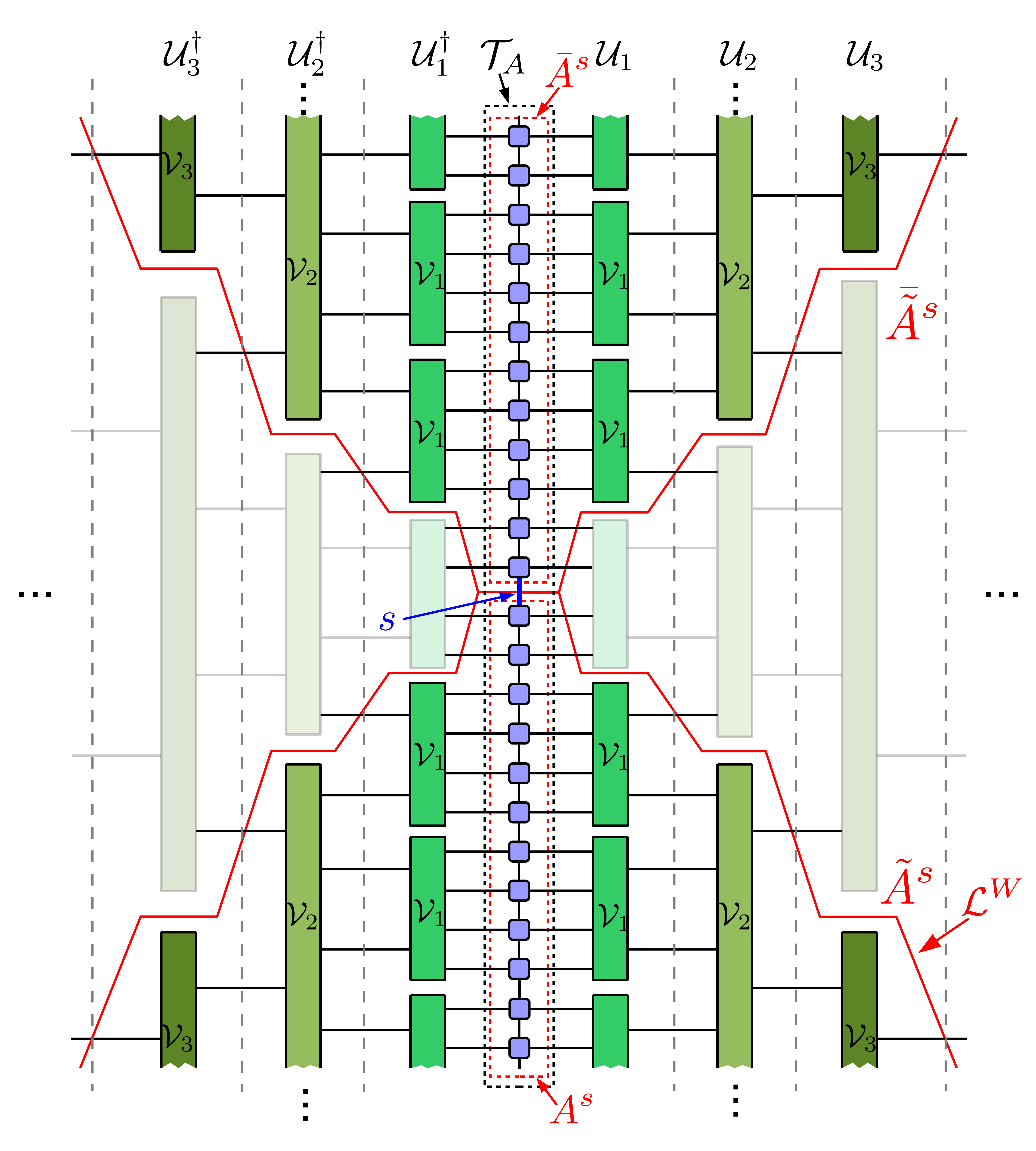}
 \caption{ 
 The transfer matrix $\mpstm{}{}$ is renormalized onto the space spanned by its dominant eigenvectors (i.e. the low energy subspace of the effective Hamiltonian $\tilde{H}$) through successive RG transformations $\mathcal{U}_{r}$ that can e.g. be represented by local tensors $\mathcal{V}_{s}$ (consisting of disentanglers and isometries) forming a MERA. The red solid line denotes the causal cone of the physical degree of freedom $s$, i.e. the boundary between the two semi-infinite parts of $\mpstm{}{}$, corresponding to the exact ground state MPS representation $A^{s}$. This boundary defines the Wilson chain $\mathcal{L}^{W}$, along which an effective impurity problem arises for the physical degree of freedom $s$. The finite $D$ approximation $\tilde{A}^{s}$ is obtained by contracting the RG network outside the causal cone (non-shaded region) to retain only degrees of freedom along the Wilson chain which are relevant for $s$.}
 \label{fig:TM_RG}
\end{figure}

The following construction starts from the assumption that there exists a sequence of real space RG transformations $\mathcal{U}_{r}$ that renormalizes $\tilde{H}$ onto its low-energy subspace, where $r$ labels the layers of successively applied RG transformations. Equivalently, $\mathcal{U}_{r}$ renormalizes $\mpstm{}{}$ onto the subspace spanned by its dominant eigenvectors. For concreteness, we consider a real space RG procedure that coarse grains four neighboring sites into two renormalized sites, which can be realized e.g. by a modified binary MERA. If $\tilde{H}$ is scale-invariant, the transformations become independent of the layer index $r$ after some amount of initial layers $r^{*}$. At this point all RG irrelevant terms have been removed and the renormalized Hamiltonian is a fixed point of the scale invariant RG transformation $\mathcal{U}$. For non-scale-invariant Hamiltonians the series of RG transformations terminates after $r_{\rm max}\approx\log(\mathcal{R})$ layers, 
where $\mathcal{R}$ is the 
dominant length scale of the Hamiltonian. At this point there are no relevant DOF left. These two cases can e.g. be represented by a scale-invariant or a finite-range MERA respectively \cite{MERA_rev}, an example of which is shown in \Fig{fig:TM_RG}.

The MERA construction allows to conclude that any perturbation at position $\tau=0$ can only affect the degrees of freedom living at its causal cone \cite{MERA_impurity2}. This causal cone is shown as red solid lines in \Fig{fig:TM_RG}. In particular, let us focus on one side of this causal cone, e.g. the lower half of \Fig{fig:TM_RG}. It is apparent that only one index of the RG network protrudes the boundary of the causal cone for each layer. We can therefore interpret these legs as the sites of an effective lattice system $\mathcal{L}^{W}$ defined along the boundary of the causal cone, which we call the \textit{Wilson chain}. We label the sites along this chain by the layer index $r$. Note that the $r$-th site on this chain is an effective renormalized description of $N_{r}\approx 2^{r}$ sites of the original lattice, reminiscent of the logarithmic discretization introduced by Wilson, hence the chosen nomenclature \cite{MERA_impurity1,MERA_impurity2,WilsonRG}. 

Contracting the RG network outside the causal cone (non-shaded region in \Fig{fig:TM_RG}) allows to renormalize the MPS-TM $\mpstm{A}{}$ into a new transfer matrix $\mpstm{}{W}$ along the Wilson chain $\mathcal{L}^{W}$ \cite{MERA_impurity1,MERA_impurity2}. It is immediately clear that $\mpstm{}{W}$ corresponds to the transfer matrix $\mpstm{\tilde{A}}{}$ of a new MPS with matrices $\tilde{A}^{s}$, which are obtained by projecting the exact MPS representation $A^{s}$ exactly onto this subspace of relevant DOF along the Wilson chain. For non-critical systems, the Wilson chain is of finite length $L$ and we obtain a virtual system with finite dimension $D\approx\chi^{r_{\rm max}}$, where $\chi$ is the bond dimension of the RG network. For critical, scale-invariant systems however, the Wilson chain is still infinite and any truncation to a finite system introduces some error. 

It is of course well known that ground states with finite correlation length $\xi$ can be well approximated by a finite $D$ MPS \cite{faithful}. By virtue of the previous sections, the effective Hamiltonian $\tilde{H}$ is gapped, where the gap is exactly equal to $\xi^{-1}$. It is interesting to contrast this interpretation with the more usual considerations regarding the gap of the actual physical Hamiltonian $H$ of the system. For Lorentz-invariant systems, $H$ and $\tilde{H}$ are equal and the correlation length is directly given by the inverse of the gap in units where the speed of light $c=1$. Without Lorentz-invariance, there is no obvious relation between $H$ and its Euclidean rotation $\tilde{H}$. The result of \cite{HastingsLR} allows to bound $\xi$ (and thus the gap of $\tilde{H}$) in terms of the gap $\Delta$ of $H$. When the low-energy behavior of $H$ still allows for an effective relativistic description, it can indeed be expected that $\xi\sim v/\Delta$ where $v$ replaces the speed 
of light with some characteristic speed of the system. However, there is also the possibility of an inherently non-relativistic low energy behavior, resulting in e.g. gapless hamiltonians $H$ for which the ground state is an exact MPS \cite{MPS-PhaseTrans} so that the corresponding $\tilde{H}$ is gapped.

Finally, we can obtain an alternative interpretation by applying the Jamio{\l}kowski isomorphism to map the pure state defined on the infinite one-dimensional virtual system of the previous discussion to the density matrix of a half-infinite one-dimensional virtual system with a boundary at $\tau=0$. By again interchanging the roles of real space and imaginary time, this half-infinite system undergoes dissipative evolution in the new time-direction $x$, corresponding to a Hamiltonian containing the terms of $\tilde{H}$ with support on the region $\tau>0$, and additional Lindblad operators corresponding to the action of $\tilde{H}$ across the boundary and the additional action of an operator $O$ inserted at certain ``times'' $x$. Hence, all the Lindblad operators are acting near the boundary, and the truncation of the bond dimension corresponds to selecting the relevant degrees of freedom to describe the boundary of the system. It is a virtue of the MERA that it naturally unifies the process of selecting the 
relevant degrees of freedom for boundaries and impurities \cite{MERA_impurity2}. The resulting virtual system becomes effectively zero-dimensional, and can be interpreted as providing a holographic description of the physical system \cite{cMPS-O}.

\section{Conclusions}
\label{s:conclusions}
In this paper we have investigated how much information about the excitation spectrum of a local translation invariant Hamiltonian can be obtained from local information and static correlations in the ground state. We have approached this question using the formalism of tensor network states in particular, but have also established several general results not restricted to tensor network formulations.

We have started by defining the regular and mixed tensor network transfer matrix for lattice and continuum models in \Sec{ss:tm} and \Sec{ss:syms_mixedtm}. We have then obtained tensor network approximations for the ground states of various prototypical quantum models on a lattice in one and two dimensions and (1+1)-dimensional field theories and studied the spectrum of eigenvalues $\lambda$ of the tensor network transfer matrix in \Sec{s:numerics}. There we have observed that the the complex arguments $\phi$ of the dominant eigenvalues correspond to the momenta $k_{\rm min}$ of the minima in the low energy dispersion of the system. Especially for critical models one can therefore easily determine the momenta for which there exist gapless excitations directly from ground state properties. We have used this to determine the value of the Fermi-momentum in the Kondo Lattice model studied in \cite{KLM_kfermi}, where no claims for zero temperature were possible. We have also related the logarithm of the absolute values $\varepsilon=-\log|\lambda|$ to the minimum excitation energies $E_{\rm min}$ by some characteristic velocity $E_{\rm min}=v\,\varepsilon$, which can be estimated by e.g. 
assuming a Lorentz-invariant low energy behavior (c.f. \Sec{ss:kl}). 

These observations are of large practical importance in the context of simulating quantum many body systems using tensor network techniques: it implies that already fairly accurate information about the structure of the low energy excitation spectrum can be obtained just from a variational ground state calculation. We have demonstrated that this is especially useful for two-dimensional systems using the PEPS formalism, for which no other efficient methods are presently known to extract information about excited states beyond the value of the gap. In particular we have investigated the AKLT model on a square and hexagonal lattice cylinder, where we have obtained a first approximation of the dispersion of the elementary excitations, for which there currently exist no other competitive numerical methods. Some of the authors have subsequently used this approach to study topological phase transitions and anyon condensation in the Toric Code model with string tension \cite{ShadowsAnyons}.

In \Sec{s:statcorr} we have gathered several arguments to explain how the eigenvalue spectrum of the transfer matrix affect the low-energy excitations of the model and vice versa. We have explained how a clustering of eigenvalues along lines of constant complex phase allows to recover the Ornstein-Zernike form of correlations in gapped phases if the distribution of eigenvalues becomes sufficiently dense in the limit $D\to\infty$
in \Sec{ss:oz}. 
Using the single mode approximation and a recent proof about the locality of elementary excitations \cite{JuthoLocalized}, we have argued why the phase of the eigenvalues along these lines correspond to minima of the dispersion relation of the elementary excitations in those models
in \Sec{ss:sma}. 
There we have also discussed that the full spectrum of the transfer matrix generally contains more information than static correlation functions of specific operators, as the transfer matrix is completely independent of the choice of operators. 
To approach this connection from the reverse direction
in \Sec{ss:kl}, 
we have first called on the assumption of a Lorentz-invariant low-energy behavior to identify the spectrum of the transfer matrix with a discrete version of the K\"{a}ll\'{e}n-Lehmann representation of correlation functions. Finally,
in \Sec{ss:corr_momentum} 
we have introduced momentum-resolved correlation functions by defining gaussian wave-packets of operators centered around a certain momentum $k$ in order to obtain a momentum resolved refinement of the celebrated result of Hastings \cite{HastingsLR} relating the correlation length in the system to the gap of the Hamiltonian. 

Furthermore we have identified the \textit{tensor network transfer matrix} of an exact tensor network representation of the ground state with the \textit{quantum transfer matrix}, appearing in path integral formulations of partition functions or ground states of quantum many-body systems, in \Sec{ss:imtime}. We have then argued how the quantum transfer matrix is related to the original Hamiltonian for system with and without Lorentz-invariance in \Sec{ss:qtm_ham}. We have demonstrated that for systems without Lorentz-invariance the quantum transfer matrix can be written in terms of an effective Hamiltonian with effective parameters which are related to the original Hamiltonian. These parameters can in principle also be complex, thus yielding non-hermitian effective Hamiltonians and non-hermitian transfer matrices. For systems with commensurate order, a hermitian effective Hamiltonian can be obtained by blocking several sites in constructing the transfer matrix. Based on this construction some of the authors have subsequently obtained an analytic form of such an exact MPS ground state representation for the case of the $S=1/2$ XY model in \cite{MarekXY}.

As a final point we have demonstrated in \Sec{ss:RG} how a tensor network approximation of the ground state with finite bond dimension $D$ can be obtained from the exact quantum transfer matrix through a renormalization process where the physical system acts as an impurity. More specifically, the tensor network transfer matrix stemming from a finite $D$ tensor network ground state approximation is a low energy representation of the exact quantum transfer matrix after applying several renormalization group transformations. Further details of this relation will be published elsewhere.

\section*{Acknowledgements}
We thank B.~Descamps, G.~Vidal, H.G.~Evertz and V.~Korepin
for inspiring and helpful discussions. 
We gratefully acknowledge support by EC grants SIQS and QUERG, the Austrian Science Fund (FWF):
F4104 SFB ViCoM and F4104 SFB FoQuS,
and an Odysseus grant by the Research Foundation Flanders. We further acknowledge support by
NCN grant 2013/09/B/ST3/01603 (M.M.R.), by the Alexander von Humboldt foundation and JARA-HPC through grant JARA0084 (N.S.),
the EPSRC under grant number EP/L001578/1 (V.S.)
and a Doctoral Scholarship by the Research Foundation Flanders (L.V.).

\newpage
\appendix

\section{Derivation of the bound on the decay of momentum-filtered correlation functions}
\label{a:proof}
This section contains the proof for \eqref{eq:corrk_bound1} and \eqref{eq:corrk_bound2} in \Sec{ss:corr_momentum}.
The following derivation is general for any spatial dimension and lattice geometry.

Let us first introduce the relevant notations and conventions. Throughout this appendix, we assume to be working on a $d$-dimensional lattice $\Lambda\subset \mathbb{R}^d$ generated by the primitive translation vectors $a_1,\ldots,a_d$. The unit cell has a volume $V_{\rm cell}=| \det[a_1| a_2| \cdots |a_d]|$. Arbitrary lattice sites are denoted as $x,y,\ldots\in\Lambda$; sets of sites are denoted as $X,Y,\ldots$ and the cardinality of a set $X$ is denoted as $| X |$. To every lattice site $x\in\Lambda$ we associate identical Hilbert spaces $\hilbert_x$; the Hilbert space of the whole system is $\hilbert_{\Lambda}=\bigotimes_{x\in\Lambda}
\hilbert_x$. This requires that we are working with a finite lattice, and we assume periodic boundary conditions with a period $p_i\in\mathbb{N}$ in the direction of lattice vector $a_i$, such that a site $x$ and $x+p_i a_i$ are identified for any $i$ (no summation). The lattice is thus given by the set of points $\Lambda=\{n_1 a_1+\ldots +n_d a_d \mid n_i =0,1,\ldots,p_i-1,\forall i=1,\ldots,d\}$. We will however be interested in the thermodynamic limit $p_i\to\infty, \forall\, i=1,\ldots,d$, since this scenario was used throughout the main text.

The reciprocal lattice $\tilde{\Lambda}$ consists of all vectors $K$ such that $\exp(\ic K \cdot x)=1$, $\forall x\in\Lambda$. In particular, we can define the reciprocal basis vectors $b_i$ satisfying $b_i \cdot a_j = 2\pi \delta_{i,j}$. The Fourier transform of a lattice function $f:\Lambda\to\mathbb{C}$ is defined as
\begin{equation}
F(k)=\sum_{x\in\Lambda} \ec^{-\ic k\cdot x} f(x)
\end{equation}
and satisfies $F(k+K)=F(k)$ for any $K\in\tilde{\Lambda}$. Hence, we can restrict to momenta $k\in\mathcal{B}$, where in the Brillouin zone $\mathcal{B}$ is the Wigner-Seitz unit cell of $\tilde{\Lambda}$. Because of the periodic boundary conditions, momentum space is discretized and can be identified with $\tilde{\Lambda}=\{n_1/p_1 b_1+\ldots+n_d/p_d b_d | n_1=0,\ldots,p_i-1,\forall i=1,\ldots,d\}$. Anticipating the thermodynamic limit and in order to harmonize the notation with the main text, we nevertheless denote the inverse Fourier transformation as
\begin{equation}
f(x)=\frac{1}{V_{\mathcal{B}}}\int_{\mathcal{B}} F(k)\ec^{\ic k \cdot x}\,\rmd k= \frac{V_{\rm cell}}{(2\pi)^d}\int_{\mathcal{B}} F(k)\ec^{\ic k \cdot x}\,\rmd k
\end{equation}
with $V_{\mathcal{B}}=(2\pi)^d/V_{\rm cell}$ the volume of the Brillouin zone.

By using the Euclidean scalar product to define $p\cdot x$, we can use the Euclidean distance as compatible lattice metric $\dist(x,y)=\| x-y\|$; the distance between two sets $X$, $Y$ is defined as $\dist(X,Y)=\min_{x\in X,y\in Y} \dist(x,y)$ and the diameter of a set $X$ is defined as $\diam(X)=\max_{x,y\in X}\dist(x,y)$. We introduce a shift operator $T^x$ for all $x \in\Lambda$ that shifts a state $\ket{\Psi}\in\hilbert_{\Lambda}$ over the lattice vector $x$. The Hamiltonian is given by $H_{\Lambda}=\sum_{X\subset \Lambda} H_X$ where the terms $H_X$ are supported on a subset $X$, such that $H_{\Lambda}$ is translation invariant
\begin{equation}
\forall x \in\Lambda: [T^x, H_{\Lambda}]=0
\end{equation}
and local, i.e., there exist positive constants $\mu$, $s$ for which
\begin{equation}
\sum_{X\ni x} \| H_X\| | X| \exp[\mu \diam(X)]\leq s\leq \infty.
\end{equation}
This allows to use Lieb-Robinson bounds \cite{LR,Nachtergaele}
\begin{equation}
\| [A_X(t),B_Y]\| \leq 2\| A_X\| \| B_Y \| | X| \rme^{-\mu \dist(X,Y)} \big(\rme^{2 s | t|}-1\big)
\label{eq:lrbound}
\end{equation}
for two operators $A_X$ and $B_Y$ supported on disjoint sets $X$ and $Y$. Furthermore, we assume that $H_{\Lambda}$ has a unique, translation invariant (i.e. momentum $k=0$) ground state $\ket{\Psi_0}\in\hilbert_{\Lambda}$ with ground state energy $0$. All eigenstates of $H$ with can be labeled by a momentum vector $k\in \mathcal{B}$ and an index $\alpha$ that labels all eigenstates within a given momentum sector. We denote these energy-momentum eigenstates as $\ket{\Phi_{k,\alpha}}$, with eigenenergies $E_{k,\alpha}$. The lowest excitation energy at momentum $k$ is given by $E(k)$. Note that $T^x \ket{\Phi_k}=\ec^{-\ic k \cdot x} \ket{\Phi_k}$ for every vector $\ket{\Phi_k}$ in the sector of momentum $k$.

For two operators $A_X$ and $B_Y$ supported on disjoint finite subsets $X,Y$, we define the static connected correlation function as
\begin{equation}
C=\braket{\Psi_0|A_X B_Y |\Psi_0}
\end{equation}
where we assume $A_X$ and $B_Y$ to have zero vacuum expectation value. We attempt to filter momentum-space information from this static correlation function by replacing $B_Y$ with a wave packet
\begin{equation}
\tilde{B}_Y(k) =  N_r \sum_{x\in\Lambda}\ec^{-\frac{\| x \|^2}{2 r}} \ec^{\ic k \cdot x} T^x B_Y T^{-x}
\end{equation}
where the normalization factor $N_r$ is given by $N_r= \left(\sum_{x\in\Lambda} \ec^{-\frac{\| x\|^2}{2 r}}\right)^{-1}$. We thus attempt to bound the magnitude of the momentum-filtered correlation function
\begin{equation}
C(k)=\braket{\Psi_0|A_X \tilde{B}_Y(k) |\Psi_0}
\end{equation} 
in the regime where $\dist(X,Y)$ is large.

\subsection{Proof}
We start by defining $\tilde{A}_X$ as
\begin{equation}
\tilde{A}_{X}=\frac{1}{2\pi} \int_{-\infty}^{+\infty}\frac{\ec^{\ic H t} A_X \ec^{-\ic H t}}{-\ic t+\epsilon} \ec^{-\frac{t^2}{2 q}}\,\rmd t
\end{equation}
and first show that
\begin{eqnarray}
\fl| \braket{\Phi_{k',\alpha}|\tilde{A}_X|\Psi_0}|&=
\left|\braket{\Phi_{k',\alpha}|A_X|\Psi_0}\right|\left| \frac{1}{2\pi}\int_{-\infty}^{+\infty}\rmd t\, \frac{\ec^{\ic E_{k',\alpha} t - \frac{t^2}{2q}} }{-it+\epsilon}\right|\nonumber\\
&=\left|\braket{\Phi_{k',\alpha}|A_X|\Psi_0}\right|\left| \frac{1}{2\pi}\int_{-\infty}^{-E_{k',\alpha}} \rmd E \int_{-\infty}^{+\infty}\rmd t\, \ec^{(-\ic t+\epsilon) E}\ec^{-\frac{t^2}{2q}}\,d t\right|\nonumber\\ 
&=\left|\braket{\Phi_{k',\alpha}|A_X|\Psi_0}\right|\left|\int_{-\infty}^{-E_{k',\alpha}} \rmd E \sqrt{\frac{q}{2\pi}} \ec^{-\frac{q
E^2}{2}}\right|\nonumber\\
&\leq c_{\rm erf} \exp\left(-\frac{q E_{k',\alpha}^2}{2}\right) \left|\braket{\Phi_{k',\alpha}|A_X|\Psi_0}\right|
\end{eqnarray}
and
\begin{eqnarray}
\fl| \braket{\Psi_0|A_X-\tilde{A}_X|\Phi_{k',\alpha}}|&=
|\braket{\Psi_0|A_X|\Phi_{k',\alpha}}|\left| 1- \frac{1}{2\pi}\int_{-\infty}^{+\infty}\rmd t\, \frac{\ec^{-\ic E_{k',\alpha} t - \frac{t^2}{2q}}}{-it+\epsilon}\right|\nonumber\\
&=|\braket{\Psi_0|A_X|\Phi_{k',\alpha}}|\left| 1- \frac{1}{2\pi}\int_{-\infty}^{E_{k',\alpha}} \rmd E \int_{-\infty}^{+\infty}\rmd t\, \ec^{(-\ic t+\epsilon) E}\ec^{-\frac{t^2}{2q}}\right|\nonumber\\
&=|\braket{\Psi_0|A_X|\Phi_{k',\alpha}}|\left|\int_{E_{k',\alpha}}^{+\infty} \rmd E \sqrt{\frac{q}{2\pi}} \ec^{-\frac{q E^2}{2}}\right|\nonumber\\
&\leq c_{\rm erf} \exp\left(-\frac{q E_{k',\alpha}^2}{2}\right) |\braket{\Psi_0|A_X|\Phi_{k',\alpha}}|,
\end{eqnarray}
where $c_{\rm erf}$ is a constant that allows to bound the error function by a Gaussian. We can use these inequalities to show that
\begin{eqnarray}
\| \tilde{A}_{X}\ket{\Psi_0} \| &= \left(\int_\mathcal{B}\rmd k' \sum_{\alpha} | \braket{\Phi_{k',\alpha}|\tilde{A}_X|\Psi_0}|^2\right)^{1/2}\nonumber\\
&\leq c_{\rm erf}\| A_X\ket{\Psi_0}\| \leq c_{\rm erf}\| A_X\|\label{eq:norm1}
\end{eqnarray}
and similarly using the triangle inequality
\begin{equation}
\| \bra{\Psi_0} \tilde{A}_{X} \| \leq \| \bra{\Psi_0} (\tilde{A}_{X}-A_X) \| +\| \bra{\Psi_0} A_{X} \| \leq (c_{\rm erf}+1)\|
A_X\|.\label{eq:norm2}
\end{equation}

We now write $A_X \tilde{B}_Y(k)$ as $[\tilde{A}_X,\tilde{B}_Y(k)]+(A_X-\tilde{A}_X) \tilde{B}_Y(k) + \tilde{B}_Y(k) \tilde{A}_X$ and again use the triangle inequality to bound $|
C(k) |$ as
\begin{equation}
| C(k)| \leq | \tilde{C}(k) | + | \braket{\Psi_0|(A_X-\tilde{A}_X) \tilde{B}_Y(k)|\Psi_0}|+ | \braket{\Psi_0| \tilde{B}_Y(k)
\tilde{A}_X|\Psi_0}|,
\label{eq:bound1}
\end{equation}
where we have defined a new correlator $\tilde{C}(k)=\braket{\Psi_0|[\tilde{A}_X, \tilde{B}_Y(k)]|\Psi_0}$.

For both the second and third term on the right hand side of \eq{eq:bound1}, we introduce a resolution of the identity, which we separate into two parts, one coming from momentum sectors with momentum $k'$ satisfying $\| k'-k\| \leq \delta$ (for the second term) or $\| k'+k\| \leq \delta$ (for the third term), and one coming from the rest. For the latter contribution, we use the Cauchy-Schwarz inequality to write for e.g. the third term
\begin{eqnarray*}
\left| \int_{\| k'+k\| > \delta} \rmd k' \sum_{\alpha}\, \braket{\Psi_0| \tilde{B}_Y(k)|\Phi_{k',\alpha}}\braket{\Phi_{k',\alpha}|
\tilde{A}_X|\Psi_0}\right|\\
\fl\leq \left( \int_{\| k'+k\| > \delta} \rmd k' \sum_{\alpha}\, |\braket{\Psi_0| \tilde{B}_Y(k)|\Phi_{k',\alpha}}|^{2}\right)^{1/2} \left( \int_{\|
k'+k\| > \delta} \rmd k' \sum_{\alpha}\, |\braket{\Phi_{k',\alpha}| \tilde{A}_X|\Psi_0}|^2\right)^{1/2}.
\end{eqnarray*}
By observing that
\begin{eqnarray}
|\braket{\Psi_0| \tilde{B}_Y(k)|\Phi_{k',\alpha}}|&=|\braket{\Psi_0| B_Y|\Phi_{k',\alpha}}| \left| N_r \sum_{x\in\Lambda}\ec^{\ic (k'+k)\cdot x -
\frac{\| x\|^2}{2 r}}\right|\nonumber \\
&\leq c_{\rm gauss}\,\ec^{-\frac{r\| k+k'\|^2}{2}} |\braket{\Psi_0| B_Y|\Phi_{k',\alpha}}|,
\end{eqnarray}
where $c_{\rm gauss}$ is defined in \ref{ss:gauss_bound}, the first factor can be bounded as
\begin{eqnarray*}
\int_{\| k'+k\| > \delta} \rmd k' \sum_{\alpha}\, |\braket{\Psi_0| \tilde{B}_Y(k)|\Phi_{k',\alpha}}|^{2}\\
\leq c_{\rm gauss}^2 \int_{\| k'+k\| >
\delta} \rmd k' \sum_{\alpha}\, \ec^{-r\| k+k'\|^2} |\braket{\Psi_0| B_Y|\Phi_{k',\alpha}}|^{2}\\
\leq c_{\rm gauss}^2 \,\ec^{-r \delta^2} \int_{\| k'+k\| > \delta} \rmd k' \sum_{\alpha}\, |\braket{\Psi_0| B_Y|\Phi_{k',\alpha}}|^{2}\\
\leq c_{\rm gauss}^2\, \ec^{-r \delta^2} \| B_Y\|^2,
\end{eqnarray*}
whereas for the second factor we use
\begin{eqnarray*}
\int_{\| k'+k\| > \delta} \rmd k' \sum_{\alpha}\, |\braket{\Phi_{k',\alpha}| \tilde{A}_X|\Psi_0}|^2\\ 
\leq c_{\rm erf}^2 \int_{\| k'+k\| > \delta} \rmd k' \sum_{\alpha}\, \ec^{-q E_{k',\alpha}^2} |\braket{\Phi_{k',\alpha}| A_X|\Psi_0}|^2\\
\leq c_{\rm erf}^2 \,\| A_X\|^2.
\end{eqnarray*}
From the above we thus conclude that
\begin{equation}
\fl\left| \int_{\| k'+k\| > \delta} \rmd k' \sum_{\alpha}\, \braket{\Psi_0| \tilde{B}_Y(k)|\Phi_{k',\alpha}}\braket{\Phi_{k',\alpha}|
\tilde{A}_X|\Psi_0}\right|\leq c_{\rm gauss}\, c_{\rm erf}\, \ec^{-\frac{r\delta^2}{2}}\| A_X\| \| B_Y\|.
\end{equation}
An identical contribution is obtained from the momentum region $\| k'-k\|>\delta$ if the resolution of the identity is inserted in the second term of the right hand side of \eq{eq:bound1}. 

For the momentum region $\| k'+k\| \leq \delta$ in the third term, we use the same approach with the Cauchy-Schwarz inequality but we now bound the first factor by
\begin{equation*}
\int_{\| k'+k\| < \delta} \rmd k' \sum_{\alpha}\, |\braket{\Psi_0| \tilde{B}_Y(k)|\Phi_{k',\alpha}}|^{2}\leq c_{\rm gauss}^2\| B_Y\|^2
\end{equation*}
and the second factor by
\begin{eqnarray*}
\int_{\| k'+k\| > \delta} \rmd k' \sum_{\alpha}\, |\braket{\Phi_{k',\alpha}| \tilde{A}_X|\Psi_0}|^2\\
\leq c_{\rm erf}^2 \int_{\| k'+k\| > \delta}
\rmd k' \sum_{\alpha}\, \ec^{-q E_{k',\alpha}^2} |\braket{\Phi_{k',\alpha}| A_X|\Psi_0}|^2\leq c_{\rm erf}^2 \ec^{-q E^{\ast}(-k,\delta)^2} \|
A_X\|^2,
\end{eqnarray*}
where $E^{\ast}(k,\delta)=\min_{\| k'-k\|\leq \delta} E(k')$. Again, the contribution of $\| k'-k\| <\delta$ in the second term is evaluated completely analogously. 

We can thus rewrite \eq{eq:bound1} as
\begin{equation}
| C(k)| \leq | \tilde{C}(k) | + c_{\rm gauss}\, c_{\rm erf}\, \left[ 2\ec^{-\frac{r \delta^2}{2}} + \ec^{-\frac{q E^{\ast}(-k,\delta)^2}{2}}+
\ec^{-\frac{q E^{\ast}(k,\delta)^2}{2}}\right].
\end{equation}

To bound the new correlator $\tilde{C}(k)$, we replace the wave packet $\tilde{B}_Y(k)$ by a completely local version $\tilde{\tilde{B}}_Y(k)$ defined as
\begin{equation}
\tilde{\tilde{B}}_Y(k)=\sum_{x\in\Lambda \atop \| x\| \leq \ell } N_r \ec^{-\frac{\| x \|^2}{2 r}} \ec^{-\ic k \cdot x} T^x B_Y T^{-x},
\end{equation}
and thus define yet another correlator $\tilde{\tilde{C}}(k)$ as
\begin{equation}
\tilde{\tilde{C}}(k)=\braket{\Psi_0|[\tilde{A}_X, \tilde{\tilde{B}}_Y(k)]|\Psi_0}.
\end{equation}
The error in operator norm can be bounded by the triangle inequality as
\begin{eqnarray}
\| \tilde{B}_Y(k) - \tilde{\tilde{B}}_Y(k)\| &= \left\|\sum_{x\in\Lambda\atop \| x\| > \ell } N_r \ec^{-\frac{\| x \|^2}{2 r}} \ec^{-\ic k \cdot x} T^x B_Y T^{-x}\right\|,\nonumber\\
&\leq N_r \| B_{Y} \| \sum_{x\in\Lambda\atop \| x\| > \ell } \exp\left[-\frac{\| x\|^2}{2 r}\right]\nonumber\\
&\leq c_{\Lambda} \| B_Y\| \exp\left(-\frac{\ell^2}{2r}\right) 
\end{eqnarray}
where an accurate determination of $c_{\Lambda}$ requires detailed knowledge about the structure of the lattice $\Lambda$. We can thus write
\begin{eqnarray}
| \tilde{C}(k)|&\leq |\tilde{\tilde{C}}(k)| +| \braket{\Psi_0|\tilde{A}_Y\Big(\tilde{B}_Y(k)-\tilde{\tilde{B}}_Y(k)\Big)|\Psi_0}|\nonumber\\
&+\braket{\Psi_0|\Big(\tilde{B}_Y(k)-\tilde{\tilde{B}}_Y(k)\Big)\tilde{A}_Y|\Psi_0}|
\end{eqnarray}
and using the Cauchy-Schwarz inequality in the different terms and \eq{eq:norm1},\eq{eq:norm2}, we obtain
\begin{equation}
| \tilde{C}(k)| \leq |\tilde{\tilde{C}}(k)| +c_{\Lambda}(2 c_{\rm erf}+1)\ec^{-\frac{\ell^2}{2r}} \| A_X\| \| B_Y\|.
\end{equation}

Finally, to also bound $\tilde{\tilde{C}}(k)$, we separate the time integral in the definition of $\tilde{A}_X$ into two pieces. For $t<T$, we obtain
\begin{eqnarray}
\left| \frac{1}{2\pi} \int_{-T}^{+T} \rmd t\, \frac{\braket{\Psi_0|[\ec^{\ic H t} A_X \ec^{-\ic H t}, \tilde{\tilde{B}}_Y(k)]|\Psi_0}}{-\ic t+\epsilon}\ec^{-\frac{t^2}{2q}}\right| \nonumber\\
\leq \frac{2\| A_X\| \| \tilde{\tilde{B}}_Y(k)\| | X| \ec^{-\mu(\dist(X,Y)-\ell)}}{2\pi} \int_{-T}^{+T} \rmd t \,\frac{\ec^{2 s| t|}-1}{-\ic t+\epsilon}\ec^{-\frac{t^2}{2q}}\nonumber\\
\leq \frac{4\| A_X\| \| \tilde{\tilde{B}}_Y(k)\| | X| \ec^{-\mu(\dist(X,Y)-\ell)}}{2\pi} \int_{0}^{T} \rmd t \,\frac{\ec^{2 s t}-1}{t} \nonumber\\
\leq \frac{4\| A_X\| \| \tilde{\tilde{B}}_Y(k)\| | X| \ec^{-\mu(\dist(X,Y)-\ell)}}{2\pi} \int_{0}^{T} \rmd t \, 2s \ec^{2s t}\nonumber\\
\leq \frac{2}{\pi} \| A_X\| \| B_Y\| | X| \ec^{2s T-\mu(\dist(X,Y)-\ell)},
\end{eqnarray}
where we have also used
\begin{equation*}
\| \tilde{\tilde{B}}_Y(k)\| \leq \| B_Y\| \,| N_r\sum_{x\in\Lambda\atop \| x\| \leq \ell }  \ec^{-\frac{\| x \|^2}{2 r}}|\leq \|
B_Y \|.
\end{equation*}
For the contribution of $| t| >T$, we use the Gaussian in the integrand to bound
\begin{eqnarray*}
\| \tilde{A}_X^>\| &=\left\| \frac{1}{2\pi} \int_{| t|>T}\rmd t \, \frac{\ec^{\ic H t} A_X \ec^{-\ic H t}}{-\ic t+\epsilon} \ec^{-\frac{t^2}{2q}}\right\|\nonumber\\
&\leq \frac{\| A_X\|}{\pi T} \int_{T}^{+\infty} \rmd t \,\ec^{-\frac{t^2}{2q}} \leq \| A_X\|
\sqrt{\frac{2q}{\pi}}\frac{c_{\rm erf}}{T} \ec^{-\frac{T^2}{2q}} .
\end{eqnarray*}
The bound on $\tilde{\tilde{C}}(k)$ is thus given by
\begin{equation}
| \tilde{\tilde{C}}(k)| \leq \| A_X\| \| B_Y\|\left[ \frac{2}{\pi}| X| \ec^{2s T-\mu(\dist(X,Y)-\ell)} + \frac{2 c_{\rm erf}}{T}
\sqrt{\frac{2q}{\pi}} \ec^{-\frac{T^2}{2q}}\right].
\end{equation}

Finally, putting everything together, we obtain
\begin{eqnarray}
\frac{| C(k)|}{\| A_X\| \| B_Y\|} \leq c_{\rm gauss}\, c_{\rm erf}\, \left[ 2\ec^{-\frac{r \delta^2}{2}} + \ec^{-\frac{q E^{\ast}(-k,\delta)^2}{2}}+
\ec^{-\frac{q E^{\ast}(k,\delta)^2}{2}}\right]\nonumber\\
+c_{\Lambda}(2c_{\rm erf}+1) \ec^{-\frac{\ell^2}{2r}}+\frac{2}{\pi}| X| \ec^{2s T-\mu(\dist(X,Y)-\ell)} + \frac{2 c_{\rm erf}}{T} \sqrt{\frac{2r}{\pi}} \ec^{-\frac{T^2}{2q}}.
\end{eqnarray}
In this expression, we can tune the constants $r$, $\delta$, $q$, $\ell$ and $T$ as a function of $\dist(X,Y)$ and the characteristics of $E(k)$ around $k$, i.e. $E^{\ast}(k,\delta)$. It is clear that we want to impose the restrictions:
\begin{eqnarray}
\ell& < \dist(X,Y), & 2s T\leq \mu\Big(\dist(X,Y)-\ell\Big).
\end{eqnarray}
Let us start by setting $\ell = \alpha \dist(X,Y)$ with $\alpha <1$ and $T=\dist(X,Y)/v$ with $v>\frac{2s}{\mu(1-\alpha)}$. In addition, we choose $q=\beta \dist(X,Y)$ and $r=\gamma \dist(X,Y)$ so that
\begin{eqnarray}
\frac{| C(k)|}{\| A_X\| \| B_Y\|} \nonumber\\
\leq c_{\rm gauss}\, c_{\rm erf}\, \left[ 2\ec^{-\frac{\gamma \delta^2}{2}\dist(X,Y)} +
\ec^{-\frac{\beta E^{\ast}(-k,\delta)^2}{2}\dist(X,Y)}+ \ec^{-\frac{\beta E^{\ast}(k,\delta)^2}{2}\dist(X,Y)}\right]\nonumber\\
+c_{\Lambda}(2c_{\rm erf}+1) \ec^{-\frac{\alpha^2}{2\gamma}\dist(X,Y)}+\frac{2}{\pi}| X| \ec^{-[\mu(1-\alpha)-\frac{2s}{v}]\dist(X,Y)}\nonumber\\
+ \frac{2 c_{\rm erf}}{T} \sqrt{\frac{2r}{\pi}} \ec^{-\frac{1}{2\beta v^2}\dist(X,Y)}
\end{eqnarray}
We can now fine tune the remaining constant so as to have a similar decay in all exponentials. Assuming the system has reflection invariance, i.e. $E^{\ast}(k,\delta)=E^{\ast}(-k,\delta)$, we can therefore choose the constants such that
\begin{equation}
\mu(1-\alpha)-\frac{2s}{v}=\frac{1}{2\beta v^2}=\frac{\alpha^2}{2\gamma}=\frac{\gamma \delta^2}{2} =\frac{\beta E^{\ast}(k,\delta)^2}{2}.
\end{equation}
If there is no reflection invariance, the smaller of both $E^{\ast}(k,\delta)$ and $E^{\ast}(-k,\delta)$ will determine the slowest exponential decay and should be used in the equations above.

Clearly, we should set $\beta=[v E^{\ast}(k,\delta)]^{-1}$. We then obtain $\mu(1-\alpha)-2s/v= E^{\ast}(k,\delta)/(2 v)$ from which we determine the optimal velocity $v=[4s+E^{\ast}(k,\delta)]/[\mu(1-\alpha)]$. From the second equality we can fix $\gamma=\alpha/\delta$. The remaining equation is therefore $E^{\ast}(k,\delta)/(2v)=\alpha \delta/2$. Inserting the velocity, we obtain
\begin{equation*}
2\mu(1-\alpha) E^{\ast}(k,\delta)=\alpha \delta [4s+E^{\ast}(k,\delta)]
\end{equation*}
which determines $\alpha$ as
\begin{equation}
\alpha^{-1}=1+\frac{\delta}{2\mu} + \frac{2\delta s}{E^{\ast}(k,\delta)\mu}
\end{equation} 
and only leaves $\delta$ to be determined. Note that the restriction $\alpha<1$ is satisfied. The fastest exponential decay is obtained by maximizing
\begin{equation}
\alpha\delta=\delta\left(1+\frac{\delta}{2\mu} + \frac{2\delta s}{E^{\ast}(k,\delta)\mu}\right)^{-1}=\left(\frac{1}{\delta}+\frac{2s}{\mu E^{\ast}(k,\delta)}+\frac{1}{2\mu}\right)^{-1}.
\label{eq:xip_nonoptimized}
\end{equation}
If we want to minimize the denominator, there is a clear tradeoff since increasing $\delta$ decreases the first term and increases the second. If $k$ corresponds to a minimum of the dispersion relation $E(k)$, then the function $E^{\ast}(k,\delta)$ will be insensitive to $\delta$ in some region, and we can choose $\delta$ as large as possible within this region. However, if $k$ corresponds to a regular point where $\frac{\rmd E(k)}{\rmd k}(k)\neq 0$, then there is a direct effect from increasing $k$ to decreasing $E^{\ast}(k,\delta)$.

A more intuitive result is obtained if we treat the term coming from the Lieb-Robinson bound separately, as the decay properties of this term are specific to the details of the Hamiltonian. Let us assume that we only know about the existence of some maximal velocity of propagation $v_{\rm LR}$, such that for any $v_{\rm LR}| t |\leq
\dist(X,Y)$, we can write
\begin{equation}
\| [A_X(t),B_Y]\| \leq c_{\rm LR} \| A_X\| \| B_Y\| \exp\left(-\frac{\dist(X,Y)}{\xi}\right).\label{eq:lrbound2}
\end{equation}
Clearly, choosing $v_{\rm LR}$ larger results in a smaller $\xi$ (a quicker exponential decay of the Lieb-Robinson bound) and vice versa. In order for the Lieb-Robinson bound to be the smallest error in the proof in \cite{HastingsLR} of the exponential decay of correlations, we need to choose $v_{\rm LR}$ at least large enough such that
\begin{equation}
\frac{1}{\xi} \geq \frac{\Delta E}{2v_{\rm LR}}.\label{eq:lrvelocity}
\end{equation}

If we now use the new Lieb-Robinson bound of \eq{eq:lrbound2} in the above, we would again choose $\ell=\alpha \dist(X,Y)$ but we would need to fix $T=(1-\alpha)\dist(X,Y)/v_{\rm LR}$. Hence, we use a fixed velocity of propagation and do not optimize over it (essentially the parameter $\mu$ in \eqref{eq:xip_nonoptimized}). We will again try to have an equal decay in all exponentials, except for the one coming from the Lieb-Robinson bound, which we allow to decay faster. We thus obtain
\begin{eqnarray}
\frac{1-\alpha}{\xi} \geq \frac{(1-\alpha)^2}{2\beta v_{\rm LR}^2}=\frac{\alpha^2}{2\gamma}=\frac{\gamma \delta^2}{2}=\frac{\beta E^{\ast}(k,\delta)^2}{2}.
\end{eqnarray}
Setting $\gamma=\alpha/\delta$ and $\beta=(1-\alpha)/[v_{\rm LR}E^{\ast}(k,\delta)]$ reduces these equations down to
\begin{eqnarray}
\frac{1-\alpha}{\xi} \geq \frac{(1-\alpha) E^{\ast}(k,\delta)}{2 v_{\rm LR}}=\frac{\alpha \delta }{2}.
\end{eqnarray}
Clearly, the first inequality is trivially satisfied, since $E^{\ast}(k,\delta)\geq \Delta E$ and the fixed velocity $v_{\rm LR}$ satisfies \eq{eq:lrvelocity}. From the last equation, we obtain \begin{equation}
\alpha^{-1}=1+\frac{v_{\rm LR}\delta}{E^{\ast}(k,\delta)}
\end{equation} 
and the rate of the exponential decay in $\dist(X,Y)$ is given by
\begin{equation}
\alpha\delta=\left(\frac{1}{\delta}+\frac{v_{\rm LR}}{E^{\ast}(k,\delta)}\right)^{-1}.
\end{equation}
We can then optimize over $\delta$ to find an optimal decay.

\subsection{Bounds on Fourier Transforms of Gaussians}
\label{ss:gauss_bound}
We compute the discrete Fourier transform of a sampled Gaussian by inserting the inverse continuous Fourier transform
\begin{eqnarray}
G(k)&=N_r\sum_{x\in\Lambda} \rme^{-\ic k \cdot x} \exp\left(-\frac{\| x\|^2}{2 r}\right)\nonumber\\
&=N_r\left(\frac{r}{2\pi}\right)^{d/2} \int_{\mathbb{R}^d} \rmd k'\, \exp\left(-\frac{r\| k'\|^2}{2}\right)\sum_{x\in\Lambda} \rme^{\ic (k'-k) \cdot x}\nonumber\\
&=N_r\left(\frac{r}{2\pi}\right)^{d/2} \int_{\mathbb{R}^d} \rmd k'\, \exp\left(-\frac{r\| k'\|^2}{2}\right)V_{\mathcal{B}}\sum_{K\in\tilde{\Lambda}} \delta(k'-k+K)\nonumber\\
&=N_r\left(\frac{r}{2\pi}\right)^{d/2} V_{\mathcal{B}} \sum_{K\in\tilde{\Lambda}}\exp\left(-\frac{r\| k-K\|^2}{2}\right)
\end{eqnarray}
with $\tilde{\Lambda}$ the reciprocal lattice and $V_{\mathcal{B}}$ the volume of the Brillouin zone (which is the unit cell of $\tilde{\Lambda}$). Hence, the Fourier transform of
the sampled Gaussian is a sum of Gaussians centered around the different lattice points of the reciprocal lattice $\tilde{\Lambda}$. We are only interested in the value of $G(k)$
for $k\in\mathcal{B}$, so if $r$ is sufficiently large the contributions of the Gaussians around the points $K\neq 0$ will be very small. In general, there exists a constant
$c_{\rm gauss}$ so that we can bound $G(k)$ by
\begin{eqnarray}
| G(k)| &=N_r\left(\frac{r}{2\pi}\right)^{d/2} V_{\mathcal{B}} \exp(-r \| k\|^2/2) \sum_{K\in\tilde{\Lambda}}\exp\left(r k\cdot K-\frac{r\| K\|^2}{2}\right)\nonumber\\
&\leq N_r \left(\frac{r}{2\pi}\right)^{d/2} V_{\mathcal{B}} \exp(-r \| k\|^2/2) \sum_{K\in\tilde{\Lambda}}\exp\left[- \frac{r\| K\|^2}{2}+r \| K \| \max_{k\in \mathcal{B}} \| k\| \right]\nonumber\\
&\leq c_{\rm gauss} \exp(-r \| k \|^2/2).
\end{eqnarray}



\section*{References}
\begin{thebibliography}{999}
\bibitem{LR} E.H.~Lieb, D.~Robinson, 
\textit{The finite group velocity of quantum spin systems}, 
Comm. Math. Phys. {\bf 28}, 251 (1972)

\bibitem{Nachtergaele} B.~Nachtergaele, R.~Sims, 
\textit{Lieb-Robinson Bounds and the Exponential Clustering Theorem}, 
Comm. Math. Phys. {\bf 265}, 119 (2006)

\bibitem{HastingsLR} M.B.~Hastings, 
Phys. Rev. Lett. {\bf 93}, 140402 (2004)


\bibitem{faithful} F.~Verstraete, J.I.~Cirac, 
Phys. Rev. B {\bf 73}, 094423 (2006)

\bibitem{Nrepre} Y.K.~Liu, M.~Christandl, F.~Verstraete, 
Phys. Rev. Lett. {\bf 98}, 110503 (2007)

\bibitem{Suzuki} M.~Suzuki, 
Prog. Theor. Phys. {\bf 56}, 1454 (1976)

\bibitem{Betsuyaku} H.~Betsuyaku, 
Phys. Rev. Lett. {\bf 53}, 629 (1984); 
Prog. Theor. Phys. {\bf 73}, 319 (1985)

\bibitem{Nishino_2D1} T.~Nishino, 
J. Phys. Soc. Jpn. {\bf 64}, 3598 (1995)

\bibitem{Nishino_2D2} T.~Nishino, K.~Okunishi, 
J. Phys. Soc. Jpn. {\bf 65}, 891 (1996)

\bibitem{Nishino_3D} T.~Nishino, K.~Okunishi, 
J. Phys. Soc. Jpn. {\bf 67}, 3066 (1998)

\bibitem{Bursill} R.J.~Bursill, T.~Xiang, G.A.~Gehring, 
J. Phys. Condens. Matter {\bf 8}, L583 (1996)

\bibitem{WangXiang} X.~Wang, T.~Xiang, 
Phys. Rev. B {\bf 56}, 5061 (1997)

\bibitem{Shibata} N.~Shibata, 
J. Phys. Soc. Jpn. {\bf 66}, 2221 (1997)

\bibitem{Sirker} J.~Sirker, A.~Kl\"umper, 
Europhys. Lett. {\bf 60}, 262 (2002);
Phys. Rev. B {\bf 71}, 241101(R) (2005)

\bibitem{Huang} Y.-K.~Huang, P.~Chen, Y.-J.~Kao, T.~Xiang, 
Phys. Rev. B {\bf 89}, 201102(R) (2014)

\bibitem{MERA} G.~Vidal, 
Phys. Rev. Lett. {\bf 99}, 220405 (2007)

\bibitem{MERA_rev} G.~Evenbly, G.~Vidal, 
Phys. Rev. B {\bf 79}, 144108 (2009)

\bibitem{MERA_criticality} R.N.C.~Pfeifer, G.~Evenbly, G.~Vidal,
Phys. Rev. A {\bf 79}, 040301 (2009)

\bibitem{MERA_impurity1} G.~Evenbly et al., 
Phys. Rev. B {\bf 82}, 161107(R) (2010)

\bibitem{MERA_impurity2} G.~Evenbly, G.~Vidal,
J. Stat. Phys. {\bf 157}, 931 (2014)


\bibitem{FNW} M.~Fannes, B.~Nachtergaele, R.~Werner, 
Comm. Math. Phys. {\bf 144}, 443 (1992)

\bibitem{MPS-P} D.~P\'erez-Garc\'ia, F.~Verstraete, M.M.~Wolf, J.I.~Cirac,  
Quantum Inf. Comput. {\bf 7}, 401 (2007)

\bibitem{MPS-V} F.~Verstraete, V.~Murg, J.I.~Cirac, 
Adv. Phys. {\bf 57}, 143 (2008)

\bibitem{MPS-Scholl2} U.~Schollw\"ock, 
Ann. Phys. {\bf 326}, 96 (2011)

\bibitem{cMPS-F} F.~Verstraete, J.I.~Cirac, 
Phys. Rev. Lett. {\bf 104}, 190405 (2010)

\bibitem{cMPS-O} T.J.~Osborne, J.~Eisert and F.~Verstraete, 
Phys. Rev. Lett. {\bf 105}, 260401 (2010)

\bibitem{cMPS-J} J.~Haegeman, J.I.~Cirac, T.~Osborne, F.~Verstraete, 
Phys. Rev. B {\bf 88}, 085118 (2013)  

\bibitem{PEPS} F.~Verstraete, J.I.~Cirac, 
arXiv:cond-mat/0407066 (2004)

\bibitem{iTEBD} G.~Vidal, 
Phys. Rev. Lett. {\bf 98}, 070201 (2007)

\bibitem{TDVP} J.~Haegeman et al., 
Phys. Rev. Lett. {\bf 107}, 070601 (2011)

\bibitem{iPEPS} J.~Jordan et al., 
Phys. Rev. Lett. {\bf 101}, 250602 (2008)

\bibitem{syms} D.~P\'erez-Garc\'ia et al., 
Phys. Rev. Lett. {\bf 100}, 167202 (2008)

\bibitem{MerminWagner} N.D.~Mermin, H.~Wagner, 
Phys. Rev. Lett. {\bf 17}, 1133 (1966)

\bibitem{Coleman} S.~Coleman, 
Commun. Math. Phys. {\bf 31}, 259 (1973)

\bibitem{WangBreak} H.-L.~Wang, J.-H.~Zhao, B.~Li, H.-Q.~Zhou,
J. Stat. Mech. (2011) L10001

\bibitem{DamianExcitations} D.~Draxler et al., 
Phys. Rev. Lett. {\bf 111}, 020402 (2013)

\bibitem{JuthoExcitations} J.~Haegeman et al., 
Phys. Rev. B {\bf 85}, 100408 (2012)

\bibitem{XY_Katsura} S.~Katsura, 
Phys. Rev. {\bf 127}, 1508 (1962)

\bibitem{XY_Barouch} E.~Barouch, B.M.~McCoy, M.~Dresden, 
Phys. Rev. A {\bf 2}, 1075 (1970); 
E.~Barouch, B.M.~McCoy, 
Phys. Rev. A {\bf 3}, 786 (1971)

\bibitem{XY_BunderMcKenzie} J.E.~Bunder, R.H.~McKenzie, 
Phys. Rev. B {\bf 60}, 344 (1999)

\bibitem{Bethe} H.~Bethe, 
Z. Phys. A {\bf 71}, 205 (1931)

\bibitem{XXZBethe1} M.~Takahashi, M.~Suzuki, 
Prog. Theor. Phys. {\bf 48}, 2187 (1972)

\bibitem{Takahashi} M.~Takahashi, 
\textit{Thermodynamics of One-Dimensional Solvable Models}, 
Cambridge University Press, Cambridge, 1999

\bibitem{KLM} J.~Kondo,
Prog. Theor. Phys. {\bf 32}, 37 (1964)

\bibitem{LaurensScattering} L.~Vanderstraeten, J.~Haegeman, T.~Osborne, F.~Verstraete,
Phys. Rev. Lett. {\bf 112}, 257202 (2014) 

\bibitem{KLM_kfermi} Y.H.~Su, Q.H.~Xiao, T.~Xiang, Z.B.~Su,
J. Phys. Condens. Matter {\bf 16}, 5163 (2004)

\bibitem{Kemper} A.~Kemper, A.~Schadschneider,
Phys. Rev. B {\bf 68}, 235102 (2003)

\bibitem{LiebLiniger} E.H.~Lieb, W.~Liniger,
Phys. Rev. {\bf 130}, 1605 (1963)

\bibitem{VidFES} V.~Stojevic et al.,
arXiv:1401.7654 (2014)

\bibitem{PirvuFES} B.~Pirvu, G.~Vidal, F.~Verstraete, L.~Tagliacozzo,
Phys. Rev. B {\bf 86}, 075117 (2012)  

\bibitem{AKLT} I.~Affleck, T~Kennedy, E.H.~Lieb, H.~Tasaki, 
Comm. Math. Phys. {\bf 115}, 477 (1988)

\bibitem{TNRG} A.~Garcia-Saez, V.~Murg, T.-C.~Wei, 
Phys. Rev. B {\bf 88}, 245118 (2013)

\bibitem{ShadowsAnyons} J.~Haegeman, V.~Zauner, N.~Schuch, F.~Verstraete
arXiv:1410.5443 (2014)

\bibitem{OrnsteinZernike} L.S.~Ornstein, F.~Zernike, 
Proc. Acad. Sci. Amsterdam {\bf 17}, 795 (1914)

\bibitem{Kennedy} T.~Kennedy, 
Comm. Math. Phys. {\bf 137}, 599 (1991)

\bibitem{MarekXY} M.M.~Rams, V.~Zauner, J.~Haegeman, F.~Verstraete,
arXiv:1411.2607 (2014)

\bibitem{SMA1} R.P.~Feynman, 
Phys. Rev. {\bf 91},  1291 (1953); 
ibid. {\bf 91}, 1301 (1953); 
ibid. {\bf 94}, 262 (1954)

\bibitem{SMA2} R.P.~Feynman and M.~Cohen, 
Phys. Rev.  {\bf 102}, 1189 (1956) 

\bibitem{SMA3} S.M.~Girvin, A.H.~MacDonald, P.M.~Platzman, 
Phys. Rev. Lett. {\bf 54}, 581 (1985);
Phys. Rev. B {\bf 33}, 2481 (1986)

\bibitem{SMA4} D.P.~Arovas, A.~Auerbach and F.D.M.~Haldane, 
Phys. Rev. Lett. {\bf 60}, 531 (1988)

\bibitem{SMA5} E.S.~S{\o}rensen, I.~Affleck, 
Phys. Rev. B {\bf 49}, 15771 (1994)

\bibitem{JuthoLocalized} J.~Haegeman et al., 
Phys. Rev. Lett. {\bf 111}, 080401 (2013)

\bibitem{SchollIncom} U. Schollw\"ock, Th.~Jolicoeur, T.~Garel, 
Phys. Rev. B {\bf 53}, 3304 (1996)

\bibitem{Nomura1} K.~Nomura, 
J. Phys. Soc. Jpn. {\bf 72}, 476 (2003)

\bibitem{Nomura2} T.~Murashima, K.~Nomura, 
Phys. Rev. B {\bf 73}, 214431 (2006)

\bibitem{LaiSutherland1} G.V.~Uimin, 
JETP Lett. {\bf 12}, 225 (1970)

\bibitem{LaiSutherland2} C.K.~Lai, 
J. Math. Phys. {\bf 15}, 1675 (1974)

\bibitem{LaiSutherland3} B.~Sutherland, 
Phys. Rev. B {\bf 12}, 3795 (1975)

\bibitem{Susskind} L.~Susskind, 
Phys. Rev. D {\bf 16}, 3031 (1977)

\bibitem{Trotter} H.~F.~Trotter, 
Proc. Am. Math. Soc. {\bf 10}, 545 (1959)

\bibitem{TITrotter1} B. Pirvu, V. Murg, J.I. Cirac, F. Verstraete, 
New J. Phys. {\bf 12}, 025012 (2010)


\bibitem{Baxter} R.J.~Baxter, 
\textit{Exactly Solved Models in Statistical Mechanics}, 
Academic Press, London, 1982

\bibitem{WilsonRG} K.G.~Wilson, 
Rev. Mod. Phys. {\bf 47}, 773 (1975)





\bibitem{MPS-PhaseTrans} M.M.~Wolf, G.~Ortiz, F.~Verstraete, J.I.~Cirac, 
Phys. Rev. Lett. {\bf 97}, 110403 (2006)


\bibitem{CT1} M.~Suzuki, M.~Inoue, 
Prog. Theor. Phys. {\bf 78}, 787 (1987)

\bibitem{CT2} T.~Koma, 
Prog. Theor. Phys. {\bf 81}, 783 (1989)

\bibitem{CT3} E.~Farhi, S.~Gutmann, 
Ann. Phys. {\bf 213}, 182 (1992)

\end{thebibliography}
\end{document}